\documentclass[aps,prb,amsmath,amssymb,twocolumn,superscriptaddress,showpacs]{revtex4-2}
\usepackage{amsmath,amssymb}
\usepackage{graphicx}
\usepackage{appendix}
\usepackage{subcaption}
\usepackage[usenames,dvipsnames]{color}
\usepackage[colorlinks=true, citecolor=blue, linkcolor=blue, urlcolor=blue]{hyperref}
\usepackage{hyphenat}
\usepackage{ulem}
\usepackage{dsfont}

\usepackage{txfonts}
\usepackage{enumitem}
\usepackage[utf8]{inputenc}

\usepackage[nottoc,numbib]{tocbibind}

\newcommand{\ep}{{\epsilon}}

\newcommand{\omL}{{\omega}}

\allowdisplaybreaks

\begin{document}

\title{
Floquet Schrieffer-Wolff transform based on Sylvester equations
}

\author{Xiao Wang}
\affiliation 
{Department of Physics, University of Oxford, UK}
\affiliation 
{Max Planck Institute for the Structure and Dynamics of Matter, Luruper Chaussee 149, 22761 Hamburg, Germany}

\author{Fabio Pablo Miguel M\'endez-C\'ordoba}
\affiliation{The Hamburg Centre for Ultrafast Imaging, Luruper Chaussee 149, Hamburg D-22761, Germany}
\affiliation{Universität Hamburg, Luruper Chaussee 149, Gebäude 69, D-22761 Hamburg, Germany}

\author{Dieter Jaksch}
\affiliation 
{Department of Physics, University of Oxford, UK}
\affiliation{The Hamburg Centre for Ultrafast Imaging, Luruper Chaussee 149, Hamburg D-22761, Germany}
\affiliation{Universität Hamburg, Luruper Chaussee 149, Gebäude 69, D-22761 Hamburg, Germany}

\author{Frank Schlawin}
\email{frank.schlawin@mpsd.mpg.de}  
\affiliation 
{Max Planck Institute for the Structure and Dynamics of Matter, Luruper Chaussee 149, 22761 Hamburg, Germany}
\affiliation{The Hamburg Centre for Ultrafast Imaging, Luruper Chaussee 149, Hamburg D-22761, Germany}
\affiliation{Universität Hamburg, Luruper Chaussee 149, Gebäude 69, D-22761 Hamburg, Germany}

\date{\today}

\begin{abstract}
We present a Floquet Schrieffer Wolff transform (FSWT) to obtain effective Floquet Hamiltonians and micro-motion operators of periodically driven many-body systems for any non-resonant driving frequency. 
The FSWT perturbatively eliminates the oscillatory components in the driven Hamiltonian by solving operator-valued Sylvester equations with systematic approximations.  
It goes beyond various high-frequency expansion methods commonly used in Floquet theory, as we demonstrate with the example of the driven Fermi-Hubbard model. 
In the limit of high driving frequencies, the FSWT Hamiltonian reduces to the widely used Floquet-Magnus result.  
We anticipate this method will be useful for designing Rydberg multi-qubit gates, controlling correlated hopping in quantum simulations in optical lattices, and describing multi-orbital and long-range interacting systems driven in-gap. 

\end{abstract}
\maketitle

\bigskip

\section{Introduction}

External driving has become an irreplaceable tool 
for quantum information processing \cite{wendin2017quantum} and quantum simulations \cite{georgescu2014quantum}, as it allows to modify qubit interactions in platforms ranging from Rydberg atoms to Josephson junctions. 
In addition, it has also become invaluable for the control of many-body dynamics in 
cold atoms \cite{Eckardt2017,weitenberg2021tailoring}, correlated materials \cite{oka2019floquet,RevModPhys.93.041002}, multi-band electrons~\cite{rudner2020band,Lindner2010FloquetTI,PhysRevB.79.081406}, photonic wave-guides \cite{RevModPhys.91.015006}, and hybrid quantum systems \cite{mivehvar2021cavity,schlawin2022cavity}. 
Seminal experimental observations in cold atoms include, inter alia, 
light-induced gauge fields \cite{goldman2014light,RevModPhys.83.1523} and topology \cite{goldman2016topological,jotzu2014experimental}, 
Floquet-induced many-body localisation \cite{bordia2017periodically,singh2019quantifying}, 
discrete time crystals \cite{zhang2017observation,choi2017observation}, 
driving-induced superfluid-Mott transitions \cite{PhysRevLett.102.100403}, 
tuning of exchange interactions~\cite{gorg2018enhancement} and occupation-dependent tunneling \cite{meinert2016floquet} in cold atoms. 
In driven correlated solids, fascinating effects such as  
transient superconductivity \cite{doi:10.1126/science.1197294,mitrano2016possible, Budden2021, Rowe2023, doi:10.1080/00107514.2017.1406623, Buzzi2020, Buzzi2021}, 
dressed surface states in topological insulators \cite{wang2013observation,mahmood2016selective}, or 
the light-induced anomalous Hall effect \cite{mciver2020light} were shown to emerge.
Floquet topological insulators can be engineered in acoustic \cite{fleury2016floquet} and photonic \cite{rechtsman2013photonic} systems. 

Though not all of these effects can be captured by it, Floquet theory remains the central theoretical framework to understand driven materials: 
When the driving frequency is sufficiently detuned from any resonance of the many-body system, the system can only exchange energy virtually with the drive. 
In this situation, heating can be suppressed and long-lived pre-thermal phases with new functionalities stabilized~\cite{RevModPhys.93.041002}. 
When the driven Hamiltonian $\hat{H}_t$ is time-periodic, we can use Floquet theory \cite{bukov2015universal,Giovannini_2020,oka2019floquet,rodriguez2021low,mori2023floquet} to describe the secular dynamics and the pre-thermal state of the driven system.
A large number of expansion techniques have been derived within the framework of Floquet theory to describe driven many-body systems. These include 
the high-frequency expansion of Van-Vleck perturbation theory in Sambe space in Ref.~\cite{Eckardt_2015}, the Floquet-Magnus expansion \cite{casas2001floquet,mananga2011introduction}, Brillioun-Wigner theory~\cite{PhysRevB.93.144307},
the effective Hamiltonian method for multi-step driving sequences~\cite{goldman2014periodically}, the Sambe space flow equation approach \cite{PhysRevLett.111.175301} for driven Boson-Hubbard model, and the Schrieffer-Wolff transform 
of the driven Fermi-Hubbard model in the strongly correlated regime \cite{PhysRevLett.116.125301,PhysRevLett.120.197601,PhysRevB.98.035116}. 
All these methods are essentially high-frequency expansions (HFE) in orders of the inverse driving frequency $\omega$. 
When the energy scales in the undriven system are comparable to, or larger than, the driving frequency, the high-frequency expansion becomes inaccurate, necessitating the development of Floquet theories beyond HFE. So far, only a few Floquet methods beyond the HFE have been constructed ~\cite{Rodriguez_Vega_2018,Eckardt_2015} based on the perturbative (block-) diagonalisation of the Floquet Hamiltonian in Sambe space. 
Crucially, these approaches require knowledge of the eigenbasis of the undriven system and thus cannot be applied to interacting many-body systems. The Sambe-space Gaussian elimination method in Ref.~\cite{PhysRevB.101.024303} suffers from the same applicability issue.
Recently, a Flow equation approach beyond HFE was developed~\cite{PhysRevX.9.021037},
which doesn't require knowledge of the eigenbasis. Instead, the Floquet Hamiltonian is constructed based on consecutive infinitesimal unitary transforms. However, the flow-truncation error becomes unpredictable when multiple fixed points \cite{claassen2021flow} exist, and the micro-motion is hard to obtain, as it requires evaluating the unitary flow.

\begin{figure}
    \centering
    \includegraphics[width=0.45\textwidth]{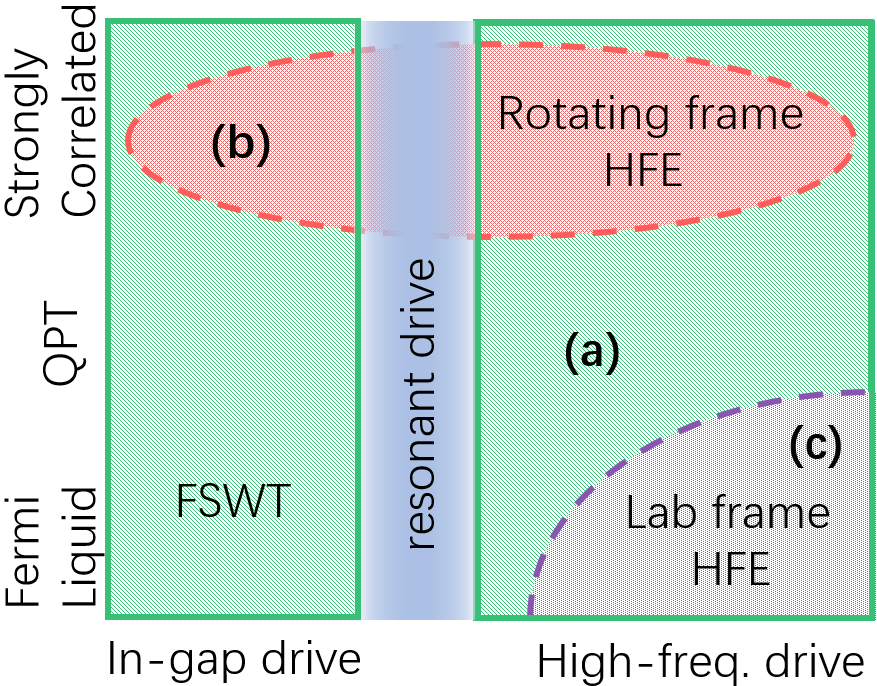}
    \caption{
    Sketch of the application regimes of different Floquet theories in terms of the driven system's interaction strength and the driving frequency relative to a system resonance: 
    Our lab-frame FSWT, marked by (a), provides a unified result that converges for all non-resonant driving regimes. The high-frequency expansion 
    encounters convergence issues when the oscillation of the Hamiltonian 
    becomes slow. The rotating-frame HFE in Ref.~\cite{PhysRevLett.116.125301}, labelled (b), diverges when the correlations are weak. The lab-frame HFE, labelled (c), diverges under the in-gap drive and converges slowly in the strongly correlated case when the interaction is comparable to the driving frequency. Dashed lines contain the regimes where the HFE methods have better convergence. Note, however, that HFE may still be valid outside these regimes, provided that the driving frequency is the largest energy scale and higher HFE orders are included.  
    }
    \label{fig:illustration}
\end{figure}

In this work, we develop a Floquet Schrieffer Wolff transform (FSWT) which can be applied beyond the HFE. This method constructs an effective Floquet Hamiltonian which is generated perturbatively in the driving strength $g$ rather than the inverse driving frequency. 
It is, therefore, particularly well suited to describe low-frequency or in-gap driving of multi-orbital systems, as illustrated in Fig.~\ref{fig:illustration}. 
The FSWT derived in this work is conceptually equivalent to a time-dependent transform previously used in circuit QED systems \cite{blais2007quantum,ann2022two}, which perturbatively eliminates the time-dependence in the two-body Hamiltonian of the driven Jaynes-Cummings model. 
Here, we use the derivative of the exponential map to formulate this transform as a systematic many-body perturbation theory based on Sylvester equations, which we can solve with systematic approximations for driven many-body systems without knowledge of their eigenbasis. 
We illustrate our method with the example of a driven Hubbard chain. We demonstrate the importance of correlated hopping terms, whose strength is underestimated in HFE. They can substantially change the Floquet-induced metal-insulator transition when the driving frequency is not too large. 
Using infinite density matrix renormalization group (DMRG) simulations~\cite{orus2019tensor} and exact diagonalisation, we benchmark the FSWT method and find substantial improvement compared to HFE Floquet approaches in large regions of parameter space (corresponding to the right part of Fig.~\ref{fig:illustration}). 

The paper is organised as follows: In Section \ref{Floq-Schrieffer-Wolff}, we construct our Floquet Schrieffer Wolff transform for general periodically driven systems. 
We perturbatively generate the underlying Sylvester equations and explain how to find the Floquet Hamiltonian and micro-motion operators in our FSWT. In Section \ref{compare}, we compare the FSWT with other Floquet methods, such as the HFE, 
the Sambe space van Vleck block-diagonalisation, and the Gaussian-elimination methods. 
We comment on the occurrence of Floquet-mediated interactions, which appear in other methods suitable for low-frequency driving. 
To demonstrate the usefulness of our method, in Section \ref{example}, we apply the FSWT to a monochromatically driven Hubbard system and find a Floquet Hamiltonian which remains valid for all ratios of $U/J$. We conclude and discuss the possible future applications of FSWT in Section \ref{conclude}.

\section{The Floquet Schrieffer-Wolff transform for multi-frequency driving}\label{Floq-Schrieffer-Wolff}

We first outline the general procedure for time-dependent unitary transform, which forms the basis of our FSWT. The evolution operator from time $t_0$ to $t$ under any time-dependent Hamiltonian $\hat{H}_t$ is denoted by $\hat{\mathcal{U}}_{t,t_0}$. It satisfies the Schr\"odinger equation $i\partial_t \hat{\mathcal{U}}_{t,t_0} = \hat{H}_t \hat{\mathcal{U}}_{t,t_0}$ where we set $\hbar=1$. Using an arbitrary time-dependent unitary transform $\hat{U}_t$, the evolution operator can be decomposed as $\hat{\mathcal{U}}_{t,t_0} = \hat{U}_t^{\dag} \hat{\mathcal{U}}'_{t,t_0} \hat{U}_{t_0}$. The evolution operator $\hat{\mathcal{U}}'_{t,t_0}$ satisfies another Schrodinger equation $i\partial_t \hat{\mathcal{U}}'_{t,t_0} = \hat{H}'_t \hat{\mathcal{U}}'_{t,t_0}$ with the transformed Hamiltonian $\hat{H}'_t = \hat{U}_t \hat{H}_t \hat{U}_t^{\dag} + i (\partial_t \hat{U}_t)\hat{U}_t^{\dag}$. By properly choosing $\hat{U}_t$, the evolution can be greatly simplified, such that $\hat{H}'_t$ becomes (approximately) time-independent, and the time-ordering becomes unnecessary when computing $\hat{\mathcal{U}}_{t,t_0}$. 

\subsection{General Formalism}\label{subsec:general-formalism}
We consider a general driven system described by the following time-dependent Hamiltonian
\begin{equation}\label{H_t}
    \hat{H}_{t} = \hat{H}^{(0)} + \sum\limits_{n=1}^{\infty} \hat{H}_t^{(n)} 
\end{equation}
Here, we assume the driving strength is characterized by a quantity $g$ (which can be proportional to the electric field's amplitude during the laser pulse), 
and $\hat{H}_t^{(n)} \sim g^n$. 
Each $\hat{H}_t^{(n)}$ term may contain both time-dependent and static terms (which appear in, e.g. the co-rotating frame). 
$\hat{H}^{(0)}$ is the undriven Hamiltonian with $\mathcal{O}(g^0)$, which is time-independent. Thus the time-dependence starts at order $\mathcal{O}(g^1)$ in $\hat{H}_t$.

In the following, we aim to find a time-dependent unitary transform $\hat{U}_t$ which eliminates the time-dependency 
in $\hat{H}_{t}$ order by order in $g$, such that the remaining time-dependence in the transformed Hamiltonian $\hat{H}'_t$ only contains higher orders of driving strength $g$. Therefore, the residual time-dependence is much weaker than its original strength $g^1$ and can be neglected, $H'_t \simeq H'$. 
The evolution operator of the driven system, 
\begin{align} \label{eq.propagator}
\hat{\mathcal{U}}_{t,t_0} = \hat{U}_t^{\dag} e^{-i\hat{H}'(t-t_0)} \hat{U}_{t_0}, 
\end{align}
is greatly simplified.
For time-periodic driving, which we analyze in Sec.~\ref{sec.time-periodic}, this procedure is equivalent to block-diagonalizing the Sambe space Hamiltonian.

Analogous to the Schrieffer-Wolff transform in a time-independent system, the form of the time-dependent unitary transform $\hat{U}_t$ is chosen to be
\begin{equation}
    \hat{U}_t = e^{\hat{F}_t}
\end{equation}
where $\hat{F}_t$ is an anti-Hermitian operator. Using the derivative of the exponential map, the transformed Hamiltonian $\hat{H}'_t$ reads
\begin{equation}\label{H'}
\begin{split}
        \hat{H}'_t &= \hat{U}_t \hat{H}_{t} \hat{U}_t^{\dag} + i (\partial_t \hat{U}_{t}) \hat{U}_t^{\dag} \\
        &= \hat{H}_{t} + \hat{G}_{t} + \frac{1}{2!} [\hat{F}_t,\hat{G}_t] + \frac{1}{3!} [\hat{F}_t,[\hat{F}_t,\hat{G}_t]] + ...
\end{split}
\end{equation}
where the operator $\hat{G}_t$ is defined as
\begin{equation}\label{G_t}
    \hat{G}_{t} = [\hat{F}_{t},\hat{H}_{t}] + i \partial_t \hat{F}_{t}.
\end{equation}
Eq.~(\ref{H'}) has been used in previous high-frequency expansions
\cite{goldman2014periodically,abanin2017effective}; however, below, we do not make use of the inverse-frequency expansion therein. 
Instead, we expand $\hat{F}_t$ in orders of driving strength $g$
\begin{equation}\label{Ft-expand}
    \hat{F}_t = \sum\limits_{n=1}^{\infty}  \hat{F}_t^{(n)} ,
\end{equation}
where $\hat{F}_t^{(n)} \sim g^n$. 
We require the unitary $\hat{U}_t$ generating the Floquet micro-motion to ($\bf{i}$) preserve the undriven dynamics and ($\bf{ii}$) preserve the macro-motion
\footnote{
In the evolution operator described by Eq.~(\ref{eq.propagator}), the middle part, $e^{-i\hat{H}'(t-t_0)}$, will be identified in Floquet theory as the stroboscopic dynamics, i.e., the Floquet macro-motion.
}
. Thus, we set
\begin{equation}\label{ansatz}
    \hat{F}^{(0)}_t = 0 ~\text{ and } \lim\limits_{T\to\infty} \frac{1}{2T} \int_{-T}^{T} dt ~ \hat{F}^{(n)}_t = 0 ,
\end{equation}
which guarantees that $\hat{F}_t$ vanishes if there is no drive, i.e. $g\to0$, and that $\hat{F}_t$ has no static part.
We next expand the transformed Hamiltonian (\ref{H'}) in powers of $g$, 
\begin{equation}
\hat{H}'_t = \sum\limits_{n=0}^{\infty} \hat{H}'^{(n)}_t    
\end{equation}
where $\hat{H}'^{(n)}_t \sim g^n$. Inserting Eqs.~(\ref{G_t}) and (\ref{Ft-expand}) into (\ref{H'}) we collect $\hat{H}'^{(n)}_t$ order by order
\begin{widetext}

\begin{subequations}\label{H'-expansion}
\begin{align}
\hat{H}'^{(0)}_t &= \hat{H}^{(0)} \label{H'-expansion-0} \\
\hat{H}'^{(1)}_t &= \hat{H}^{(1)}_t + [\hat{F}^{(1)}_t,\hat{H}^{(0)}] + i \partial_t \hat{F}^{(1)}_t \label{H'-expansion-1} \\
\hat{H}'^{(2)}_t &= \hat{H}^{(2)}_t + \frac{1}{2}[\hat{F}^{(1)}_t,\hat{H}^{(1)}_t + \hat{H}'^{(1)}_t] + [\hat{F}^{(2)}_t,\hat{H}^{(0)}] + i \partial_t \hat{F}^{(2)}_t \label{H'-expansion-2} \\
\hat{H}'^{(3)}_t &= \hat{H}^{(3)}_t + \frac{1}{2}[\hat{F}^{(1)}_t,\hat{H}^{(2)}_t + \hat{H}'^{(2)}_t] + \frac{1}{2}[\hat{F}^{(2)}_t,\hat{H}^{(1)}_t + \hat{H}'^{(1)}_t]  
+ \frac{1}{12} [\hat{F}^{(1)}_t,[\hat{F}^{(1)}_t, \hat{H}^{(1)}_t -\hat{H}'^{(1)}_t  ]] \label{H'-expansion-3} \\
& 
+ [\hat{F}^{(3)}_t,\hat{H}^{(0)}] + i \partial_t \hat{F}^{(3)}_t  \notag  \\
\hat{H}'^{(4)}_t  &= \hat{H}^{(4)}_t + \frac{1}{2}[\hat{F}^{(1)}_t,\hat{H}^{(3)}_t + \hat{H}'^{(3)}_t]  
+ \frac{1}{12} [\hat{F}^{(1)}_t,[\hat{F}^{(1)}_t, \hat{H}^{(2)}_t -\hat{H}'^{(2)}_t  ]]
+ \frac{1}{2}[\hat{F}^{(2)}_t,\hat{H}^{(2)}_t + \hat{H}'^{(2)}_t] \label{H'-expansion-4}
 \\
&+ \frac{1}{2}  [\hat{F}^{(3)}_t  ,\hat{H}^{(1)}_t + \hat{H}'^{(1)}_t] 
+ \frac{1}{12} [\hat{F}^{(1)}_t,[\hat{F}^{(2)}_t, \hat{H}^{(1)}_t -\hat{H}'^{(1)}_t  ]]
+ \frac{1}{12} [\hat{F}^{(2)}_t,[\hat{F}^{(1)}_t, \hat{H}^{(1)}_t -\hat{H}'^{(1)}_t  ]]
+ [\hat{F}^{(4)}_t,\hat{H}^{(0)}] + i \partial_t \hat{F}^{(4)}_t  \notag 
\end{align}
\end{subequations}
\end{widetext}
Now, we can determine the functions
$\hat{F}_t^{(n)}$ for each order $n=1,2,3,...$, which eliminates the time-dependence at order $g^n$ in the transformed Hamiltonian $\hat{H}'$. This is achieved iteratively: the time-
requirement $\partial_t \hat{H}'^{(n)}_t =  0$ determines $\hat{F}_t^{(n)}$, and $\hat{F}_t^{(n)}$ in turn determines the static part of $\hat{H}'^{(n+1)}_t$~\footnote{
To detail this iteration, we start at $n=1$ in Eq.~(\ref{H'-expansion}): The requirement
$\partial_t \hat{H}'^{(1)}_t =  0$ on (\ref{H'-expansion-1})
eliminates the time-dependence to order $g^1$ in $\hat{H}'$, from which $\hat{F}_t^{(1)}$ is determined. Inserting $\hat{F}_t^{(1)}$ back to Eq.~(\ref{H'-expansion}) not only renders the entire $\hat{H}'^{(1)}_t$ static in (\ref{H'-expansion-1}), but also completely determines the static part of $\hat{H}'^{(2)}_t$ in (\ref{H'-expansion-2}). Then we proceed to $n=2$: Using the function $\hat{F}_t^{(1)}$ derived above, the equation
$\partial_t \hat{H}'^{(2)}_t =  0$ for (\ref{H'-expansion-2})
eliminates the time-dependence to order $g^2$ in $\hat{H}'$, determining $\hat{F}_t^{(2)}$. 
This elimination procedure can be repeated iteratively. Once the $n$-th order equation in Eq.~(\ref{H'-expansion}) is solved, the function $\hat{F}_t^{(n)}$ will be determined which fixes the static part of $\hat{H}'^{(n+1)}_t$.}.

\subsection{Time-periodic driving}
\label{sec.time-periodic}
The above perturbative procedure works for arbitrary time-dependent Hamiltonians. Below, we assume the driven Hamiltonian $\hat{H}_t = \hat{H}_{t+2\pi/\omega}$ is time-periodic with angular frequency $\omega$. 
We can Fourier transform each order of its driving terms in Eq.~(\ref{H_t}), such that (for all $n\geq1$)
\begin{equation}\label{Ht-Fourier}
\hat{H}^{(n)}_t = \sum\limits_{j=-\infty}^{\infty} \hat{H}^{(n)}_j e^{i j\omega t}
\end{equation}
where we allow $\hat{H}^{(n)}_t$ to contain multi frequency components at $j \omega$, with $j$ an arbitrary integer. 
Here $\hat{H}^{(n)}_{j=0}$ denotes the time-independent driving terms (e.g. the resonant driving term in the rotating frame). 
In this case, we find $\hat{F}_t$ also becomes time-periodic, $\hat{F}_t = \hat{F}_{t+2\pi/\omega}$, and we expand its $n$-th order into a Fourier series
\begin{equation}\label{Ft-Fourier}
    \hat{F}^{(n)}_t = \sum\limits_{j=-\infty}^{\infty}  \hat{f}^{(n)}_j e^{i j \omega t},
\end{equation}
where $\hat{f}_{j}^{(n)} \propto (1-\delta_{n,0}) (1-\delta_{j,0})$ according to our ansatz (\ref{ansatz}), and the anti-Hermiticity of $\hat{F}_t$ means $\hat{f}_j^{(n)} = - (\hat{f}^{(n)}_{-j})^{\dag}$. 

Inserting Eqs.~(\ref{Ht-Fourier}) and (\ref{Ft-Fourier}) into Eq.~(\ref{H'-expansion}), we can determine $\hat{f}_j^{(n)}$ for each order $n=1,2,3,...$, which eliminates the oscillatory component in $\hat{H}'^{(n)}_t$. 

\subsubsection{lowest order Floquet Hamiltonian}
For $n=1$, the equation $\partial_t \hat{H}'^{(1)}_t =0$ becomes, $\forall j\neq0$, 
\begin{equation}\label{formula-for-f_1^1}
    \hat{H}^{(1)}_{j} + [\hat{f}_j^{(1)},\hat{H}^{(0)}] - j\omega \hat{f}_j^{(1)} = 0,
\end{equation}
which eliminates the term oscillating at $e^{i j\omega t}$ in $\hat{H}'^{(1)}_t$. 
Equations of the form~(\ref{formula-for-f_1^1}) are known as Sylvester equations in control theory \cite{bhatia1997and}. 

Eq.~(\ref{formula-for-f_1^1}) decides the solution of $\hat{f}_j^{(1)}$ for all $j$. 
Based on the unitary transform $\hat{U}_t = \exp\big( \hat{F}^{(1)}_t \big) = \exp\big( \sum_{j=1}^{\infty}  \hat{f}^{(1)}_j e^{i j \omega t} - H.c. \big)$, we find a transformed Hamiltonian $\hat{H}'_t$ with leading oscillatory contribution $\mathcal{O} (g^2)$.
Based on the ansatz~(\ref{ansatz}), we know the solution $\hat{f}_j^{(1)}$ completely determines the static part in $\hat{H}'^{(2)}_t$ in Eq.~(\ref{H'-expansion-2}). 
Thus, we can directly pick out the static part in the transformed Hamiltonian (\ref{H'-expansion}) up to order $g^2$, which gives 
\begin{equation}\label{H'-lowest}
    \hat{H}' = \hat{H}'^{(0)} + \hat{H}'^{(1)} + \hat{H}'^{(2)}
\end{equation}
with the zeroth order $\hat{H}'^{(0)}=\hat{H}^{(0)}$, the first order $\hat{H}'^{(1)} =\hat{H}^{(1)}_0$, and the second order contribution
\begin{equation}\label{H'2}
\begin{split}
\hat{H}'^{(2)}=\hat{H}^{(2)}_0 + \frac{1}{2}\sum\limits_{j\neq0} [\hat{f}_j^{(1)},\hat{H}^{(1)}_{-j}].
\end{split}
\end{equation}
This $\hat{H}'^{(2)}$ is identified as the lowest order Floquet Hamiltonian.

\subsubsection{higher order Floquet Hamiltonians}
For $n=2$, using $\hat{f}_j^{(1)}$ and $\hat{H}'^{(1)}=\hat{H}^{(1)}_0$ determined above, the requirement $\partial_t \hat{H}'^{(2)}_t =0$ becomes, $\forall j\neq0$,
\begin{widetext}
\begin{equation}\label{formula-for-f_1^2}
\begin{split}
\hat{H}^{(2)}_j + \frac{1}{2} \sum\limits_{j' \neq 0} [\hat{f}_{j'}^{(1)},\hat{H}_{j-j'}^{(1)}] + \frac{1}{2} [\hat{f}_j^{(1)},\hat{H}_0^{(1)}] + [\hat{f}_j^{(2)},\hat{H}^{(0)}]   - j \omega \hat{f}_j^{(2)} = 0,
\end{split}
\end{equation}
which eliminates the $\mathcal{O} (g^2)$ terms oscillating at $e^{i j \omega t}$ in $\hat{H}'^{(2)}_t$. These are again Sylvester equations, determining $\hat{f}_j^{(2)}$ 
and thus the static part of $\hat{H}'^{(3)}_t$. 
Collecting the static terms in Eq.~(\ref{H'-expansion-3}), we find this static part in $\mathcal{O} (g^3)$ to be
\begin{equation}\label{H'3}
\begin{split}
\hat{H}'^{(3)} &= \hat{H}^{(3)}_0 + \frac{1}{2}\sum\limits_{j\neq0} [\hat{f}_j^{(1)},\hat{H}^{(2)}_{-j}]
+ \frac{1}{2}\sum\limits_{j\neq0} [\hat{f}_j^{(2)},\hat{H}^{(1)}_{-j}]
+\frac{1}{12} \sum\limits_{j\neq0} \sum\limits_{j'\neq0} [\hat{f}^{(1)}_j,[\hat{f}^{(1)}_{j'}, \hat{H}^{(1)}_{-j-j'} ]]
-\frac{1}{12} \sum\limits_{j\neq0} [\hat{f}^{(1)}_j,[\hat{f}^{(1)}_{-j}, \hat{H}^{(1)}_0  ]]
.
\end{split}
\end{equation}
\end{widetext}
The corresponding transformed Hamiltonian $\hat{H}' = \hat{H}'^{(0)} + \hat{H}'^{(1)} + \hat{H}'^{(2)} + \hat{H}'^{(3)}$  is generated by the transform $\hat{U}_t = \exp\big( \hat{F}^{(1)}_t + \hat{F}^{(2)}_t \big) = \exp\big( \sum_{j=1}^{\infty}  (\hat{f}^{(1)}_j +\hat{f}^{(2)}_j) e^{i j \omega t} - H.c. \big)$. 

This elimination can be extended to arbitrary orders of $g$. In the $n$-th order perturbation process, we first transform $\partial_t \hat{H}'^{(n)}_t =0$ into Sylvester equations, from which $\hat{f}_j^{(n)}$ are determined. 
Then we use the solution $\hat{f}_j^{(n)}$ to obtain the static term in $\hat{H}'^{(n+1)}_t$, which represents the correction to the Floquet Hamiltonian $\hat{H}'$ at $g^{n+1}$ order. The corresponding transform to get this $\hat{H}'$ is given by $\hat{U}_t = \exp \big( \sum_{j=1}^{\infty}   \sum_{k=1}^{n}\hat{f}^{(k)}_j  e^{i j \omega t} - H.c. \big)$.

\subsubsection{The Floquet micro-motion}\label{sec:micro-motion}
The time-periodic operator $\hat{F}_t$, found in FSWT by solving the Sylvester equations, is also known as the Floquet micro-motion operator \cite{Eckardt_2015}. 
The formal solution to our Sylvester equations is given in Appendix \ref{linear-response}, which expresses the solution $\hat{f}^{(n)}_j$ as a Laplace-transformed Heisenberg operator. For example, the formal solution to the lowest order Sylvester equation (\ref{formula-for-f_1^1}) is, $\forall j \neq 0$,
\begin{equation}\label{formal-solution}
    \hat{f}_j^{(1)} = -i \int_0^{\infty} dt ~ e^{i j \omL t} ~ e^{- 0^+ t} ~ e^{i \hat{H}^{(0)} t} \hat{H}_j^{(1)}  e^{-i \hat{H}^{(0)} t}
\end{equation}
which directly relates the micro-motion $\hat{f}_j^{(1)}$ to the retarded Green function of the undriven system, and thus to the linear response. This link stems from the construction of $\hat{F}_t$ in orders of $g$, instead of $1/\omL$, in our FSWT. 


\section{Comparison to other Floquet methods}\label{compare}
\subsection{Comparison to High-frequency expansion}

For this comparison, we consider a system driven by a single frequency,
where $\hat{H}_t = \hat{H}^{(0)} + \hat{H}_0^{(1)} + \hat{H}^{(1)}_{-1}e^{-i\omega t} + \hat{H}^{(1)}_1e^{i\omega t}$. We find that the HFE 
expansion in Ref.~\cite{casas2001floquet} can be recovered by solving the Sylvester equation in orders of inverse driving frequency $1/\omL$ (see Appendix \ref{appendix:compare_HFE}). 
This confirms that the low-order Floquet Hamiltonians given by our FSWT in Eqs.~ (\ref{H'2}) and (\ref{H'3}), in the high-frequency limit, reduces to the HFE results in Eq.~(\ref{van-Vlek-result}). 

\subsection{Equivalence of FSWT to the Sambe space van Vleck block-diagonalisation}

Vogl et al. recently derived a low-frequency Floquet expansion~\cite{PhysRevB.101.024303}, which relies on the downfolding (i.e. Gaussian elimination) of the Sambe space Hamiltonian to the zeroth Floquet mode. 
The self-consistency requirement therein stems from the fact that the downfolding to the zeroth Floquet mode is different from the block-diagonalisation of each Floquet sector of the Sambe space Hamiltonian. 
This method becomes impractical in a many-body system as one has to solve the quasi-eigenenergy self-consistently. In our previous publication~\cite{PhysRevB.109.115137}, we used this Gaussian method without the self-consistency (to the lowest order driving strength) 
using a many-body projector. Proceeding to higher-order driving effects becomes unfeasible, as divergences appear in the self-consistent determination of the many-body quasi-eigenenergy, which have to be cancelled at each order. 

The Floquet Schrieffer-Wolff transform presented here, on the other hand, is equivalent to block-diagonalising the Sambe space Floquet Hamiltonian
(see Appendix \ref{appendix:equivalence} for details). 
We find the following connection between our FSWT approach and the previous Sambe space block-diagonalisation method in Ref.~\cite{Eckardt_2015}: 
The latter method results in an expression for the micro-motion operators in the eigenbasis of the undriven Sambe-space Hamiltonian (see Eq.C.40 therein). This expression can be obtained in our approach by expanding the Sylvester equations~ (\ref{formula-for-f_1^1}) and (\ref{formula-for-f_1^2}) in the eigenbasis of $\hat{H}^{(0)} = \sum_{l} \ep_l \vert l \rangle \langle l \vert$, which provides the formal solution to $\hat{f}^{(n)}_j$. For example, the expansion of Eq.~(\ref{formula-for-f_1^1}) yields
\begin{equation*}
\langle l \vert \hat{f}^{(1)}_j \vert l' \rangle = \frac{ \langle l \vert \hat{H}^{(1)}_j \vert l' \rangle }{j \omega -(\ep_{l'}-\ep_{l})},
\end{equation*}
which coincides with Eq. (C.40) of Ref.~\cite{Eckardt_2015}.
Although our FSWT provides the same formal solution given by the previous van Vleck method, in our case, the Sylvester equations can be solved (at least perturbatively) without knowledge of the eigenstates $\vert l \rangle$ and eigenenergies $\ep_l$. This enables us 
to describe many-body systems under in-gap drive, where the eigenbasis cannot be obtained, 
as we will demonstrate in the following section.


\subsection{Absence of spurious Floquet-mediated interactions}
For a driven many-body system, even in the absence of intrinsic interactions, several methods find spurious Floquet-mediated interactions, 
as reported in 
Ref.~\cite{PhysRevB.93.144307}.
We also find them in the Gaussian elimination method \cite{PhysRevB.101.024303}.
These Floquet-mediated interactions account for the correlations found in a specific Fourier frequency component (e.g. the 0-th Floquet sector) of the driven many-body wavefunction, 
as the wavefunction remains a product state in time domain. 
When this Floquet-mediated interaction appears in an interacting system, 
it is difficult to separate it from the higher-order (non-linear) driving effects. 
The exponential structure $\exp (\hat{F}_t)$ in our FSWT properly allocates this fake correlation into the Floquet micro-motion: Given by the transform $\exp (\hat{F}_t)$, the Floquet Hamiltonian $\hat{H}'$ is expressed entirely in terms of commutators, see Eq.~(\ref{H'3}), which cannot create Floquet-mediated interactions from a non-interacting system. Consequently, 
these correlations do not appear
in the macro-motion described by our FSWT.



\section{The Driven single-band Hubbard model}\label{example}

We next explore the applicability of the FWST with the example of the
one-band Hubbard model with on-site repulsion $U$ and nearest neighbour hopping $J$. It is driven monochromatically with a 
driving frequency $\omL$ which is much higher than the hopping energy $J$ (i.e. $\omL\gg J$). 
The FWST provides an effective Floquet Hamiltonian that remains accurate for arbitrary ratios of $U/J$. 

The Hamiltonian for this driven system reads
\begin{equation}\label{Ht-example}
\hat{H}_t = \hat{H}^{(0)} + \hat{H}^{(1)}_{-1}e^{-i\omL t} + \hat{H}^{(1)}_1e^{i\omL t},
\end{equation}
where $\hat{H}_1^{(1)} =  (\hat{H}^{(1)}_{-1})^{\dag}$ due to the Hermiticity of $\hat{H}_t$. The undriven Hamiltonian is given by
\begin{equation}\label{H^0-example}
\hat{H}^{(0)} = \hat{h} + \hat{U} - \mu \hat{N},
\end{equation}
with kinetic energy 
\begin{equation}
\hat{h} = -J \sum\limits_{s} \sum\limits_{j=1}^{L-1} (\hat{c}_{j+1,s}^{\dag} \hat{c}_{j,s} + \hat{c}_{j,s}^{\dag} \hat{c}_{j+1,s} ),
\end{equation}
local repulsion 
\begin{equation}\label{U}
\hat{U} = U \sum\limits_{j=1}^{L} \hat{n}_{j,\uparrow} \hat{n}_{j,\downarrow},
\end{equation} 
and chemical potential $- \mu \hat{N}$, where $\hat{N}= \sum_{j, s} \hat{n}_{j,s}$ is the total particle number. 
We consider the electric dipole driving in the long-wavelength limit, 
\begin{equation}\label{H_1-example}
\hat{H}^{(1)}_{1} = g \sum\limits_{j=1}^{L} 
j ~ \sum\limits_{s}  \hat{n}_{j,s}. 
\end{equation}
which can be realised, e.g., in cold atom platforms using shaken lattices ~\cite{PhysRevLett.99.220403} or in electronic platforms ~\cite{hensgens2017quantum}. 
The total particle number commutes with the driving term, and therefore, the chemical potential is not altered by the driving. In the following, we will set $\mu = 0$, unless specified otherwise. It can added to the effective Hamiltonian, if necessary. 

In Hamiltonian~(\ref{Ht-example}), we encounter the particularly simple situation $\hat{H}^{(n\geq1)}_j \propto \delta_{n,1}\delta_{j,\pm1}$, i.e. all driving terms other than $\hat{H}^{(1)}_{\pm1}$ vanish identically. This implies $\hat{H}'^{(1)}=0$, i.e. there is no 
linear contribution to the Floquet Hamiltonian. To the lowest order of driving strength ($n=1$), in Eq.~(\ref{formula-for-f_1^1}), there is only one Sylvester equation for $\hat{f}^{(1)}_j$ which needs to be solved. It reads
\begin{equation}\label{FSWT-f^1_1}
\hat{H}^{(1)}_{1} + [\hat{f}^{(1)}_1,\hat{H}^{(0)}] - \omL \hat{f}^{(1)}_1 = 0,
\end{equation}
and the corresponding Floquet Hamiltonian~(\ref{H'-lowest}) 
to quadratic order in $g$ 
simplifies to 
\begin{equation}\label{H'-1st}
\begin{split}
\hat{H}' &= \hat{H}'^{(0)} + \hat{H}'^{(2)} \\
&= \hat{H}^{(0)}  + \frac{1}{2} \big( [\hat{f}^{(1)}_1, \hat{H}^{(1)}_{-1}] + H.c. \big) .
\end{split}
\end{equation}
We next solve the 
Sylvester equation~(\ref{FSWT-f^1_1}). 
Without any approximation, 
the exact solution to Eq.~(\ref{FSWT-f^1_1}) in a general driven many-body system with interactions will contain infinitely long operator-product terms, as does the Floquet Hamiltonian given by Eq.~(\ref{H'-1st}). This is shown in Appendix~\ref{appendix:first-order-example} and was also noticed in other Floquet methods~\cite{PhysRevX.9.021037,mori2023floquet,PhysRevX.4.041048}.
Below, we show how to find approximate solutions of the Sylvester equations for the driven Hubbard model, using neither the inverse-frequency expansion in HFE, nor the eigenbasis of the undriven many-body Hamiltonian. 
Instead, since $J\ll \omL$, we perturbatively solve the lowest order Sylvester equation (\ref{FSWT-f^1_1}) by expanding $\hat{f}^{(1)}_1$ in orders of the hopping $J$. We will see that the result constructed in this way remains applicable regardless of the ratio of $U/J$, ranging from strongly correlated to weakly correlated cases. This expansion in $J$ still converges when $U>\omL$, 
in sharp contrast with the HFE results in Appendix~\ref{appendix:compare_HFE}.

\subsection{driven Hubbard dimer}\label{subsec:driven-dimer}
We start our discussion with a Hubbard dimer, with site index $j=1,2$. 
We expand the function $\hat{f}^{(1)}_1$ in orders of $J$, such that $\hat{f}^{(1)}_1 = \sum_{n=0}^{\infty} \hat{y}_n$ where $\hat{y}_n \sim J^n$, and insert it into Eq.~(\ref{FSWT-f^1_1}). This results in a set of equations for different orders of $J$,
\begin{subequations}\label{y}
    \begin{align}
& \hat{H}^{(1)}_{1} + [\hat{y}_0,\hat{U}] - \omL \hat{y}_0 = 0  \label{y0} \\
& [\hat{y}_0,\hat{h}] + [\hat{y}_1,\hat{U}] - \omL \hat{y}_1 = 0 \label{y1} \\
& [\hat{y}_1,\hat{h}] + [\hat{y}_2,\hat{U}] - \omL \hat{y}_2 = 0, 
\label{y2}
    \end{align}
\end{subequations} 
and so on. This system can be solved order by order.
The solution, including the $\hat{y}_0 \sim J^0$ and $\hat{y}_1 \sim J^1$ order, is 
\begin{equation}
\begin{split}
& \hat{f}^{(1)}_1 = \frac{g}{\omL} \sum\limits_{s} \sum\limits_{j=1,2} j ~ \hat{n}_{j,s} \\
&+ \frac{g J}{\omL^2} \sum\limits_{s}   
\hat{c}_{1,s}^{\dag} \hat{c}_{2,s} ( 1 + \beta' \hat{n}_{1,\bar{s}} + \gamma' \hat{n}_{2,\bar{s}} + \delta' \hat{n}_{1,\bar{s}} \hat{n}_{2,\bar{s}} ) \\
&- \frac{g J}{\omL^2} \sum\limits_{s}   
\hat{c}_{2,s}^{\dag} \hat{c}_{1,s} ( 1 + \beta' \hat{n}_{2,\bar{s}} + \gamma' \hat{n}_{1,\bar{s}} + \delta' \hat{n}_{2,\bar{s}} \hat{n}_{1,\bar{s}} ) \\
&+ \mathcal{O} \left( J^2 \right)
\end{split}
\end{equation}
\\
where the coefficients are given by
\begin{equation}\label{parameters'}
\begin{split}
\beta' &= \frac{-U}{\omL+U}    ~~~~~~~~ 
\gamma' = \frac{U}{\omL-U} ~~~~~~~
\delta' =  - \beta' - \gamma'  
\end{split}
\end{equation}
We stress that the form of $\hat{y}_1$ is not determined using the eigenbasis of the undriven Hamiltonian, but instead by symmetries: in Eq.~ (\ref{y1}), we find $[\hat{y}_0,\hat{h}]$ contains hopping between site 1 and 2, while $\hat{U}$ cannot create hopping, thus $\hat{y}_1$ must contain hopping. And since $\hat{U}$ correlates two spins, the hopping in $\hat{y}_1$ must be accompanied by local terms $\hat{n}_{j,\bar{s}}$ in the opposite spin. This symmetry argument completely determines the form of $\hat{y}_1$, and then the coefficients are obtained from Eq.~(\ref{y1}).

\subsection{driven Hubbard chain}

In an $L$-site Hubbard chain, following the same perturbation procedure (in orders of $J$) and the same symmetry argument (for details, see Appendix \ref{appendix:first-order-example}), we find the solution of Eq.~(\ref{FSWT-f^1_1}), which reads
\begin{widetext}
\begin{equation}\label{FSWT-H'-chain-g^2}
\begin{split}
& \hat{f}^{(1)}_1 = \frac{g}{\omL} \sum\limits_{s} \sum\limits_{j=1}^{L} j ~ \hat{n}_{j,s} 
+ \frac{g J}{\omL^2} \sum\limits_{s} \sum\limits_{i,j} \left( \delta_{i-j,1} - \delta_{j-i,1} \right) \hat{c}_{j,s}^{\dag} \hat{c}_{i,s} 
( 1 + \beta' \hat{n}_{j,\bar{s}} + \gamma' \hat{n}_{i,\bar{s}} + \delta' \hat{n}_{j,\bar{s}} \hat{n}_{i,\bar{s}} ) 
~~ + \hat{y}_2 + \mathcal{O} ( J^3 )
\end{split}
\end{equation}
where the parameters $\beta'$, $\gamma'$ and $\delta'$ are given in Eq.~(\ref{parameters'}), and the $\mathcal{O} ( J^2 )$ contribution $\hat{y}_2$ is given in Eq.~(\ref{y2-full-result}).
According to Eq.~(\ref{H'-1st}), the corresponding lowest-order corrected Floquet Hamiltonian is given by
\begin{equation}\label{Floquet-H'}
\begin{split}
\hat{H}'^{(0)} &+ \hat{H}'^{(2)}= -J(1-\frac{g^2}{\omL^2}) \sum\limits_{s} \sum\limits_{j=1}^{L-1} ( \hat{c}_{j+1,s}^{\dag} \hat{c}_{j,s} + \hat{c}_{j,s}^{\dag} \hat{c}_{j+1,s} )  ~~~~~ + ~~~~ U \sum\limits_{j=1}^{L} \hat{n}_{j,\uparrow} \hat{n}_{j,\downarrow} \\
&~~~~~~  - J ~ \frac{g^2 U}{\omL^2} \left( \frac{1}{U-\omL} + \frac{1}{\omL+U} \right) \sum\limits_{s} \sum\limits_{j=1}^{L-1}  ( \hat{c}_{j,s}^{\dag} \hat{c}_{j+1,s} + \hat{c}_{j+1,s}^{\dag} \hat{c}_{j,s}   )
\left( \frac{ \hat{n}_{j,\bar{s}} +  \hat{n}_{j+1,\bar{s}} }{2} - \hat{n}_{j,\bar{s}} \hat{n}_{j+1,\bar{s}} \right) \\
&~~~~~~  + \frac{1}{2}([\hat{y}_2,\hat{H}^{(1)}_{-1}]+ H.c.)  + \mathcal{O} ( J^3 ).
\end{split}
\end{equation}
\end{widetext}
The $\mathcal{O} ( J^2 )$ term, $\frac{1}{2}([\hat{y}_2,\hat{H}^{(1)}_{-1}]+ H.c.)$ in the third line, is given explicitly in Eq.~(\ref{Floquet-H'2-t2}).
The non-interacting part of the Floquet Hamiltonian~(\ref{Floquet-H'}) shows the lowest order driving-induced bandwidth renormalisation (dynamical localisation), as predicted theoretically in, e.g.  Refs.~\cite{PhysRevB.34.3625,PhysRevLett.95.260404,PhysRevB.78.235124} and realised in e.g. Refs.~\cite{PhysRevLett.99.220403,PhysRevLett.81.5093}. The interacting part of Eq.~(\ref{Floquet-H'}) 
contains a driving-induced interaction in the second line of Eq.~(\ref{Floquet-H'}), which modifies the interaction term $\hat{U}$ in the undriven system and renders the overall interaction non-local. This driving-induced interaction in Eq.~(\ref{Floquet-H'}) has the form of correlated hopping \cite{PhysRevB.88.115115}.
Similar correlated hopping was previously identified in the effective Hamiltonians of undriven systems (e.g. in cold atoms near Feshbach resonances \cite{Duan_2008} and in cavity-coupled quantum materials \cite{PhysRevB.99.085116}). 
The correlated hopping may lead to eta-pairing scars in one dimension~\cite{PhysRevB.102.075132} and superconductivity in two dimensions~\cite{PhysRevB.39.11653,Duan_2008}. It may also modify the Kibble-Zurek mechanism \cite{PhysRevB.95.104306} when the driving quenches the system across a quantum phase transition (QPT) \cite{PhysRevLett.120.127601}.
In our driven system, as given by FSWT,
both the near-resonant contribution $1 / ( U - \omL )$ and the off-resonant contribution $1 / (\omL + U)$ can be found in the coefficient of the correlated hopping in Eq.~(\ref{Floquet-H'}). 

The HFE result (see Appendix \ref{appendix:compare_HFE}) is equivalent to expanding our FSWT result Eq.~(\ref{Floquet-H'}) in orders of $1/\omL$. In HFE, this correlated hopping term only occurs to order $\omL^{-2} (U^2 - \omL^2)^{-1} \sim \mathcal{O}(\omL^{-4})$, and then we might expect it to be much weaker than the bandwidth renormalisation effect which is of order $\omL^{-2} $. Thus, in the lowest order HFE, one typically ignores the entire correlated hopping term. 
However, in our FSWT result Eq.~(\ref{Floquet-H'}), we find the correlated hopping may be of similar strength as the bandwidth renormalization. 

As an example, in Fig.~\ref{fig:MIT} (see Appendix~\ref{appendix:MIT} for details),  we carry out infinite DMRG simulations of the ground state occupancy, $\langle \hat{n} \rangle = \lim_{L\to\infty}  \langle \hat{N} \rangle / L$, in a one-dimensional Hubbard chain using the TeNPy library~\cite{tenpy}.
In these simulations, we assume a non-zero chemical potential $\mu$. This is necessary to observe 
the equilibrium metal-insulator transition in one-dimensional Fermi Hubbard model~\cite{essler2005one}, and it does not alter the solution of any of the Sylvester equations. 
We rescale the horizontal axis to $\mu-U/2$ to match the convention of Lieb-Wu equation. 
Near criticality, where the charge excitation gap closes and the DMRG algorithm no longer converges, we extrapolate the DMRG result using the Lieb-Wu equation~\cite{essler2005one}. 
Our simulations show how the correlated hopping term in our FSWT result Eq.~(\ref{Floquet-H'}), ignored in the lowest order HFE, modifies the occupancy near the Mott transition. 
We see in Fig.~\ref{fig:MIT} that it weakens the Mott insulating phase. It changes the phase boundary as well as the charge susceptibility of the metallic phase (i.e., the slope of the lines) close to the boundary. 
The change is comparable in size to the dynamical localisation effect itself, i.e., the shift from the undriven case to the HFE result. 
This modification, arising solely from the $\mathcal{O}(J)$ and $\mathcal{O}(J^2)$ correlated hopping terms, is expected to be observable in electronic platforms, for instance, in quantum dot Fermi-Hubbard simulators~\cite{hensgens2017quantum} where $\mu$ is controlled by gating.

\begin{figure}[h!]
    \centering
    \includegraphics[width=0.45\textwidth]{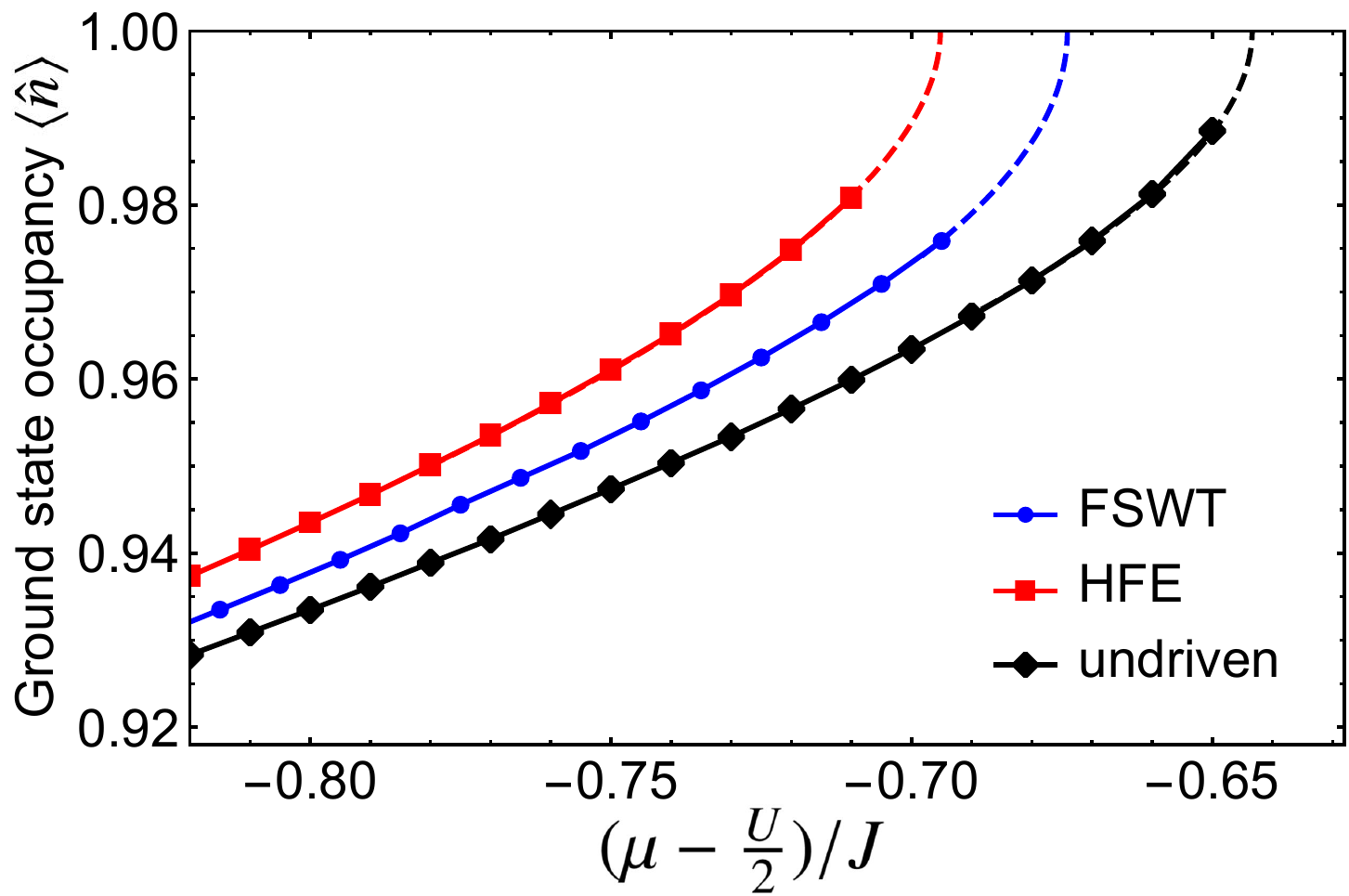}
    \caption{
    The ground state occupancy $\langle \hat{n} \rangle $ is shown as a function of the rescaled chemical potential $\mu - U/2$.
    iDMRG simulations with bond dimension 600 are indicated by markers. The extrapolations are obtained from the Lieb-Wu equation. 
    The parameters are chosen as $ U=4J, \omL=12J, g=3J$. 
    }

    \label{fig:MIT}
\end{figure}

\subsection{Comparison to exact dynamics}
\begin{figure}[h!]
    \centering
        \includegraphics[width=0.45\textwidth]{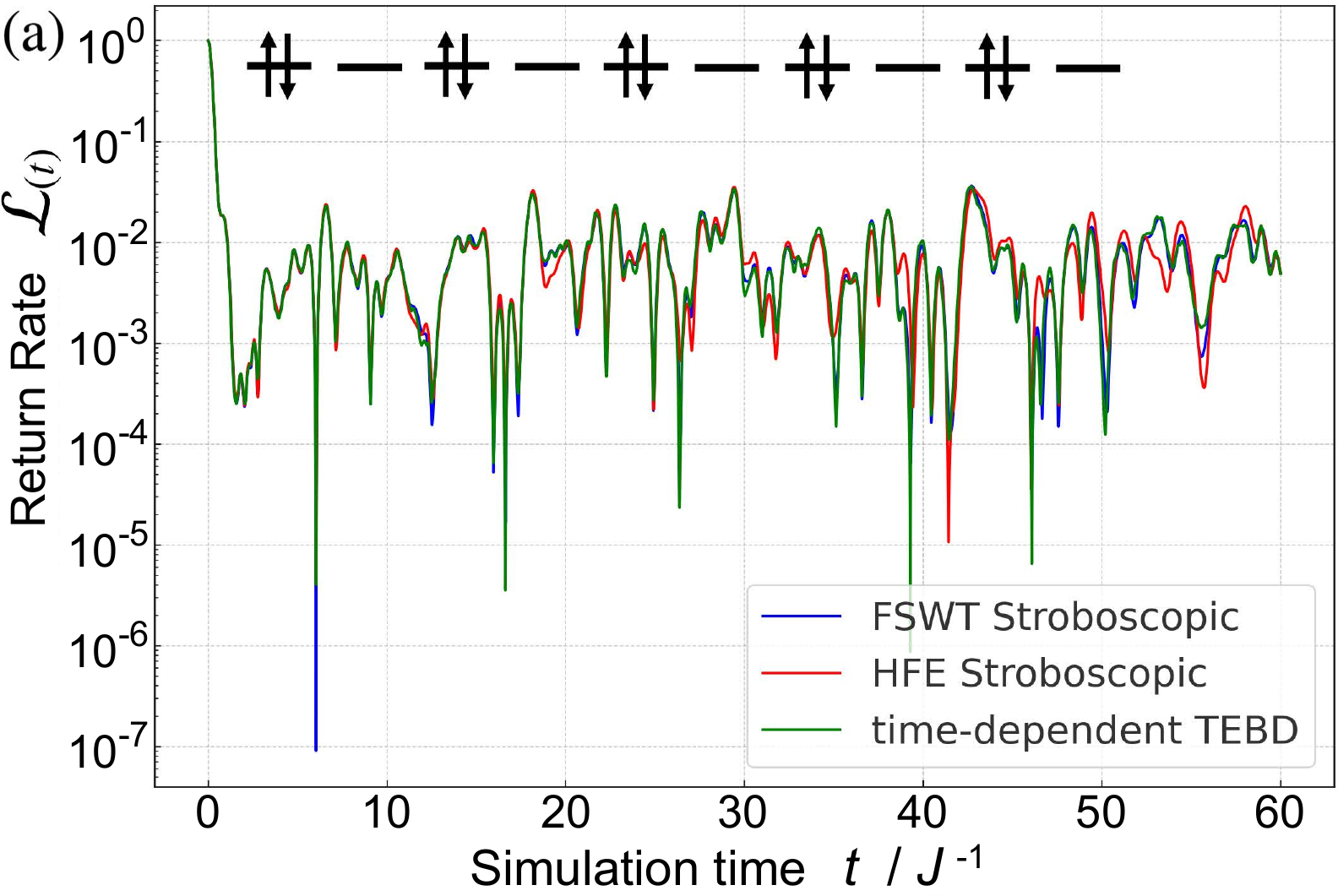}
        \label{fig:subfig1}
\vspace{0.2cm} 
        \includegraphics[width=0.45\textwidth]{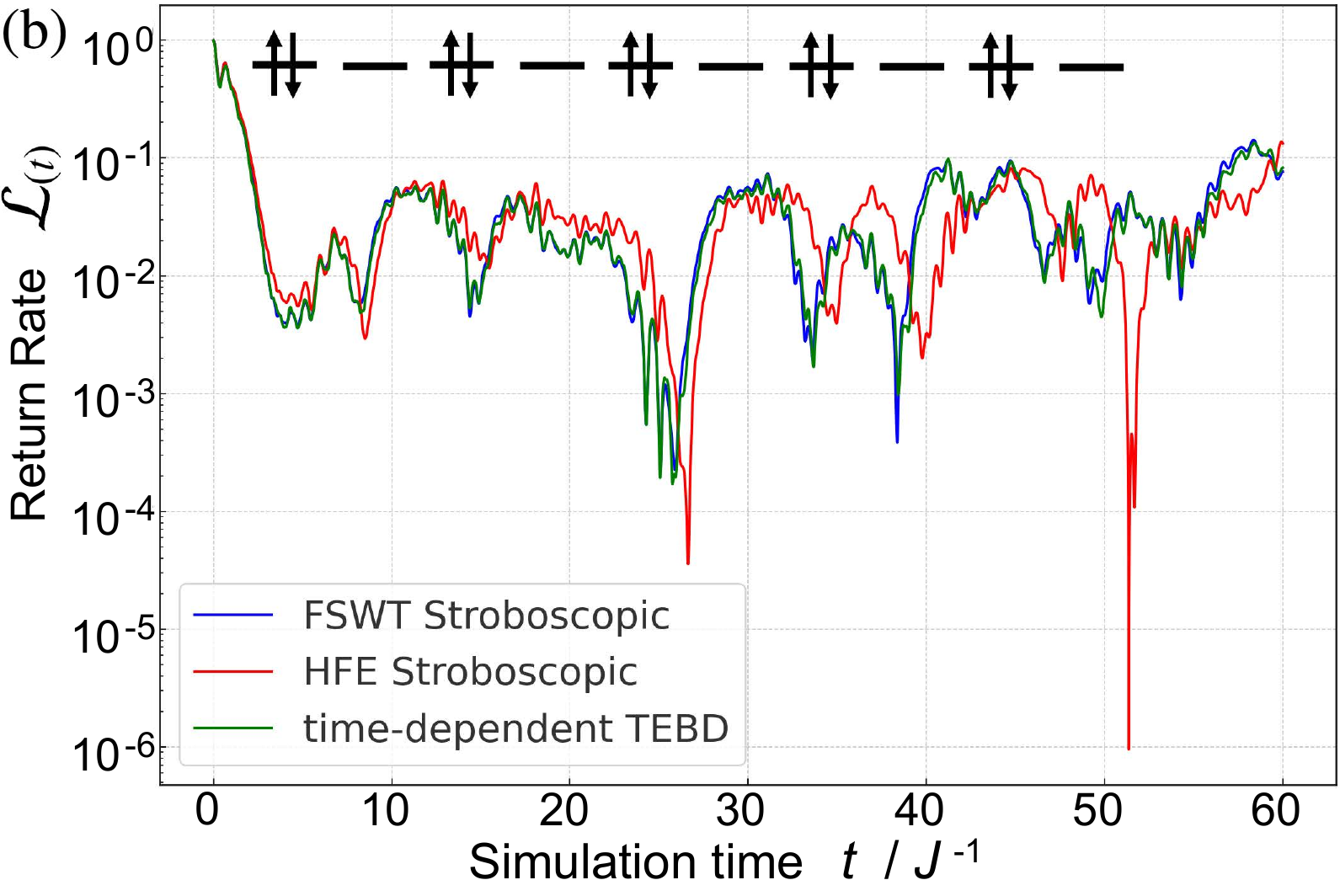}
        \label{fig:subfig2}
\vspace{0.2cm} 
        \includegraphics[width=0.45\textwidth]{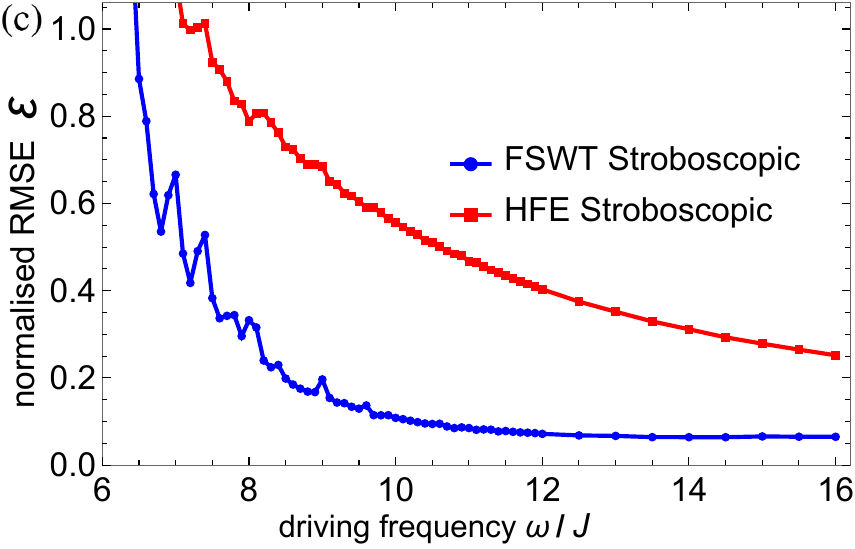}
        \label{fig:subfig3}
    \caption{
    (a) Return rate of CDW state under driving, Eq.~(\ref{eq.Loschmidt}), in an $L=10$ site lattice at driving frequency $\omL = 16 J$ and repulsion $U=3J$, with driving strength fixed at $g = \omL/4$. 
    The Floquet Hamiltonian dynamics are given by the time-dependent Variational principle (TDVP) \cite{PhysRevLett.107.070601,PhysRevB.94.165116}, and the exact time-dependent Hamiltonian dynamics are given by time-evolving block-decimation (TEBD), with bond-dimension $\chi=300$. 
    (b) Same as (a) but with increased repulsion $U=8J$. 
    (c) The RMSE of the return rate normalised by its mean value, Eq.~(\ref{eq.NRMSE}), for $U = 3 J$, $g = \omega /4$, $L=6$ lattice sites, and a simulation time $t J = 60$. }
    \label{fig:Evo}
\end{figure}

To gauge the accuracy of our effective Hamiltonian~(\ref{Floquet-H'}), we have carried out extensive simulations comparing it to the exact diagonalisation of small systems and DMRG-based simulations of larger chains. 
We follow Ref.~ \cite{Mendoza-Arenas_2022} and simulate the driven dynamics starting from a product charge density wave (CDW) state of a finite lattice, $\vert \psi_{CDW} \rangle = \Pi_{j=1,3,5,...} \hat{c}_{j,\uparrow}^{\dag} \hat{c}_{j,\downarrow}^{\dag} \vert vac \rangle $, and plot the time-evolution of the return rate~\footnote{The Floquet micro-motion will be ignored in the following, which is a good approximation since the return rate $\mathcal{L}_{(t)}$ we simulate represents an expectation value (of the observable $ \vert \psi_{CDW} \rangle \langle \psi_{CDW} \vert $). If we instead simulated the Floquet fidelity, i.e. the wavefunction overlap between the Floquet evolution and the exact dynamics, then we would have to take the micro-motion into account. },
\begin{align} \label{eq.Loschmidt}
\mathcal{L}_{(t)} &\equiv\big\vert \langle \psi_{CDW} \vert \hat{\mathcal{U}}_{t,t_0} \vert \psi_{CDW} \rangle \big\vert ^2, 
\end{align}
which is closely related to the Loschmidt echo \cite{Sharma_2014,Mendoza-Arenas_2022}. 
The results are shown in Fig.~\ref{fig:Evo}. 
We compare the stroboscopic Floquet Hamiltonian dynamics [where we ignore the micro-motion operators in Eq.~(\ref{eq.propagator})] given by FSWT (\ref{Floquet-H'}) and HFE, with the exact time-dependent Hamiltonian dynamics according to Eq.~(\ref{Ht-example}). 

In Fig.~\ref{fig:Evo}(a), we chose parameters such that $U\sim J \ll \omL$. Hence, both HFE and FSWT agree very well with the exact dynamics, even though some deviations of the HFE dynamics can be observed after sufficiently long propagation times. 
In Fig.~\ref{fig:Evo}(b), the onsite repulsion is increased, such that $U\sim\omL$, and 
the FSWT dynamics still show excellent agreement with the exact result, whereas the HFE expansion deviates substantially~\footnote{We note that the $\mathcal{O}(J^2)$ part of the FSWT Floquet Hamiltonian (\ref{Floquet-H'}), given by $\frac{1}{2}([\hat{y}_2,\hat{H}^{(1)}_{-1}]+ H.c.)$ in Eq.~(\ref{Floquet-H'2-t2}), is vital for accurately describing the dynamics involving the doublon excitations.}. 
To make this comparison quantitative, we measure the evolution error of FSWT and HFE Hamiltonians, compared to the exact evolution $\mathcal{L}^{exact}_{(t)}$ given by Eq.~(\ref{Ht-example}), using 
\begin{align}\label{eq.NRMSE}
    \mathcal{E} \equiv \bigg( \int_0^{t_f} \frac{\mathrm{d}t}{t_f} ( \mathcal{L}_{(t)} - \mathcal{L}^{exact}_{(t)} )^2 \bigg)^{1/2}  \bigg/ ~  \bigg( \int_0^{t_f} \frac{\mathrm{d}t}{t_f} \mathcal{L}^{exact}_{(t)} \bigg) ,
\end{align}
i.e. the ratio between the averaged root mean squared error (RMSE) of the return rate and the averaged return rate over the simulation time $t_f$. 
The results are shown in Fig.~\ref{fig:Evo}(c), where we fix $U=3J$ and vary the driving frequency $\omL$ under fixed ratio $g/\omL=1/4$ (which keeps the HFE Hamiltonian unchanged). 
Our FSWT result is nearly exact when $\omL\gtrsim 9J$, i.e. $\gamma' J / \omL \lesssim 0.06 $, with an error almost an order of magnitude smaller than that of the HFE result. 
For smaller $\omL$, our FSWT result begins to break down, which we attribute to the above-Mott-gap resonances~\cite{okamoto2021floquet} and the $\mathcal{O}(J^3)$ contribution ignored in Eq.~(\ref{Floquet-H'}) [see the discussion in Appendix~\ref{appendix:breakdown} for $\omega=8.5 J$ when the relative error reaches $\mathcal{E} \sim 0.2$]. In the HFE result where the $\mathcal{O}(J)$ and $\mathcal{O}(J^2)$ correlated hopping terms are absent, such a low error can only be reached at a much higher driving frequency $\omL\sim 20 J$.

\vspace{1cm}

As the driving frequency decreases further, such that the ratio $ J / \omL$ or $\gamma' J / \omL $ increases to $\mathcal{O} (1)$, the $\hat{y}_3 \sim \mathcal{O}(J^3)$ correction must be included in our FSWT Hamiltonian (\ref{Floquet-H'}). Since hundreds of terms appear in the analytical solution of $\hat{y}_3$, numerical methods may become necessary to solve the Sylvester equation (\ref{FSWT-f^1_1}) in this near-resonant case
\footnote{We can treat $\hat{f}^{(1)}_1$ as a matrix product operator, and solve the Sylvester equation (\ref{FSWT-f^1_1}) using tensor network methods. The corresponding lowest order Floquet Hamiltonian $\hat{H}^{0} + \hat{H}'^{(2)}$ is directly given in an MPO form, which has large bond dimension so that it cannot be explicitly written down analytically, but this MPO can be directly used in the DMRG to predict the steady state under weak drive. This Sylvester-based MPO approach for Floquet Hamiltonian is free from the many-body quasi-energy issue faced by Sambe space DMRG methods like Ref~\cite{sahoo2019periodically}. }, where Floquet heating is expected to appear faster \cite{PhysRevB.93.155132,PhysRevLett.120.197601}, and the pre-thermal states need to be studied in the rotating frame \cite{PhysRevLett.116.125301,Herrmann_2017}. 
Based on the solution of $\hat{f}^{(1)}_1$ in Eq.~(\ref{FSWT-H'-chain-g^2}), we proceed to the second- and third-lowest order FSWT in Appendices~\ref{appendix:higher-order-example} and \ref{appendix:3rd-order-example}, which results in the $\mathcal{O}(g^4)$ Floquet Hamiltonian $\hat{H}'^{(4)}$. For the parameters considered here, i.e. $g/\omL < 0.5$ and the two-photon process $2\omL$ remaining off-resonant from the doublon energy $U$, we find $\hat{H}'^{(4)}$ is negligible.  

Several additional benchmarks on the FSWT result are included in Appendices \ref{appendix:breakdown} and \ref{appendix:other-ini}: The return rate in the $U>\omL$ in-gap driving regime is shown in Fig.~\ref{fig:in-gap}; the doublon-doublon correlation in the $U\sim\omL$ regime is shown in Fig.~\ref{fig:d-d-correlation}; and the return rate starting from another initial state is shown in Fig.~\ref{fig:other-initial-states}. All of our simulations start from product states, as the cold-atom platforms are commonly initialised in such states to study driving-induced dynamics \cite{Mendoza-Arenas_2022,https://doi.org/10.1002/andp.201700024}. These simulations further confirm that the FSWT Hamiltonian (\ref{Floquet-H'}) outperforms the lowest order HFE in the lab-frame, providing a unified description of the non-resonantly driven Hubbard model.  


\subsection{Comparison to Floquet methods in the strongly correlated regime}

We next compare the Floquet Hamiltonian~(\ref{Floquet-H'}) to the one derived in Ref.~\cite{PhysRevLett.116.125301}, where a  Floquet Hamiltonian in driven strongly correlated systems is constructed systematically based on an interaction-dependent frame rotation followed by a high-frequency expansion. 
In this widely used method [categorized as method (b) in Fig.~\ref{fig:illustration}], the transformation to a rotating frame 
is carried out, in which the system oscillates at new frequencies $U+l\omL$ with $l$ being an integer. 
This method converges in the strongly correlated limit and yields a Floquet Hamiltonian with coefficients $\frac{J^2}{U+l\omL} \big( \frac{g}{\omL} \big)^{2l}$. 
In this rotating frame, 
the driving terms gain a factor $J$ (see Eq.~(S4) in Ref.~\cite{PhysRevLett.116.125301}), and thus the lowest-order high-frequency expansion in this frame is quadratic in $J$. 
In contrast, the Floquet Hamiltonian (\ref{Floquet-H'}) given by our lowest order FSWT gives a contribution which is linear in $J$. 
The reason for this discrepancy is that these two Floquet methods are constructed in different frames, and we will show these two Floquet Hamiltonians are equivalent in the limit $U\gg J$, where the two can be compared. 

To show this equivalence in the $\mathcal{O}(g^2)$ Floquet Hamiltonian when $U\gg J$ in a straightforward manner, we consider the half-filling case ($\mu = 0$) and project to the lowest Mott band to obtain effective spin Hamiltonians. 
In our Floquet Hamiltonian (\ref{Floquet-H'}), where we treat the $\mathcal{O}(J)$ part of $(\hat{h}+\hat{H}'^{(2)})$ as the perturbation and $\hat{U}$ as the unperturbed term, 
degenerate perturbation theory using the projector $\mathcal{P}$ onto the zero-doublon manifold yields (see Appendix \ref{appendix:project_large_U_limit} for details)
\begin{equation}\label{project_large_U_limit}
\begin{split}
&\mathcal{P} (\hat{h}+\hat{H}'^{(2)}) \frac{1}{-U} (\hat{h}+\hat{H}'^{(2)}) \mathcal{P} \\
&\approx \frac{1}{-U} \mathcal{P} \hat{h}^2 \mathcal{P} + \frac{1}{-U} \mathcal{P} \hat{h} \hat{H}'^{(2)} \mathcal{P} + \frac{1}{-U} \mathcal{P}  \hat{H}'^{(2)} \hat{h} \mathcal{P} + \mathcal{O}(g^4)  \\
&=  \frac{4 J^2}{U} (1-2\frac{g^2}{\omL^2}) \sum\limits_{j=1}^{L-1} {\bf S}_j \cdot {\bf S}_{j+1} \\  
&~~ + 4  \frac{g^2 J^2 }{\omL^2} \left( \frac{1}{U-\omL} + \frac{1}{\omL+U} \right) \sum\limits_{j=1}^{L-1} {\bf S}_j \cdot {\bf S}_{j+1},
\end{split} 
\end{equation}
where $ {\bf S}_j = ( S_j^x , S_j^y , S_j^z  ) $, with $ S_j^x = \frac{1}{2} ( \hat{c}_{j,\uparrow}^{\dag} \hat{c}_{j,\downarrow} + \hat{c}_{j,\downarrow}^{\dag} \hat{c}_{j,\uparrow}  )$, $ S_j^y = \frac{i}{2} ( -\hat{c}_{j,\uparrow}^{\dag} \hat{c}_{j,\downarrow} + \hat{c}_{j,\downarrow}^{\dag} \hat{c}_{j,\uparrow}  )$ and $ S_j^z = \frac{1}{2} ( \hat{n}_{j,\uparrow} - \hat{n}_{j,\downarrow} ) $.
This projected Hamiltonian is equivalent to the expression
for $\hat{H}_{\text{eff}}^{(1)}$ in Ref.~\cite{PhysRevLett.116.125301}. 
We thus conclude that in the $U\gg J$ limit, we can derive the previous Hamiltonian (to order $\mathcal{O} (g^2)$) in Ref.~\cite{PhysRevLett.116.125301} from our lowest order FSWT Hamiltonian.
In other words, the Hamiltonian (\ref{Floquet-H'}) given by our FSWT is the electronic parent Hamiltonian of the spin model derived in Ref.~\cite{PhysRevLett.116.125301}, just like the 
Hubbard model 
reduces to the 
t-J model \cite{essler2005one,spalek2007tj}. 
Moreover, in a two-site lattice, the projected Hamiltonian (\ref{project_large_U_limit}) reduces exactly to the previous Floquet result given by Ref.~\cite{mentink2015ultrafast}. 
Similarly, in the sub-lattice-driven Hubbard model, the enhancement of super-exchange pairing, that was reported in~\cite{PhysRevB.96.085104}, can also be traced back to the correlated hopping terms given by FSWT.
Unlike Refs.~\cite{PhysRevLett.116.125301,mentink2015ultrafast,PhysRevB.96.085104}, our FSWT result can be applied for all ratios $U/J$, as long as $J\ll \omL, \omL/\gamma'$. Our FSWT Hamiltonian (\ref{Floquet-H'}) can be used in the zero-correlation limit $U\to 0$ and the weakly correlated case $U\sim J$, as shown in Fig.~\ref{fig:Evo}.


\subsection{Applicability of our model}

We illustrate the preceding discussion in Fig.~\ref{fig:illustration}, where we sketch the range of applicability of (a) our FSWT result Eq.~(\ref{Floquet-H'}), 
(b) the lowest order $1 / (U+l\omL)$ large-U method in Ref.~\cite{PhysRevLett.116.125301}, and (c) the lowest order HFE result given by solving the Sylvester equation in (a) in orders of $1/\omL$. 
Method (b) can be used when $U\gg J$, and method (c) can be used when $\omL \gg U$. The FWST result (a) can deal with these regimes as well. 
In addition, we identify a regime, 
$J \lesssim U \lesssim \omL$ with $J\ll \omL, \omL/\gamma'$ [e.g. $U=3J$, $\omL=3 U$ in Fig.~\ref{fig:Evo}(c)], 
where the FSWT method (a) is the only working method among these three methods: 
Here $J\ll \omL, \omL/\gamma'$ guarantees (a) to be applicable, $U \lesssim \omL$ implies that the driving frequency is not sufficiently high for method (c) to be applicable, and $J \lesssim  U$ implies that the undriven system is near the metal-insulator transition where method (b) is not accurate. 
Our method can still be applied in this regime.

\section{outlook}\label{conclude}
We have constructed a Floquet Schrieffer Wolff transform to perturbatively generate the many-body Floquet Hamiltonian and micro-motion operators in orders of the driving strength $g$. 
This transform is obtained from the solution of operator-valued Sylvester equations and does not require knowledge of the eigenbasis of the undriven many-body system.
As an example, we showed how to solve the Sylvester equations perturbatively in driven Hubbard models: By expanding the Sylvester equation in orders of hopping $J$, 
we derived a Floquet Hamiltonian that may be applied for any interaction strength of $U$. This remains true even when $U\sim J$ or $U>\omL$, provided that $J\ll \omL$ and one does not resonantly drive the Mott gap. 
We used this Floquet Hamiltonian to show that driving-induced correlated hoppings shift the Mott-insulator transition point in a one-dimensional chain. We showed that these effects, which are underestimated in high-frequency expansions, give a large contribution that can be on par with the dominant renormalization of electronic hopping. 

We envision several extensions of our formalism that can treat important scenarios of driven many-body physics: 
For near-resonant driving, it will be convenient to construct the FSWT in the co-rotating frame, where the solution to the Sylvester equation has no divergence. The corresponding Floquet Hamiltonian then describes how off-resonant driving terms, which give rise to, e.g., the Bloch-Siegert shift \cite{Sie2018} in TMDC, influence the resonant dynamics. 
For deep strong driving (relevant for rf-driven superconducting qubits \cite{doi:10.1126/science.1119678}), where $g\gg \omL$, one can still apply our FSWT by switching to a nonuniformly rotating frame (via a Peierls transform)~\cite{PhysRevA.95.023601,doi:10.1126/science.1119678,PhysRevLett.125.195301,10.21468/SciPostPhys.5.2.017,PhysRevA.101.053408}, where the oscillatory strength is transferred from $g$ to smaller parameters in the undriven system, such as the hopping $J$, see Appendix \ref{FSWT-Peierls}. In such a frame, we can construct the FSWT expansion in powers of $J$, which provides the stroboscopic Hamiltonian in the $g\gg \omL$ case. 

The FSWT expansion method presented here 
could be directly adapted to other driven many-body systems with more interaction terms (such as longer-ranged density-density repulsion or multi-orbital Kanamori interactions \cite{PhysRevB.100.220403}). 
The method is also suitable for systems driven by time-dependent interactions~\cite{PhysRevLett.109.203005,meinert2016floquet}, where the Floquet Hamiltonian can be obtained by solving Eq.~(\ref{y2-part}).
The crucial advantage of our method is that it provides a single effective Hamiltonian that is not restricted to the driving frequency being the largest energy scale involved. 
This advantage becomes more prominent when more and more interacting terms are taken into account, such as in driven Rydberg arrays \cite{zhao2023floquet,doi:10.1126/science.abg2530} relevant for multi-qubit gate-design \cite{PhysRevA.98.032306,PhysRevLett.123.170503}. 
Finally, we note that the FSWT provides a convenient and accurate method to analyze driven correlated materials near the Mott transition. This regime is particularly relevant for several unconventional superconductors, where light-induced superconductivity was observed in recent experiments in fullerides~\cite{mitrano2016possible, Budden2021, Rowe2023} and charge transfer salts~\cite{Buzzi2020, Buzzi2021}.

\section{Acknowledgement}

The authors would like to thank André Eckardt, Thomas Pohl, Francesco Petiziol, Alexander Schnell, and Li You for their helpful discussions on the spurious Floquet-mediated interactions and the relation with van Vleck theory. FPMMC,
DJ and FS acknowledge support from the Cluster of Excellence 'Advanced Imaging of Matter' of the Deutsche Forschungsgemeinschaft (DFG) - EXC 2056 - project ID 390715994.
DJ acknowledges support from the Hamburg Quantum Computing Initiative (HQIC) project EFRE. The project is co-financed by ERDF of the European Union and by “Fonds of the Hamburg Ministry of Science, Research, Equalities and Districts (BWFGB)”.

\bibliography{FSWT}

\begin{thebibliography}{106}%
\makeatletter
\providecommand \@ifxundefined [1]{%
 \@ifx{#1\undefined}
}%
\providecommand \@ifnum [1]{%
 \ifnum #1\expandafter \@firstoftwo
 \else \expandafter \@secondoftwo
 \fi
}%
\providecommand \@ifx [1]{%
 \ifx #1\expandafter \@firstoftwo
 \else \expandafter \@secondoftwo
 \fi
}%
\providecommand \natexlab [1]{#1}%
\providecommand \enquote  [1]{``#1''}%
\providecommand \bibnamefont  [1]{#1}%
\providecommand \bibfnamefont [1]{#1}%
\providecommand \citenamefont [1]{#1}%
\providecommand \href@noop [0]{\@secondoftwo}%
\providecommand \href [0]{\begingroup \@sanitize@url \@href}%
\providecommand \@href[1]{\@@startlink{#1}\@@href}%
\providecommand \@@href[1]{\endgroup#1\@@endlink}%
\providecommand \@sanitize@url [0]{\catcode `\\12\catcode `\$12\catcode `\&12\catcode `\#12\catcode `\^12\catcode `\_12\catcode `\%12\relax}%
\providecommand \@@startlink[1]{}%
\providecommand \@@endlink[0]{}%
\providecommand \url  [0]{\begingroup\@sanitize@url \@url }%
\providecommand \@url [1]{\endgroup\@href {#1}{\urlprefix }}%
\providecommand \urlprefix  [0]{URL }%
\providecommand \Eprint [0]{\href }%
\providecommand \doibase [0]{https://doi.org/}%
\providecommand \selectlanguage [0]{\@gobble}%
\providecommand \bibinfo  [0]{\@secondoftwo}%
\providecommand \bibfield  [0]{\@secondoftwo}%
\providecommand \translation [1]{[#1]}%
\providecommand \BibitemOpen [0]{}%
\providecommand \bibitemStop [0]{}%
\providecommand \bibitemNoStop [0]{.\EOS\space}%
\providecommand \EOS [0]{\spacefactor3000\relax}%
\providecommand \BibitemShut  [1]{\csname bibitem#1\endcsname}%
\let\auto@bib@innerbib\@empty
\bibitem [{\citenamefont {Wendin}(2017)}]{wendin2017quantum}%
  \BibitemOpen
  \bibfield  {author} {\bibinfo {author} {\bibfnamefont {G.}~\bibnamefont {Wendin}},\ }\bibfield  {title} {\bibinfo {title} {Quantum information processing with superconducting circuits: a review},\ }\href {https://iopscience.iop.org/article/10.1088/1361-6633/aa7e1a} {\bibfield  {journal} {\bibinfo  {journal} {Rep. Prog. Phys.}\ }\textbf {\bibinfo {volume} {80}},\ \bibinfo {pages} {106001} (\bibinfo {year} {2017})}\BibitemShut {NoStop}%
\bibitem [{\citenamefont {Georgescu}\ \emph {et~al.}(2014)\citenamefont {Georgescu}, \citenamefont {Ashhab},\ and\ \citenamefont {Nori}}]{georgescu2014quantum}%
  \BibitemOpen
  \bibfield  {author} {\bibinfo {author} {\bibfnamefont {I.~M.}\ \bibnamefont {Georgescu}}, \bibinfo {author} {\bibfnamefont {S.}~\bibnamefont {Ashhab}},\ and\ \bibinfo {author} {\bibfnamefont {F.}~\bibnamefont {Nori}},\ }\bibfield  {title} {\bibinfo {title} {Quantum simulation},\ }\href {https://journals.aps.org/rmp/abstract/10.1103/RevModPhys.86.153} {\bibfield  {journal} {\bibinfo  {journal} {Rev. Mod. Phys.}\ }\textbf {\bibinfo {volume} {86}},\ \bibinfo {pages} {153} (\bibinfo {year} {2014})}\BibitemShut {NoStop}%
\bibitem [{\citenamefont {Eckardt}(2017)}]{Eckardt2017}%
  \BibitemOpen
  \bibfield  {author} {\bibinfo {author} {\bibfnamefont {A.}~\bibnamefont {Eckardt}},\ }\bibfield  {title} {\bibinfo {title} {Colloquium: Atomic quantum gases in periodically driven optical lattices},\ }\href {https://doi.org/10.1103/RevModPhys.89.011004} {\bibfield  {journal} {\bibinfo  {journal} {Rev. Mod. Phys.}\ }\textbf {\bibinfo {volume} {89}},\ \bibinfo {pages} {011004} (\bibinfo {year} {2017})}\BibitemShut {NoStop}%
\bibitem [{\citenamefont {Weitenberg}\ and\ \citenamefont {Simonet}(2021)}]{weitenberg2021tailoring}%
  \BibitemOpen
  \bibfield  {author} {\bibinfo {author} {\bibfnamefont {C.}~\bibnamefont {Weitenberg}}\ and\ \bibinfo {author} {\bibfnamefont {J.}~\bibnamefont {Simonet}},\ }\bibfield  {title} {\bibinfo {title} {Tailoring quantum gases by floquet engineering},\ }\href {https://www.nature.com/articles/s41567-021-01316-x} {\bibfield  {journal} {\bibinfo  {journal} {Nat. Phys.}\ }\textbf {\bibinfo {volume} {17}},\ \bibinfo {pages} {1342} (\bibinfo {year} {2021})}\BibitemShut {NoStop}%
\bibitem [{\citenamefont {Oka}\ and\ \citenamefont {Kitamura}(2019)}]{oka2019floquet}%
  \BibitemOpen
  \bibfield  {author} {\bibinfo {author} {\bibfnamefont {T.}~\bibnamefont {Oka}}\ and\ \bibinfo {author} {\bibfnamefont {S.}~\bibnamefont {Kitamura}},\ }\bibfield  {title} {\bibinfo {title} {Floquet engineering of quantum materials},\ }\href {https://doi.org/10.1146/annurev-conmatphys-031218-013423} {\bibfield  {journal} {\bibinfo  {journal} {Annu. Rev. Condens. Matter Phys.}\ }\textbf {\bibinfo {volume} {10}},\ \bibinfo {pages} {387} (\bibinfo {year} {2019})}\BibitemShut {NoStop}%
\bibitem [{\citenamefont {de~la Torre}\ \emph {et~al.}(2021)\citenamefont {de~la Torre}, \citenamefont {Kennes}, \citenamefont {Claassen}, \citenamefont {Gerber}, \citenamefont {McIver},\ and\ \citenamefont {Sentef}}]{RevModPhys.93.041002}%
  \BibitemOpen
  \bibfield  {author} {\bibinfo {author} {\bibfnamefont {A.}~\bibnamefont {de~la Torre}}, \bibinfo {author} {\bibfnamefont {D.~M.}\ \bibnamefont {Kennes}}, \bibinfo {author} {\bibfnamefont {M.}~\bibnamefont {Claassen}}, \bibinfo {author} {\bibfnamefont {S.}~\bibnamefont {Gerber}}, \bibinfo {author} {\bibfnamefont {J.~W.}\ \bibnamefont {McIver}},\ and\ \bibinfo {author} {\bibfnamefont {M.~A.}\ \bibnamefont {Sentef}},\ }\bibfield  {title} {\bibinfo {title} {Colloquium: Nonthermal pathways to ultrafast control in quantum materials},\ }\href {https://link.aps.org/doi/10.1103/RevModPhys.93.041002} {\bibfield  {journal} {\bibinfo  {journal} {Rev. Mod. Phys.}\ }\textbf {\bibinfo {volume} {93}},\ \bibinfo {pages} {041002} (\bibinfo {year} {2021})}\BibitemShut {NoStop}%
\bibitem [{\citenamefont {Rudner}\ and\ \citenamefont {Lindner}(2020)}]{rudner2020band}%
  \BibitemOpen
  \bibfield  {author} {\bibinfo {author} {\bibfnamefont {M.~S.}\ \bibnamefont {Rudner}}\ and\ \bibinfo {author} {\bibfnamefont {N.~H.}\ \bibnamefont {Lindner}},\ }\bibfield  {title} {\bibinfo {title} {Band structure engineering and non-equilibrium dynamics in floquet topological insulators},\ }\href {https://doi.org/10.1038/s42254-020-0170-z} {\bibfield  {journal} {\bibinfo  {journal} {Nat. Rev. Phys}\ }\textbf {\bibinfo {volume} {2}},\ \bibinfo {pages} {229} (\bibinfo {year} {2020})}\BibitemShut {NoStop}%
\bibitem [{\citenamefont {Lindner}\ \emph {et~al.}(2010)\citenamefont {Lindner}, \citenamefont {Refael},\ and\ \citenamefont {Galitski}}]{Lindner2010FloquetTI}%
  \BibitemOpen
  \bibfield  {author} {\bibinfo {author} {\bibfnamefont {N.~H.}\ \bibnamefont {Lindner}}, \bibinfo {author} {\bibfnamefont {G.}~\bibnamefont {Refael}},\ and\ \bibinfo {author} {\bibfnamefont {V.~M.}\ \bibnamefont {Galitski}},\ }\bibfield  {title} {\bibinfo {title} {Floquet topological insulator in semiconductor quantum wells},\ }\href {https://doi.org/10.1038/nphys1926} {\bibfield  {journal} {\bibinfo  {journal} {Nat. Phys.}\ }\textbf {\bibinfo {volume} {7}},\ \bibinfo {pages} {490} (\bibinfo {year} {2010})}\BibitemShut {NoStop}%
\bibitem [{\citenamefont {Oka}\ and\ \citenamefont {Aoki}(2009)}]{PhysRevB.79.081406}%
  \BibitemOpen
  \bibfield  {author} {\bibinfo {author} {\bibfnamefont {T.}~\bibnamefont {Oka}}\ and\ \bibinfo {author} {\bibfnamefont {H.}~\bibnamefont {Aoki}},\ }\bibfield  {title} {\bibinfo {title} {Photovoltaic hall effect in graphene},\ }\href {https://doi.org/10.1103/PhysRevB.79.081406} {\bibfield  {journal} {\bibinfo  {journal} {Phys. Rev. B}\ }\textbf {\bibinfo {volume} {79}},\ \bibinfo {pages} {081406(R)} (\bibinfo {year} {2009})}\BibitemShut {NoStop}%
\bibitem [{\citenamefont {Ozawa}\ \emph {et~al.}(2019)\citenamefont {Ozawa}, \citenamefont {Price}, \citenamefont {Amo}, \citenamefont {Goldman}, \citenamefont {Hafezi}, \citenamefont {Lu}, \citenamefont {Rechtsman}, \citenamefont {Schuster}, \citenamefont {Simon}, \citenamefont {Zilberberg},\ and\ \citenamefont {Carusotto}}]{RevModPhys.91.015006}%
  \BibitemOpen
  \bibfield  {author} {\bibinfo {author} {\bibfnamefont {T.}~\bibnamefont {Ozawa}}, \bibinfo {author} {\bibfnamefont {H.~M.}\ \bibnamefont {Price}}, \bibinfo {author} {\bibfnamefont {A.}~\bibnamefont {Amo}}, \bibinfo {author} {\bibfnamefont {N.}~\bibnamefont {Goldman}}, \bibinfo {author} {\bibfnamefont {M.}~\bibnamefont {Hafezi}}, \bibinfo {author} {\bibfnamefont {L.}~\bibnamefont {Lu}}, \bibinfo {author} {\bibfnamefont {M.~C.}\ \bibnamefont {Rechtsman}}, \bibinfo {author} {\bibfnamefont {D.}~\bibnamefont {Schuster}}, \bibinfo {author} {\bibfnamefont {J.}~\bibnamefont {Simon}}, \bibinfo {author} {\bibfnamefont {O.}~\bibnamefont {Zilberberg}},\ and\ \bibinfo {author} {\bibfnamefont {I.}~\bibnamefont {Carusotto}},\ }\bibfield  {title} {\bibinfo {title} {Topological photonics},\ }\href {https://doi.org/10.1103/RevModPhys.91.015006} {\bibfield  {journal} {\bibinfo  {journal} {Rev. Mod. Phys.}\ }\textbf {\bibinfo {volume} {91}},\ \bibinfo {pages} {015006} (\bibinfo {year} {2019})}\BibitemShut {NoStop}%
\bibitem [{\citenamefont {Mivehvar}\ \emph {et~al.}(2021)\citenamefont {Mivehvar}, \citenamefont {Piazza}, \citenamefont {Donner},\ and\ \citenamefont {Ritsch}}]{mivehvar2021cavity}%
  \BibitemOpen
  \bibfield  {author} {\bibinfo {author} {\bibfnamefont {F.}~\bibnamefont {Mivehvar}}, \bibinfo {author} {\bibfnamefont {F.}~\bibnamefont {Piazza}}, \bibinfo {author} {\bibfnamefont {T.}~\bibnamefont {Donner}},\ and\ \bibinfo {author} {\bibfnamefont {H.}~\bibnamefont {Ritsch}},\ }\bibfield  {title} {\bibinfo {title} {Cavity qed with quantum gases: new paradigms in many-body physics},\ }\href {https://doi.org/10.1080/00018732.2021.1969727} {\bibfield  {journal} {\bibinfo  {journal} {Adv. Phys.}\ }\textbf {\bibinfo {volume} {70}},\ \bibinfo {pages} {1} (\bibinfo {year} {2021})}\BibitemShut {NoStop}%
\bibitem [{\citenamefont {Schlawin}\ \emph {et~al.}(2022)\citenamefont {Schlawin}, \citenamefont {Kennes},\ and\ \citenamefont {Sentef}}]{schlawin2022cavity}%
  \BibitemOpen
  \bibfield  {author} {\bibinfo {author} {\bibfnamefont {F.}~\bibnamefont {Schlawin}}, \bibinfo {author} {\bibfnamefont {D.~M.}\ \bibnamefont {Kennes}},\ and\ \bibinfo {author} {\bibfnamefont {M.~A.}\ \bibnamefont {Sentef}},\ }\bibfield  {title} {\bibinfo {title} {Cavity quantum materials},\ }\href {https://doi.org/10.1063/5.0083825} {\bibfield  {journal} {\bibinfo  {journal} {Appl. Phys. Rev.}\ }\textbf {\bibinfo {volume} {9}},\ \bibinfo {pages} {011312} (\bibinfo {year} {2022})}\BibitemShut {NoStop}%
\bibitem [{\citenamefont {Goldman}\ \emph {et~al.}(2014)\citenamefont {Goldman}, \citenamefont {Juzeli{\=u}nas}, \citenamefont {{\"O}hberg},\ and\ \citenamefont {Spielman}}]{goldman2014light}%
  \BibitemOpen
  \bibfield  {author} {\bibinfo {author} {\bibfnamefont {N.}~\bibnamefont {Goldman}}, \bibinfo {author} {\bibfnamefont {G.}~\bibnamefont {Juzeli{\=u}nas}}, \bibinfo {author} {\bibfnamefont {P.}~\bibnamefont {{\"O}hberg}},\ and\ \bibinfo {author} {\bibfnamefont {I.~B.}\ \bibnamefont {Spielman}},\ }\bibfield  {title} {\bibinfo {title} {Light-induced gauge fields for ultracold atoms},\ }\href {https://iopscience.iop.org/article/10.1088/0034-4885/77/12/126401} {\bibfield  {journal} {\bibinfo  {journal} {Rep. Prog. Phys.}\ }\textbf {\bibinfo {volume} {77}},\ \bibinfo {pages} {126401} (\bibinfo {year} {2014})}\BibitemShut {NoStop}%
\bibitem [{\citenamefont {Dalibard}\ \emph {et~al.}(2011)\citenamefont {Dalibard}, \citenamefont {Gerbier}, \citenamefont {Juzeli\ifmmode~\bar{u}\else \={u}\fi{}nas},\ and\ \citenamefont {\"Ohberg}}]{RevModPhys.83.1523}%
  \BibitemOpen
  \bibfield  {author} {\bibinfo {author} {\bibfnamefont {J.}~\bibnamefont {Dalibard}}, \bibinfo {author} {\bibfnamefont {F.}~\bibnamefont {Gerbier}}, \bibinfo {author} {\bibfnamefont {G.}~\bibnamefont {Juzeli\ifmmode~\bar{u}\else \={u}\fi{}nas}},\ and\ \bibinfo {author} {\bibfnamefont {P.}~\bibnamefont {\"Ohberg}},\ }\bibfield  {title} {\bibinfo {title} {Colloquium: Artificial gauge potentials for neutral atoms},\ }\href {https://doi.org/10.1103/RevModPhys.83.1523} {\bibfield  {journal} {\bibinfo  {journal} {Rev. Mod. Phys.}\ }\textbf {\bibinfo {volume} {83}},\ \bibinfo {pages} {1523} (\bibinfo {year} {2011})}\BibitemShut {NoStop}%
\bibitem [{\citenamefont {Goldman}\ \emph {et~al.}(2016)\citenamefont {Goldman}, \citenamefont {Budich},\ and\ \citenamefont {Zoller}}]{goldman2016topological}%
  \BibitemOpen
  \bibfield  {author} {\bibinfo {author} {\bibfnamefont {N.}~\bibnamefont {Goldman}}, \bibinfo {author} {\bibfnamefont {J.~C.}\ \bibnamefont {Budich}},\ and\ \bibinfo {author} {\bibfnamefont {P.}~\bibnamefont {Zoller}},\ }\bibfield  {title} {\bibinfo {title} {Topological quantum matter with ultracold gases in optical lattices},\ }\href {https://www.nature.com/articles/nphys3803} {\bibfield  {journal} {\bibinfo  {journal} {Nat. Phys.}\ }\textbf {\bibinfo {volume} {12}},\ \bibinfo {pages} {639} (\bibinfo {year} {2016})}\BibitemShut {NoStop}%
\bibitem [{\citenamefont {Jotzu}\ \emph {et~al.}(2014)\citenamefont {Jotzu}, \citenamefont {Messer}, \citenamefont {Desbuquois}, \citenamefont {Lebrat}, \citenamefont {Uehlinger}, \citenamefont {Greif},\ and\ \citenamefont {Esslinger}}]{jotzu2014experimental}%
  \BibitemOpen
  \bibfield  {author} {\bibinfo {author} {\bibfnamefont {G.}~\bibnamefont {Jotzu}}, \bibinfo {author} {\bibfnamefont {M.}~\bibnamefont {Messer}}, \bibinfo {author} {\bibfnamefont {R.}~\bibnamefont {Desbuquois}}, \bibinfo {author} {\bibfnamefont {M.}~\bibnamefont {Lebrat}}, \bibinfo {author} {\bibfnamefont {T.}~\bibnamefont {Uehlinger}}, \bibinfo {author} {\bibfnamefont {D.}~\bibnamefont {Greif}},\ and\ \bibinfo {author} {\bibfnamefont {T.}~\bibnamefont {Esslinger}},\ }\bibfield  {title} {\bibinfo {title} {Experimental realization of the topological haldane model with ultracold fermions},\ }\href {https://doi.org/10.1038/nature13915} {\bibfield  {journal} {\bibinfo  {journal} {Nature}\ }\textbf {\bibinfo {volume} {515}},\ \bibinfo {pages} {237} (\bibinfo {year} {2014})}\BibitemShut {NoStop}%
\bibitem [{\citenamefont {Bordia}\ \emph {et~al.}(2017)\citenamefont {Bordia}, \citenamefont {L{\"u}schen}, \citenamefont {Schneider}, \citenamefont {Knap},\ and\ \citenamefont {Bloch}}]{bordia2017periodically}%
  \BibitemOpen
  \bibfield  {author} {\bibinfo {author} {\bibfnamefont {P.}~\bibnamefont {Bordia}}, \bibinfo {author} {\bibfnamefont {H.}~\bibnamefont {L{\"u}schen}}, \bibinfo {author} {\bibfnamefont {U.}~\bibnamefont {Schneider}}, \bibinfo {author} {\bibfnamefont {M.}~\bibnamefont {Knap}},\ and\ \bibinfo {author} {\bibfnamefont {I.}~\bibnamefont {Bloch}},\ }\bibfield  {title} {\bibinfo {title} {Periodically driving a many-body localized quantum system},\ }\href {https://www.nature.com/articles/nphys4020} {\bibfield  {journal} {\bibinfo  {journal} {Nat. Phys.}\ }\textbf {\bibinfo {volume} {13}},\ \bibinfo {pages} {460} (\bibinfo {year} {2017})}\BibitemShut {NoStop}%
\bibitem [{\citenamefont {Singh}\ \emph {et~al.}(2019)\citenamefont {Singh}, \citenamefont {Fujiwara}, \citenamefont {Geiger}, \citenamefont {Simmons}, \citenamefont {Lipatov}, \citenamefont {Cao}, \citenamefont {Dotti}, \citenamefont {Rajagopal}, \citenamefont {Senaratne}, \citenamefont {Shimasaki}, \citenamefont {Heyl}, \citenamefont {Eckardt},\ and\ \citenamefont {Weld}}]{singh2019quantifying}%
  \BibitemOpen
  \bibfield  {author} {\bibinfo {author} {\bibfnamefont {K.}~\bibnamefont {Singh}}, \bibinfo {author} {\bibfnamefont {C.~J.}\ \bibnamefont {Fujiwara}}, \bibinfo {author} {\bibfnamefont {Z.~A.}\ \bibnamefont {Geiger}}, \bibinfo {author} {\bibfnamefont {E.~Q.}\ \bibnamefont {Simmons}}, \bibinfo {author} {\bibfnamefont {M.}~\bibnamefont {Lipatov}}, \bibinfo {author} {\bibfnamefont {A.}~\bibnamefont {Cao}}, \bibinfo {author} {\bibfnamefont {P.}~\bibnamefont {Dotti}}, \bibinfo {author} {\bibfnamefont {S.~V.}\ \bibnamefont {Rajagopal}}, \bibinfo {author} {\bibfnamefont {R.}~\bibnamefont {Senaratne}}, \bibinfo {author} {\bibfnamefont {T.}~\bibnamefont {Shimasaki}}, \bibinfo {author} {\bibfnamefont {M.}~\bibnamefont {Heyl}}, \bibinfo {author} {\bibfnamefont {A.}~\bibnamefont {Eckardt}},\ and\ \bibinfo {author} {\bibfnamefont {D.~M.}\ \bibnamefont {Weld}},\ }\bibfield  {title} {\bibinfo {title} {Quantifying and controlling prethermal nonergodicity in interacting floquet matter},\ }\href
  {https://journals.aps.org/prx/abstract/10.1103/PhysRevX.9.041021} {\bibfield  {journal} {\bibinfo  {journal} {Phys. Rev. X}\ }\textbf {\bibinfo {volume} {9}},\ \bibinfo {pages} {041021} (\bibinfo {year} {2019})}\BibitemShut {NoStop}%
\bibitem [{\citenamefont {Zhang}\ \emph {et~al.}(2017)\citenamefont {Zhang}, \citenamefont {Hess}, \citenamefont {Kyprianidis}, \citenamefont {Becker}, \citenamefont {Lee}, \citenamefont {Smith}, \citenamefont {Pagano}, \citenamefont {Potirniche}, \citenamefont {Potter},\ and\ \citenamefont {Vishwanath}}]{zhang2017observation}%
  \BibitemOpen
  \bibfield  {author} {\bibinfo {author} {\bibfnamefont {J.}~\bibnamefont {Zhang}}, \bibinfo {author} {\bibfnamefont {P.~W.}\ \bibnamefont {Hess}}, \bibinfo {author} {\bibfnamefont {A.}~\bibnamefont {Kyprianidis}}, \bibinfo {author} {\bibfnamefont {P.}~\bibnamefont {Becker}}, \bibinfo {author} {\bibfnamefont {A.}~\bibnamefont {Lee}}, \bibinfo {author} {\bibfnamefont {J.}~\bibnamefont {Smith}}, \bibinfo {author} {\bibfnamefont {G.}~\bibnamefont {Pagano}}, \bibinfo {author} {\bibfnamefont {I.-D.}\ \bibnamefont {Potirniche}}, \bibinfo {author} {\bibfnamefont {A.~C.}\ \bibnamefont {Potter}},\ and\ \bibinfo {author} {\bibfnamefont {A.}~\bibnamefont {Vishwanath}},\ }\bibfield  {title} {\bibinfo {title} {Observation of a discrete time crystal},\ }\href {https://www.nature.com/articles/nature21413} {\bibfield  {journal} {\bibinfo  {journal} {Nature}\ }\textbf {\bibinfo {volume} {543}},\ \bibinfo {pages} {217} (\bibinfo {year} {2017})}\BibitemShut {NoStop}%
\bibitem [{\citenamefont {Choi}\ \emph {et~al.}(2017)\citenamefont {Choi}, \citenamefont {Choi}, \citenamefont {Landig}, \citenamefont {Kucsko}, \citenamefont {Zhou}, \citenamefont {Isoya}, \citenamefont {Jelezko}, \citenamefont {Onoda}, \citenamefont {Sumiya}, \citenamefont {Khemani}, \citenamefont {von Keyserlingk}, \citenamefont {Yao}, \citenamefont {Demler},\ and\ \citenamefont {Lukin}}]{choi2017observation}%
  \BibitemOpen
  \bibfield  {author} {\bibinfo {author} {\bibfnamefont {S.}~\bibnamefont {Choi}}, \bibinfo {author} {\bibfnamefont {J.}~\bibnamefont {Choi}}, \bibinfo {author} {\bibfnamefont {R.}~\bibnamefont {Landig}}, \bibinfo {author} {\bibfnamefont {G.}~\bibnamefont {Kucsko}}, \bibinfo {author} {\bibfnamefont {H.}~\bibnamefont {Zhou}}, \bibinfo {author} {\bibfnamefont {J.}~\bibnamefont {Isoya}}, \bibinfo {author} {\bibfnamefont {F.}~\bibnamefont {Jelezko}}, \bibinfo {author} {\bibfnamefont {S.}~\bibnamefont {Onoda}}, \bibinfo {author} {\bibfnamefont {H.}~\bibnamefont {Sumiya}}, \bibinfo {author} {\bibfnamefont {V.}~\bibnamefont {Khemani}}, \bibinfo {author} {\bibfnamefont {C.}~\bibnamefont {von Keyserlingk}}, \bibinfo {author} {\bibfnamefont {N.~Y.}\ \bibnamefont {Yao}}, \bibinfo {author} {\bibfnamefont {E.}~\bibnamefont {Demler}},\ and\ \bibinfo {author} {\bibfnamefont {M.~D.}\ \bibnamefont {Lukin}},\ }\bibfield  {title} {\bibinfo {title} {Observation of discrete time-crystalline order in a disordered dipolar many-body
  system},\ }\href {https://www.nature.com/articles/nature21426} {\bibfield  {journal} {\bibinfo  {journal} {Nature}\ }\textbf {\bibinfo {volume} {543}},\ \bibinfo {pages} {221} (\bibinfo {year} {2017})}\BibitemShut {NoStop}%
\bibitem [{\citenamefont {Zenesini}\ \emph {et~al.}(2009)\citenamefont {Zenesini}, \citenamefont {Lignier}, \citenamefont {Ciampini}, \citenamefont {Morsch},\ and\ \citenamefont {Arimondo}}]{PhysRevLett.102.100403}%
  \BibitemOpen
  \bibfield  {author} {\bibinfo {author} {\bibfnamefont {A.}~\bibnamefont {Zenesini}}, \bibinfo {author} {\bibfnamefont {H.}~\bibnamefont {Lignier}}, \bibinfo {author} {\bibfnamefont {D.}~\bibnamefont {Ciampini}}, \bibinfo {author} {\bibfnamefont {O.}~\bibnamefont {Morsch}},\ and\ \bibinfo {author} {\bibfnamefont {E.}~\bibnamefont {Arimondo}},\ }\bibfield  {title} {\bibinfo {title} {Coherent control of dressed matter waves},\ }\href {https://doi.org/10.1103/PhysRevLett.102.100403} {\bibfield  {journal} {\bibinfo  {journal} {Phys. Rev. Lett.}\ }\textbf {\bibinfo {volume} {102}},\ \bibinfo {pages} {100403} (\bibinfo {year} {2009})}\BibitemShut {NoStop}%
\bibitem [{\citenamefont {G{\"o}rg}\ \emph {et~al.}(2018)\citenamefont {G{\"o}rg}, \citenamefont {Messer}, \citenamefont {Sandholzer}, \citenamefont {Jotzu}, \citenamefont {Desbuquois},\ and\ \citenamefont {Esslinger}}]{gorg2018enhancement}%
  \BibitemOpen
  \bibfield  {author} {\bibinfo {author} {\bibfnamefont {F.}~\bibnamefont {G{\"o}rg}}, \bibinfo {author} {\bibfnamefont {M.}~\bibnamefont {Messer}}, \bibinfo {author} {\bibfnamefont {K.}~\bibnamefont {Sandholzer}}, \bibinfo {author} {\bibfnamefont {G.}~\bibnamefont {Jotzu}}, \bibinfo {author} {\bibfnamefont {R.}~\bibnamefont {Desbuquois}},\ and\ \bibinfo {author} {\bibfnamefont {T.}~\bibnamefont {Esslinger}},\ }\bibfield  {title} {\bibinfo {title} {Enhancement and sign change of magnetic correlations in a driven quantum many-body system},\ }\href {https://www.nature.com/articles/nature25135} {\bibfield  {journal} {\bibinfo  {journal} {Nature}\ }\textbf {\bibinfo {volume} {553}},\ \bibinfo {pages} {481} (\bibinfo {year} {2018})}\BibitemShut {NoStop}%
\bibitem [{\citenamefont {Meinert}\ \emph {et~al.}(2016)\citenamefont {Meinert}, \citenamefont {Mark}, \citenamefont {Lauber}, \citenamefont {Daley},\ and\ \citenamefont {N{\"a}gerl}}]{meinert2016floquet}%
  \BibitemOpen
  \bibfield  {author} {\bibinfo {author} {\bibfnamefont {F.}~\bibnamefont {Meinert}}, \bibinfo {author} {\bibfnamefont {M.~J.}\ \bibnamefont {Mark}}, \bibinfo {author} {\bibfnamefont {K.}~\bibnamefont {Lauber}}, \bibinfo {author} {\bibfnamefont {A.~J.}\ \bibnamefont {Daley}},\ and\ \bibinfo {author} {\bibfnamefont {H.-C.}\ \bibnamefont {N{\"a}gerl}},\ }\bibfield  {title} {\bibinfo {title} {Floquet engineering of correlated tunneling in the bose-hubbard model with ultracold atoms},\ }\href {https://journals.aps.org/prl/abstract/10.1103/PhysRevLett.116.205301} {\bibfield  {journal} {\bibinfo  {journal} {Phys. Rev. Lett}\ }\textbf {\bibinfo {volume} {116}},\ \bibinfo {pages} {205301} (\bibinfo {year} {2016})}\BibitemShut {NoStop}%
\bibitem [{\citenamefont {Fausti}\ \emph {et~al.}(2011)\citenamefont {Fausti}, \citenamefont {Tobey}, \citenamefont {Dean}, \citenamefont {Kaiser}, \citenamefont {Dienst}, \citenamefont {Hoffmann}, \citenamefont {Pyon}, \citenamefont {Takayama}, \citenamefont {Takagi},\ and\ \citenamefont {Cavalleri}}]{doi:10.1126/science.1197294}%
  \BibitemOpen
  \bibfield  {author} {\bibinfo {author} {\bibfnamefont {D.}~\bibnamefont {Fausti}}, \bibinfo {author} {\bibfnamefont {R.~I.}\ \bibnamefont {Tobey}}, \bibinfo {author} {\bibfnamefont {N.}~\bibnamefont {Dean}}, \bibinfo {author} {\bibfnamefont {S.}~\bibnamefont {Kaiser}}, \bibinfo {author} {\bibfnamefont {A.}~\bibnamefont {Dienst}}, \bibinfo {author} {\bibfnamefont {M.~C.}\ \bibnamefont {Hoffmann}}, \bibinfo {author} {\bibfnamefont {S.}~\bibnamefont {Pyon}}, \bibinfo {author} {\bibfnamefont {T.}~\bibnamefont {Takayama}}, \bibinfo {author} {\bibfnamefont {H.}~\bibnamefont {Takagi}},\ and\ \bibinfo {author} {\bibfnamefont {A.}~\bibnamefont {Cavalleri}},\ }\bibfield  {title} {\bibinfo {title} {Light-induced superconductivity in a stripe-ordered cuprate},\ }\href {https://doi.org/10.1126/science.1197294} {\bibfield  {journal} {\bibinfo  {journal} {Science}\ }\textbf {\bibinfo {volume} {331}},\ \bibinfo {pages} {189} (\bibinfo {year} {2011})}\BibitemShut {NoStop}%
\bibitem [{\citenamefont {Mitrano}\ \emph {et~al.}(2016)\citenamefont {Mitrano}, \citenamefont {Cantaluppi}, \citenamefont {Nicoletti}, \citenamefont {Kaiser}, \citenamefont {Perucchi}, \citenamefont {Lupi}, \citenamefont {Di~Pietro}, \citenamefont {Pontiroli}, \citenamefont {Ricc{\`o}}, \citenamefont {Clark}, \citenamefont {Jaksch},\ and\ \citenamefont {Cavalleri}}]{mitrano2016possible}%
  \BibitemOpen
  \bibfield  {author} {\bibinfo {author} {\bibfnamefont {M.}~\bibnamefont {Mitrano}}, \bibinfo {author} {\bibfnamefont {A.}~\bibnamefont {Cantaluppi}}, \bibinfo {author} {\bibfnamefont {D.}~\bibnamefont {Nicoletti}}, \bibinfo {author} {\bibfnamefont {S.}~\bibnamefont {Kaiser}}, \bibinfo {author} {\bibfnamefont {A.}~\bibnamefont {Perucchi}}, \bibinfo {author} {\bibfnamefont {S.}~\bibnamefont {Lupi}}, \bibinfo {author} {\bibfnamefont {P.}~\bibnamefont {Di~Pietro}}, \bibinfo {author} {\bibfnamefont {D.}~\bibnamefont {Pontiroli}}, \bibinfo {author} {\bibfnamefont {M.}~\bibnamefont {Ricc{\`o}}}, \bibinfo {author} {\bibfnamefont {S.~R.}\ \bibnamefont {Clark}}, \bibinfo {author} {\bibfnamefont {D.}~\bibnamefont {Jaksch}},\ and\ \bibinfo {author} {\bibfnamefont {A.}~\bibnamefont {Cavalleri}},\ }\bibfield  {title} {\bibinfo {title} {Possible light-induced superconductivity in k3c60 at high temperature},\ }\href {https://www.nature.com/articles/nature16522} {\bibfield  {journal} {\bibinfo  {journal} {Nature}\ }\textbf
  {\bibinfo {volume} {530}},\ \bibinfo {pages} {461} (\bibinfo {year} {2016})}\BibitemShut {NoStop}%
\bibitem [{\citenamefont {Budden}\ \emph {et~al.}(2021)\citenamefont {Budden}, \citenamefont {Gebert}, \citenamefont {Buzzi}, \citenamefont {Jotzu}, \citenamefont {Wang}, \citenamefont {Matsuyama}, \citenamefont {Meier}, \citenamefont {Laplace}, \citenamefont {Pontiroli}, \citenamefont {Ricc{\`o}}, \citenamefont {Schlawin}, \citenamefont {Jaksch},\ and\ \citenamefont {Cavalleri}}]{Budden2021}%
  \BibitemOpen
  \bibfield  {author} {\bibinfo {author} {\bibfnamefont {M.}~\bibnamefont {Budden}}, \bibinfo {author} {\bibfnamefont {T.}~\bibnamefont {Gebert}}, \bibinfo {author} {\bibfnamefont {M.}~\bibnamefont {Buzzi}}, \bibinfo {author} {\bibfnamefont {G.}~\bibnamefont {Jotzu}}, \bibinfo {author} {\bibfnamefont {E.}~\bibnamefont {Wang}}, \bibinfo {author} {\bibfnamefont {T.}~\bibnamefont {Matsuyama}}, \bibinfo {author} {\bibfnamefont {G.}~\bibnamefont {Meier}}, \bibinfo {author} {\bibfnamefont {Y.}~\bibnamefont {Laplace}}, \bibinfo {author} {\bibfnamefont {D.}~\bibnamefont {Pontiroli}}, \bibinfo {author} {\bibfnamefont {M.}~\bibnamefont {Ricc{\`o}}}, \bibinfo {author} {\bibfnamefont {F.}~\bibnamefont {Schlawin}}, \bibinfo {author} {\bibfnamefont {D.}~\bibnamefont {Jaksch}},\ and\ \bibinfo {author} {\bibfnamefont {A.}~\bibnamefont {Cavalleri}},\ }\bibfield  {title} {\bibinfo {title} {Evidence for metastable photo-induced superconductivity in k3c60},\ }\href {https://doi.org/10.1038/s41567-020-01148-1} {\bibfield
  {journal} {\bibinfo  {journal} {Nat. Phys.}\ }\textbf {\bibinfo {volume} {17}},\ \bibinfo {pages} {611} (\bibinfo {year} {2021})}\BibitemShut {NoStop}%
\bibitem [{\citenamefont {Rowe}\ \emph {et~al.}(2023)\citenamefont {Rowe}, \citenamefont {Yuan}, \citenamefont {Buzzi}, \citenamefont {Jotzu}, \citenamefont {Zhu}, \citenamefont {Fechner}, \citenamefont {F{\"o}rst}, \citenamefont {Liu}, \citenamefont {Pontiroli}, \citenamefont {Ricc{\`o}},\ and\ \citenamefont {Cavalleri}}]{Rowe2023}%
  \BibitemOpen
  \bibfield  {author} {\bibinfo {author} {\bibfnamefont {E.}~\bibnamefont {Rowe}}, \bibinfo {author} {\bibfnamefont {B.}~\bibnamefont {Yuan}}, \bibinfo {author} {\bibfnamefont {M.}~\bibnamefont {Buzzi}}, \bibinfo {author} {\bibfnamefont {G.}~\bibnamefont {Jotzu}}, \bibinfo {author} {\bibfnamefont {Y.}~\bibnamefont {Zhu}}, \bibinfo {author} {\bibfnamefont {M.}~\bibnamefont {Fechner}}, \bibinfo {author} {\bibfnamefont {M.}~\bibnamefont {F{\"o}rst}}, \bibinfo {author} {\bibfnamefont {B.}~\bibnamefont {Liu}}, \bibinfo {author} {\bibfnamefont {D.}~\bibnamefont {Pontiroli}}, \bibinfo {author} {\bibfnamefont {M.}~\bibnamefont {Ricc{\`o}}},\ and\ \bibinfo {author} {\bibfnamefont {A.}~\bibnamefont {Cavalleri}},\ }\bibfield  {title} {\bibinfo {title} {Resonant enhancement of photo-induced superconductivity in k3c60},\ }\href {https://doi.org/10.1038/s41567-023-02235-9} {\bibfield  {journal} {\bibinfo  {journal} {Nat. Phys.}\ }\textbf {\bibinfo {volume} {19}},\ \bibinfo {pages} {1821} (\bibinfo {year}
  {2023})}\BibitemShut {NoStop}%
\bibitem [{\citenamefont {Cavalleri}(2018)}]{doi:10.1080/00107514.2017.1406623}%
  \BibitemOpen
  \bibfield  {author} {\bibinfo {author} {\bibfnamefont {A.}~\bibnamefont {Cavalleri}},\ }\bibfield  {title} {\bibinfo {title} {Photo-induced superconductivity},\ }\href {https://doi.org/10.1080/00107514.2017.1406623} {\bibfield  {journal} {\bibinfo  {journal} {Contemp. Phys.}\ }\textbf {\bibinfo {volume} {59}},\ \bibinfo {pages} {31} (\bibinfo {year} {2018})}\BibitemShut {NoStop}%
\bibitem [{\citenamefont {Buzzi}\ \emph {et~al.}(2020)\citenamefont {Buzzi}, \citenamefont {Nicoletti}, \citenamefont {Fechner}, \citenamefont {Tancogne-Dejean}, \citenamefont {Sentef}, \citenamefont {Georges}, \citenamefont {Biesner}, \citenamefont {Uykur}, \citenamefont {Dressel}, \citenamefont {Henderson}, \citenamefont {Siegrist}, \citenamefont {Schlueter}, \citenamefont {Miyagawa}, \citenamefont {Kanoda}, \citenamefont {Nam}, \citenamefont {Ardavan}, \citenamefont {Coulthard}, \citenamefont {Tindall}, \citenamefont {Schlawin}, \citenamefont {Jaksch},\ and\ \citenamefont {Cavalleri}}]{Buzzi2020}%
  \BibitemOpen
  \bibfield  {author} {\bibinfo {author} {\bibfnamefont {M.}~\bibnamefont {Buzzi}}, \bibinfo {author} {\bibfnamefont {D.}~\bibnamefont {Nicoletti}}, \bibinfo {author} {\bibfnamefont {M.}~\bibnamefont {Fechner}}, \bibinfo {author} {\bibfnamefont {N.}~\bibnamefont {Tancogne-Dejean}}, \bibinfo {author} {\bibfnamefont {M.~A.}\ \bibnamefont {Sentef}}, \bibinfo {author} {\bibfnamefont {A.}~\bibnamefont {Georges}}, \bibinfo {author} {\bibfnamefont {T.}~\bibnamefont {Biesner}}, \bibinfo {author} {\bibfnamefont {E.}~\bibnamefont {Uykur}}, \bibinfo {author} {\bibfnamefont {M.}~\bibnamefont {Dressel}}, \bibinfo {author} {\bibfnamefont {A.}~\bibnamefont {Henderson}}, \bibinfo {author} {\bibfnamefont {T.}~\bibnamefont {Siegrist}}, \bibinfo {author} {\bibfnamefont {J.~A.}\ \bibnamefont {Schlueter}}, \bibinfo {author} {\bibfnamefont {K.}~\bibnamefont {Miyagawa}}, \bibinfo {author} {\bibfnamefont {K.}~\bibnamefont {Kanoda}}, \bibinfo {author} {\bibfnamefont {M.-S.}\ \bibnamefont {Nam}}, \bibinfo {author} {\bibfnamefont
  {A.}~\bibnamefont {Ardavan}}, \bibinfo {author} {\bibfnamefont {J.}~\bibnamefont {Coulthard}}, \bibinfo {author} {\bibfnamefont {J.}~\bibnamefont {Tindall}}, \bibinfo {author} {\bibfnamefont {F.}~\bibnamefont {Schlawin}}, \bibinfo {author} {\bibfnamefont {D.}~\bibnamefont {Jaksch}},\ and\ \bibinfo {author} {\bibfnamefont {A.}~\bibnamefont {Cavalleri}},\ }\bibfield  {title} {\bibinfo {title} {Photomolecular high-temperature superconductivity},\ }\href {https://doi.org/10.1103/PhysRevX.10.031028} {\bibfield  {journal} {\bibinfo  {journal} {Phys. Rev. X}\ }\textbf {\bibinfo {volume} {10}},\ \bibinfo {pages} {031028} (\bibinfo {year} {2020})}\BibitemShut {NoStop}%
\bibitem [{\citenamefont {Buzzi}\ \emph {et~al.}(2021)\citenamefont {Buzzi}, \citenamefont {Nicoletti}, \citenamefont {Fava}, \citenamefont {Jotzu}, \citenamefont {Miyagawa}, \citenamefont {Kanoda}, \citenamefont {Henderson}, \citenamefont {Siegrist}, \citenamefont {Schlueter}, \citenamefont {Nam}, \citenamefont {Ardavan},\ and\ \citenamefont {Cavalleri}}]{Buzzi2021}%
  \BibitemOpen
  \bibfield  {author} {\bibinfo {author} {\bibfnamefont {M.}~\bibnamefont {Buzzi}}, \bibinfo {author} {\bibfnamefont {D.}~\bibnamefont {Nicoletti}}, \bibinfo {author} {\bibfnamefont {S.}~\bibnamefont {Fava}}, \bibinfo {author} {\bibfnamefont {G.}~\bibnamefont {Jotzu}}, \bibinfo {author} {\bibfnamefont {K.}~\bibnamefont {Miyagawa}}, \bibinfo {author} {\bibfnamefont {K.}~\bibnamefont {Kanoda}}, \bibinfo {author} {\bibfnamefont {A.}~\bibnamefont {Henderson}}, \bibinfo {author} {\bibfnamefont {T.}~\bibnamefont {Siegrist}}, \bibinfo {author} {\bibfnamefont {J.~A.}\ \bibnamefont {Schlueter}}, \bibinfo {author} {\bibfnamefont {M.-S.}\ \bibnamefont {Nam}}, \bibinfo {author} {\bibfnamefont {A.}~\bibnamefont {Ardavan}},\ and\ \bibinfo {author} {\bibfnamefont {A.}~\bibnamefont {Cavalleri}},\ }\bibfield  {title} {\bibinfo {title} {Phase diagram for light-induced superconductivity in $\ensuremath{\kappa}\text{\ensuremath{-}}(\mathrm{ET}{)}_{2}\text{\ensuremath{-}}\mathrm{X}$},\ }\href
  {https://doi.org/10.1103/PhysRevLett.127.197002} {\bibfield  {journal} {\bibinfo  {journal} {Phys. Rev. Lett.}\ }\textbf {\bibinfo {volume} {127}},\ \bibinfo {pages} {197002} (\bibinfo {year} {2021})}\BibitemShut {NoStop}%
\bibitem [{\citenamefont {Wang}\ \emph {et~al.}(2013)\citenamefont {Wang}, \citenamefont {Steinberg}, \citenamefont {Jarillo-Herrero},\ and\ \citenamefont {Gedik}}]{wang2013observation}%
  \BibitemOpen
  \bibfield  {author} {\bibinfo {author} {\bibfnamefont {Y.}~\bibnamefont {Wang}}, \bibinfo {author} {\bibfnamefont {H.}~\bibnamefont {Steinberg}}, \bibinfo {author} {\bibfnamefont {P.}~\bibnamefont {Jarillo-Herrero}},\ and\ \bibinfo {author} {\bibfnamefont {N.}~\bibnamefont {Gedik}},\ }\bibfield  {title} {\bibinfo {title} {Observation of floquet-bloch states on the surface of a topological insulator},\ }\href {https://doi.org/10.1126/science.1239834} {\bibfield  {journal} {\bibinfo  {journal} {Science}\ }\textbf {\bibinfo {volume} {342}},\ \bibinfo {pages} {453} (\bibinfo {year} {2013})}\BibitemShut {NoStop}%
\bibitem [{\citenamefont {Mahmood}\ \emph {et~al.}(2016)\citenamefont {Mahmood}, \citenamefont {Chan}, \citenamefont {Alpichshev}, \citenamefont {Gardner}, \citenamefont {Lee}, \citenamefont {Lee},\ and\ \citenamefont {Gedik}}]{mahmood2016selective}%
  \BibitemOpen
  \bibfield  {author} {\bibinfo {author} {\bibfnamefont {F.}~\bibnamefont {Mahmood}}, \bibinfo {author} {\bibfnamefont {C.-K.}\ \bibnamefont {Chan}}, \bibinfo {author} {\bibfnamefont {Z.}~\bibnamefont {Alpichshev}}, \bibinfo {author} {\bibfnamefont {D.}~\bibnamefont {Gardner}}, \bibinfo {author} {\bibfnamefont {Y.}~\bibnamefont {Lee}}, \bibinfo {author} {\bibfnamefont {P.~A.}\ \bibnamefont {Lee}},\ and\ \bibinfo {author} {\bibfnamefont {N.}~\bibnamefont {Gedik}},\ }\bibfield  {title} {\bibinfo {title} {Selective scattering between floquet--bloch and volkov states in a topological insulator},\ }\href {https://www.nature.com/articles/nphys3609} {\bibfield  {journal} {\bibinfo  {journal} {Nat. Phys.}\ }\textbf {\bibinfo {volume} {12}},\ \bibinfo {pages} {306} (\bibinfo {year} {2016})}\BibitemShut {NoStop}%
\bibitem [{\citenamefont {McIver}\ \emph {et~al.}(2020)\citenamefont {McIver}, \citenamefont {Schulte}, \citenamefont {Stein}, \citenamefont {Matsuyama}, \citenamefont {Jotzu}, \citenamefont {Meier},\ and\ \citenamefont {Cavalleri}}]{mciver2020light}%
  \BibitemOpen
  \bibfield  {author} {\bibinfo {author} {\bibfnamefont {J.~W.}\ \bibnamefont {McIver}}, \bibinfo {author} {\bibfnamefont {B.}~\bibnamefont {Schulte}}, \bibinfo {author} {\bibfnamefont {F.-U.}\ \bibnamefont {Stein}}, \bibinfo {author} {\bibfnamefont {T.}~\bibnamefont {Matsuyama}}, \bibinfo {author} {\bibfnamefont {G.}~\bibnamefont {Jotzu}}, \bibinfo {author} {\bibfnamefont {G.}~\bibnamefont {Meier}},\ and\ \bibinfo {author} {\bibfnamefont {A.}~\bibnamefont {Cavalleri}},\ }\bibfield  {title} {\bibinfo {title} {Light-induced anomalous hall effect in graphene},\ }\href {https://www.nature.com/articles/s41567-019-0698-y} {\bibfield  {journal} {\bibinfo  {journal} {Nat. Phys.}\ }\textbf {\bibinfo {volume} {16}},\ \bibinfo {pages} {38} (\bibinfo {year} {2020})}\BibitemShut {NoStop}%
\bibitem [{\citenamefont {Fleury}\ \emph {et~al.}(2016)\citenamefont {Fleury}, \citenamefont {Khanikaev},\ and\ \citenamefont {Alu}}]{fleury2016floquet}%
  \BibitemOpen
  \bibfield  {author} {\bibinfo {author} {\bibfnamefont {R.}~\bibnamefont {Fleury}}, \bibinfo {author} {\bibfnamefont {A.~B.}\ \bibnamefont {Khanikaev}},\ and\ \bibinfo {author} {\bibfnamefont {A.}~\bibnamefont {Alu}},\ }\bibfield  {title} {\bibinfo {title} {Floquet topological insulators for sound},\ }\href {https://www.nature.com/articles/ncomms11744} {\bibfield  {journal} {\bibinfo  {journal} {Nat. Commun.}\ }\textbf {\bibinfo {volume} {7}},\ \bibinfo {pages} {11744} (\bibinfo {year} {2016})}\BibitemShut {NoStop}%
\bibitem [{\citenamefont {Rechtsman}\ \emph {et~al.}(2013)\citenamefont {Rechtsman}, \citenamefont {Zeuner}, \citenamefont {Plotnik}, \citenamefont {Lumer}, \citenamefont {Podolsky}, \citenamefont {Dreisow}, \citenamefont {Nolte}, \citenamefont {Segev},\ and\ \citenamefont {Szameit}}]{rechtsman2013photonic}%
  \BibitemOpen
  \bibfield  {author} {\bibinfo {author} {\bibfnamefont {M.~C.}\ \bibnamefont {Rechtsman}}, \bibinfo {author} {\bibfnamefont {J.~M.}\ \bibnamefont {Zeuner}}, \bibinfo {author} {\bibfnamefont {Y.}~\bibnamefont {Plotnik}}, \bibinfo {author} {\bibfnamefont {Y.}~\bibnamefont {Lumer}}, \bibinfo {author} {\bibfnamefont {D.}~\bibnamefont {Podolsky}}, \bibinfo {author} {\bibfnamefont {F.}~\bibnamefont {Dreisow}}, \bibinfo {author} {\bibfnamefont {S.}~\bibnamefont {Nolte}}, \bibinfo {author} {\bibfnamefont {M.}~\bibnamefont {Segev}},\ and\ \bibinfo {author} {\bibfnamefont {A.}~\bibnamefont {Szameit}},\ }\bibfield  {title} {\bibinfo {title} {Photonic floquet topological insulators},\ }\href {https://www.nature.com/articles/nature12066} {\bibfield  {journal} {\bibinfo  {journal} {Nature}\ }\textbf {\bibinfo {volume} {496}},\ \bibinfo {pages} {196} (\bibinfo {year} {2013})}\BibitemShut {NoStop}%
\bibitem [{\citenamefont {Bukov}\ \emph {et~al.}(2015)\citenamefont {Bukov}, \citenamefont {D'Alessio},\ and\ \citenamefont {Polkovnikov}}]{bukov2015universal}%
  \BibitemOpen
  \bibfield  {author} {\bibinfo {author} {\bibfnamefont {M.}~\bibnamefont {Bukov}}, \bibinfo {author} {\bibfnamefont {L.}~\bibnamefont {D'Alessio}},\ and\ \bibinfo {author} {\bibfnamefont {A.}~\bibnamefont {Polkovnikov}},\ }\bibfield  {title} {\bibinfo {title} {Universal high-frequency behavior of periodically driven systems: from dynamical stabilization to floquet engineering},\ }\href {https://doi.org/10.1080/00018732.2015.1055918} {\bibfield  {journal} {\bibinfo  {journal} {Adv. Phys.}\ }\textbf {\bibinfo {volume} {64}},\ \bibinfo {pages} {139} (\bibinfo {year} {2015})}\BibitemShut {NoStop}%
\bibitem [{\citenamefont {Giovannini}\ and\ \citenamefont {Hübener}(2019)}]{Giovannini_2020}%
  \BibitemOpen
  \bibfield  {author} {\bibinfo {author} {\bibfnamefont {U.~D.}\ \bibnamefont {Giovannini}}\ and\ \bibinfo {author} {\bibfnamefont {H.}~\bibnamefont {Hübener}},\ }\bibfield  {title} {\bibinfo {title} {Floquet analysis of excitations in materials},\ }\href {https://doi.org/10.1088/2515-7639/ab387b} {\bibfield  {journal} {\bibinfo  {journal} {J. Phys.: Mater.}\ }\textbf {\bibinfo {volume} {3}},\ \bibinfo {pages} {012001} (\bibinfo {year} {2019})}\BibitemShut {NoStop}%
\bibitem [{\citenamefont {Rodriguez-Vega}\ \emph {et~al.}(2021)\citenamefont {Rodriguez-Vega}, \citenamefont {Vogl},\ and\ \citenamefont {Fiete}}]{rodriguez2021low}%
  \BibitemOpen
  \bibfield  {author} {\bibinfo {author} {\bibfnamefont {M.}~\bibnamefont {Rodriguez-Vega}}, \bibinfo {author} {\bibfnamefont {M.}~\bibnamefont {Vogl}},\ and\ \bibinfo {author} {\bibfnamefont {G.~A.}\ \bibnamefont {Fiete}},\ }\bibfield  {title} {\bibinfo {title} {Low-frequency and moir{\'e}--floquet engineering: A review},\ }\href {https://doi.org/10.1016/j.aop.2021.168434} {\bibfield  {journal} {\bibinfo  {journal} {Ann. Phys.}\ }\textbf {\bibinfo {volume} {435}},\ \bibinfo {pages} {168434} (\bibinfo {year} {2021})}\BibitemShut {NoStop}%
\bibitem [{\citenamefont {Mori}(2023)}]{mori2023floquet}%
  \BibitemOpen
  \bibfield  {author} {\bibinfo {author} {\bibfnamefont {T.}~\bibnamefont {Mori}},\ }\bibfield  {title} {\bibinfo {title} {Floquet states in open quantum systems},\ }\href {https://doi.org/10.1146/annurev-conmatphys-040721-015537} {\bibfield  {journal} {\bibinfo  {journal} {Annu. Rev. Condens. Matter Phys.}\ }\textbf {\bibinfo {volume} {14}},\ \bibinfo {pages} {35} (\bibinfo {year} {2023})}\BibitemShut {NoStop}%
\bibitem [{\citenamefont {Eckardt}\ and\ \citenamefont {Anisimovas}(2015)}]{Eckardt_2015}%
  \BibitemOpen
  \bibfield  {author} {\bibinfo {author} {\bibfnamefont {A.}~\bibnamefont {Eckardt}}\ and\ \bibinfo {author} {\bibfnamefont {E.}~\bibnamefont {Anisimovas}},\ }\bibfield  {title} {\bibinfo {title} {High-frequency approximation for periodically driven quantum systems from a floquet-space perspective},\ }\href {https://doi.org/10.1088/1367-2630/17/9/093039} {\bibfield  {journal} {\bibinfo  {journal} {New J. Phys.}\ }\textbf {\bibinfo {volume} {17}},\ \bibinfo {pages} {093039} (\bibinfo {year} {2015})}\BibitemShut {NoStop}%
\bibitem [{\citenamefont {Casas}\ \emph {et~al.}(2001)\citenamefont {Casas}, \citenamefont {Oteo},\ and\ \citenamefont {Ros}}]{casas2001floquet}%
  \BibitemOpen
  \bibfield  {author} {\bibinfo {author} {\bibfnamefont {F.}~\bibnamefont {Casas}}, \bibinfo {author} {\bibfnamefont {J.}~\bibnamefont {Oteo}},\ and\ \bibinfo {author} {\bibfnamefont {J.}~\bibnamefont {Ros}},\ }\bibfield  {title} {\bibinfo {title} {Floquet theory: exponential perturbative treatment},\ }\href {https://iopscience.iop.org/article/10.1088/0305-4470/34/16/305} {\bibfield  {journal} {\bibinfo  {journal} {J. Phys. A: Math. Gen.}\ }\textbf {\bibinfo {volume} {34}},\ \bibinfo {pages} {3379} (\bibinfo {year} {2001})}\BibitemShut {NoStop}%
\bibitem [{\citenamefont {Mananga}\ and\ \citenamefont {Charpentier}(2011)}]{mananga2011introduction}%
  \BibitemOpen
  \bibfield  {author} {\bibinfo {author} {\bibfnamefont {E.~S.}\ \bibnamefont {Mananga}}\ and\ \bibinfo {author} {\bibfnamefont {T.}~\bibnamefont {Charpentier}},\ }\bibfield  {title} {\bibinfo {title} {Introduction of the floquet-magnus expansion in solid-state nuclear magnetic resonance spectroscopy},\ }\href {https://doi.org/10.1063/1.3610943} {\bibfield  {journal} {\bibinfo  {journal} {J. Chem. Phys.}\ }\textbf {\bibinfo {volume} {135}} (\bibinfo {year} {2011})}\BibitemShut {NoStop}%
\bibitem [{\citenamefont {Mikami}\ \emph {et~al.}(2016)\citenamefont {Mikami}, \citenamefont {Kitamura}, \citenamefont {Yasuda}, \citenamefont {Tsuji}, \citenamefont {Oka},\ and\ \citenamefont {Aoki}}]{PhysRevB.93.144307}%
  \BibitemOpen
  \bibfield  {author} {\bibinfo {author} {\bibfnamefont {T.}~\bibnamefont {Mikami}}, \bibinfo {author} {\bibfnamefont {S.}~\bibnamefont {Kitamura}}, \bibinfo {author} {\bibfnamefont {K.}~\bibnamefont {Yasuda}}, \bibinfo {author} {\bibfnamefont {N.}~\bibnamefont {Tsuji}}, \bibinfo {author} {\bibfnamefont {T.}~\bibnamefont {Oka}},\ and\ \bibinfo {author} {\bibfnamefont {H.}~\bibnamefont {Aoki}},\ }\bibfield  {title} {\bibinfo {title} {Brillouin-wigner theory for high-frequency expansion in periodically driven systems: Application to floquet topological insulators},\ }\href {https://doi.org/10.1103/PhysRevB.93.144307} {\bibfield  {journal} {\bibinfo  {journal} {Phys. Rev. B}\ }\textbf {\bibinfo {volume} {93}},\ \bibinfo {pages} {144307} (\bibinfo {year} {2016})}\BibitemShut {NoStop}%
\bibitem [{\citenamefont {Goldman}\ and\ \citenamefont {Dalibard}(2014)}]{goldman2014periodically}%
  \BibitemOpen
  \bibfield  {author} {\bibinfo {author} {\bibfnamefont {N.}~\bibnamefont {Goldman}}\ and\ \bibinfo {author} {\bibfnamefont {J.}~\bibnamefont {Dalibard}},\ }\bibfield  {title} {\bibinfo {title} {Periodically driven quantum systems: effective hamiltonians and engineered gauge fields},\ }\href {https://journals.aps.org/prx/abstract/10.1103/PhysRevX.4.031027} {\bibfield  {journal} {\bibinfo  {journal} {Phys. Rev. X}\ }\textbf {\bibinfo {volume} {4}},\ \bibinfo {pages} {031027} (\bibinfo {year} {2014})}\BibitemShut {NoStop}%
\bibitem [{\citenamefont {Verdeny}\ \emph {et~al.}(2013)\citenamefont {Verdeny}, \citenamefont {Mielke},\ and\ \citenamefont {Mintert}}]{PhysRevLett.111.175301}%
  \BibitemOpen
  \bibfield  {author} {\bibinfo {author} {\bibfnamefont {A.}~\bibnamefont {Verdeny}}, \bibinfo {author} {\bibfnamefont {A.}~\bibnamefont {Mielke}},\ and\ \bibinfo {author} {\bibfnamefont {F.}~\bibnamefont {Mintert}},\ }\bibfield  {title} {\bibinfo {title} {Accurate effective hamiltonians via unitary flow in floquet space},\ }\href {https://doi.org/10.1103/PhysRevLett.111.175301} {\bibfield  {journal} {\bibinfo  {journal} {Phys. Rev. Lett.}\ }\textbf {\bibinfo {volume} {111}},\ \bibinfo {pages} {175301} (\bibinfo {year} {2013})}\BibitemShut {NoStop}%
\bibitem [{\citenamefont {Bukov}\ \emph {et~al.}(2016{\natexlab{a}})\citenamefont {Bukov}, \citenamefont {Kolodrubetz},\ and\ \citenamefont {Polkovnikov}}]{PhysRevLett.116.125301}%
  \BibitemOpen
  \bibfield  {author} {\bibinfo {author} {\bibfnamefont {M.}~\bibnamefont {Bukov}}, \bibinfo {author} {\bibfnamefont {M.}~\bibnamefont {Kolodrubetz}},\ and\ \bibinfo {author} {\bibfnamefont {A.}~\bibnamefont {Polkovnikov}},\ }\bibfield  {title} {\bibinfo {title} {Schrieffer-wolff transformation for periodically driven systems: Strongly correlated systems with artificial gauge fields},\ }\href {https://doi.org/10.1103/PhysRevLett.116.125301} {\bibfield  {journal} {\bibinfo  {journal} {Phys. Rev. Lett.}\ }\textbf {\bibinfo {volume} {116}},\ \bibinfo {pages} {125301} (\bibinfo {year} {2016}{\natexlab{a}})}\BibitemShut {NoStop}%
\bibitem [{\citenamefont {Peronaci}\ \emph {et~al.}(2018)\citenamefont {Peronaci}, \citenamefont {Schir\'o},\ and\ \citenamefont {Parcollet}}]{PhysRevLett.120.197601}%
  \BibitemOpen
  \bibfield  {author} {\bibinfo {author} {\bibfnamefont {F.}~\bibnamefont {Peronaci}}, \bibinfo {author} {\bibfnamefont {M.}~\bibnamefont {Schir\'o}},\ and\ \bibinfo {author} {\bibfnamefont {O.}~\bibnamefont {Parcollet}},\ }\bibfield  {title} {\bibinfo {title} {Resonant thermalization of periodically driven strongly correlated electrons},\ }\href {https://doi.org/10.1103/PhysRevLett.120.197601} {\bibfield  {journal} {\bibinfo  {journal} {Phys. Rev. Lett.}\ }\textbf {\bibinfo {volume} {120}},\ \bibinfo {pages} {197601} (\bibinfo {year} {2018})}\BibitemShut {NoStop}%
\bibitem [{\citenamefont {Coulthard}\ \emph {et~al.}(2018)\citenamefont {Coulthard}, \citenamefont {Clark},\ and\ \citenamefont {Jaksch}}]{PhysRevB.98.035116}%
  \BibitemOpen
  \bibfield  {author} {\bibinfo {author} {\bibfnamefont {J.~R.}\ \bibnamefont {Coulthard}}, \bibinfo {author} {\bibfnamefont {S.~R.}\ \bibnamefont {Clark}},\ and\ \bibinfo {author} {\bibfnamefont {D.}~\bibnamefont {Jaksch}},\ }\bibfield  {title} {\bibinfo {title} {Ground-state phase diagram of the one-dimensional $t\ensuremath{-}j$ model with pair hopping terms},\ }\href {https://doi.org/10.1103/PhysRevB.98.035116} {\bibfield  {journal} {\bibinfo  {journal} {Phys. Rev. B}\ }\textbf {\bibinfo {volume} {98}},\ \bibinfo {pages} {035116} (\bibinfo {year} {2018})}\BibitemShut {NoStop}%
\bibitem [{\citenamefont {Rodriguez-Vega}\ \emph {et~al.}(2018)\citenamefont {Rodriguez-Vega}, \citenamefont {Lentz},\ and\ \citenamefont {Seradjeh}}]{Rodriguez_Vega_2018}%
  \BibitemOpen
  \bibfield  {author} {\bibinfo {author} {\bibfnamefont {M.}~\bibnamefont {Rodriguez-Vega}}, \bibinfo {author} {\bibfnamefont {M.}~\bibnamefont {Lentz}},\ and\ \bibinfo {author} {\bibfnamefont {B.}~\bibnamefont {Seradjeh}},\ }\bibfield  {title} {\bibinfo {title} {Floquet perturbation theory: formalism and application to low-frequency limit},\ }\href {https://doi.org/10.1088/1367-2630/aade37} {\bibfield  {journal} {\bibinfo  {journal} {New J. Phys.}\ }\textbf {\bibinfo {volume} {20}},\ \bibinfo {pages} {093022} (\bibinfo {year} {2018})}\BibitemShut {NoStop}%
\bibitem [{\citenamefont {Vogl}\ \emph {et~al.}(2020)\citenamefont {Vogl}, \citenamefont {Rodriguez-Vega},\ and\ \citenamefont {Fiete}}]{PhysRevB.101.024303}%
  \BibitemOpen
  \bibfield  {author} {\bibinfo {author} {\bibfnamefont {M.}~\bibnamefont {Vogl}}, \bibinfo {author} {\bibfnamefont {M.}~\bibnamefont {Rodriguez-Vega}},\ and\ \bibinfo {author} {\bibfnamefont {G.~A.}\ \bibnamefont {Fiete}},\ }\bibfield  {title} {\bibinfo {title} {Effective floquet hamiltonian in the low-frequency regime},\ }\href {https://doi.org/10.1103/PhysRevB.101.024303} {\bibfield  {journal} {\bibinfo  {journal} {Phys. Rev. B}\ }\textbf {\bibinfo {volume} {101}},\ \bibinfo {pages} {024303} (\bibinfo {year} {2020})}\BibitemShut {NoStop}%
\bibitem [{\citenamefont {Vogl}\ \emph {et~al.}(2019)\citenamefont {Vogl}, \citenamefont {Laurell}, \citenamefont {Barr},\ and\ \citenamefont {Fiete}}]{PhysRevX.9.021037}%
  \BibitemOpen
  \bibfield  {author} {\bibinfo {author} {\bibfnamefont {M.}~\bibnamefont {Vogl}}, \bibinfo {author} {\bibfnamefont {P.}~\bibnamefont {Laurell}}, \bibinfo {author} {\bibfnamefont {A.~D.}\ \bibnamefont {Barr}},\ and\ \bibinfo {author} {\bibfnamefont {G.~A.}\ \bibnamefont {Fiete}},\ }\bibfield  {title} {\bibinfo {title} {Flow equation approach to periodically driven quantum systems},\ }\href {https://doi.org/10.1103/PhysRevX.9.021037} {\bibfield  {journal} {\bibinfo  {journal} {Phys. Rev. X}\ }\textbf {\bibinfo {volume} {9}},\ \bibinfo {pages} {021037} (\bibinfo {year} {2019})}\BibitemShut {NoStop}%
\bibitem [{\citenamefont {Claassen}(2021)}]{claassen2021flow}%
  \BibitemOpen
  \bibfield  {author} {\bibinfo {author} {\bibfnamefont {M.}~\bibnamefont {Claassen}},\ }\href@noop {} {\bibinfo {title} {Flow renormalization and emergent prethermal regimes of periodically-driven quantum systems}} (\bibinfo {year} {2021}),\ \Eprint {https://arxiv.org/abs/2103.07485} {arXiv:2103.07485 [quant-ph]} \BibitemShut {NoStop}%
\bibitem [{\citenamefont {Blais}\ \emph {et~al.}(2007)\citenamefont {Blais}, \citenamefont {Gambetta}, \citenamefont {Wallraff}, \citenamefont {Schuster}, \citenamefont {Girvin}, \citenamefont {Devoret},\ and\ \citenamefont {Schoelkopf}}]{blais2007quantum}%
  \BibitemOpen
  \bibfield  {author} {\bibinfo {author} {\bibfnamefont {A.}~\bibnamefont {Blais}}, \bibinfo {author} {\bibfnamefont {J.}~\bibnamefont {Gambetta}}, \bibinfo {author} {\bibfnamefont {A.}~\bibnamefont {Wallraff}}, \bibinfo {author} {\bibfnamefont {D.~I.}\ \bibnamefont {Schuster}}, \bibinfo {author} {\bibfnamefont {S.~M.}\ \bibnamefont {Girvin}}, \bibinfo {author} {\bibfnamefont {M.~H.}\ \bibnamefont {Devoret}},\ and\ \bibinfo {author} {\bibfnamefont {R.~J.}\ \bibnamefont {Schoelkopf}},\ }\bibfield  {title} {\bibinfo {title} {Quantum-information processing with circuit quantum electrodynamics},\ }\href {https://journals.aps.org/pra/abstract/10.1103/PhysRevA.75.032329} {\bibfield  {journal} {\bibinfo  {journal} {Phys. Rev. A}\ }\textbf {\bibinfo {volume} {75}},\ \bibinfo {pages} {032329} (\bibinfo {year} {2007})}\BibitemShut {NoStop}%
\bibitem [{\citenamefont {Ann}\ \emph {et~al.}(2022)\citenamefont {Ann}, \citenamefont {Kessels},\ and\ \citenamefont {Steele}}]{ann2022two}%
  \BibitemOpen
  \bibfield  {author} {\bibinfo {author} {\bibfnamefont {B.-m.}\ \bibnamefont {Ann}}, \bibinfo {author} {\bibfnamefont {W.}~\bibnamefont {Kessels}},\ and\ \bibinfo {author} {\bibfnamefont {G.~A.}\ \bibnamefont {Steele}},\ }\bibfield  {title} {\bibinfo {title} {Two-photon sideband interaction in a driven quantum rabi model: Quantitative discussions with derived longitudinal drives and beyond the rotating wave approximation},\ }\href {https://journals.aps.org/prresearch/abstract/10.1103/PhysRevResearch.4.013005} {\bibfield  {journal} {\bibinfo  {journal} {Phys. Rev. Res}\ }\textbf {\bibinfo {volume} {4}},\ \bibinfo {pages} {013005} (\bibinfo {year} {2022})}\BibitemShut {NoStop}%
\bibitem [{\citenamefont {Or{\'u}s}(2019)}]{orus2019tensor}%
  \BibitemOpen
  \bibfield  {author} {\bibinfo {author} {\bibfnamefont {R.}~\bibnamefont {Or{\'u}s}},\ }\bibfield  {title} {\bibinfo {title} {Tensor networks for complex quantum systems},\ }\href {https://www.nature.com/articles/s42254-019-0086-7} {\bibfield  {journal} {\bibinfo  {journal} {Nat. Rev. Phys.}\ }\textbf {\bibinfo {volume} {1}},\ \bibinfo {pages} {538} (\bibinfo {year} {2019})}\BibitemShut {NoStop}%
\bibitem [{\citenamefont {Abanin}\ \emph {et~al.}(2017)\citenamefont {Abanin}, \citenamefont {De~Roeck}, \citenamefont {Ho},\ and\ \citenamefont {Huveneers}}]{abanin2017effective}%
  \BibitemOpen
  \bibfield  {author} {\bibinfo {author} {\bibfnamefont {D.~A.}\ \bibnamefont {Abanin}}, \bibinfo {author} {\bibfnamefont {W.}~\bibnamefont {De~Roeck}}, \bibinfo {author} {\bibfnamefont {W.~W.}\ \bibnamefont {Ho}},\ and\ \bibinfo {author} {\bibfnamefont {F.}~\bibnamefont {Huveneers}},\ }\bibfield  {title} {\bibinfo {title} {Effective hamiltonians, prethermalization, and slow energy absorption in periodically driven many-body systems},\ }\href {https://link.aps.org/doi/10.1103/PhysRevB.95.014112} {\bibfield  {journal} {\bibinfo  {journal} {Phys. Rev. B}\ }\textbf {\bibinfo {volume} {95}},\ \bibinfo {pages} {014112} (\bibinfo {year} {2017})}\BibitemShut {NoStop}%
\bibitem [{Note1()}]{Note1}%
  \BibitemOpen
  \bibinfo {note} {In the evolution operator described by Eq.~(\ref {eq.propagator}), the middle part, $e^{-i\protect \hat {H}'(t-t_0)}$, will be identified in Floquet theory as the stroboscopic dynamics, i.e., the Floquet macro-motion.}\BibitemShut {Stop}%
\bibitem [{Note2()}]{Note2}%
  \BibitemOpen
  \bibinfo {note} {To detail this iteration, we start at $n=1$ in Eq.~(\ref {H'-expansion}): The requirement $\partial _t \protect \hat {H}'^{(1)}_t = 0$ on (\ref {H'-expansion-1}) eliminates the time-dependence to order $g^1$ in $\protect \hat {H}'$, from which $\protect \hat {F}_t^{(1)}$ is determined. Inserting $\protect \hat {F}_t^{(1)}$ back to Eq.~(\ref {H'-expansion}) not only renders the entire $\protect \hat {H}'^{(1)}_t$ static in (\ref {H'-expansion-1}), but also completely determines the static part of $\protect \hat {H}'^{(2)}_t$ in (\ref {H'-expansion-2}). Then we proceed to $n=2$: Using the function $\protect \hat {F}_t^{(1)}$ derived above, the equation $\partial _t \protect \hat {H}'^{(2)}_t = 0$ for (\ref {H'-expansion-2}) eliminates the time-dependence to order $g^2$ in $\protect \hat {H}'$, determining $\protect \hat {F}_t^{(2)}$. This elimination procedure can be repeated iteratively. Once the $n$-th order equation in Eq.~(\ref {H'-expansion}) is solved, the function $\protect \hat
  {F}_t^{(n)}$ will be determined which fixes the static part of $\protect \hat {H}'^{(n+1)}_t$.}\BibitemShut {Stop}%
\bibitem [{\citenamefont {Bhatia}\ and\ \citenamefont {Rosenthal}(1997)}]{bhatia1997and}%
  \BibitemOpen
  \bibfield  {author} {\bibinfo {author} {\bibfnamefont {R.}~\bibnamefont {Bhatia}}\ and\ \bibinfo {author} {\bibfnamefont {P.}~\bibnamefont {Rosenthal}},\ }\bibfield  {title} {\bibinfo {title} {How and why to solve the operator equation ax- xb= y},\ }\href {https://doi.org/10.1112/S0024609396001828} {\bibfield  {journal} {\bibinfo  {journal} {Bull. London Math. Soc.}\ }\textbf {\bibinfo {volume} {29}},\ \bibinfo {pages} {1} (\bibinfo {year} {1997})}\BibitemShut {NoStop}%
\bibitem [{\citenamefont {Wang}\ \emph {et~al.}(2024)\citenamefont {Wang}, \citenamefont {Jaksch},\ and\ \citenamefont {Schlawin}}]{PhysRevB.109.115137}%
  \BibitemOpen
  \bibfield  {author} {\bibinfo {author} {\bibfnamefont {X.}~\bibnamefont {Wang}}, \bibinfo {author} {\bibfnamefont {D.}~\bibnamefont {Jaksch}},\ and\ \bibinfo {author} {\bibfnamefont {F.}~\bibnamefont {Schlawin}},\ }\bibfield  {title} {\bibinfo {title} {Excitonic enhancement of cavity-mediated interactions in a two-band hubbard model},\ }\href {https://doi.org/10.1103/PhysRevB.109.115137} {\bibfield  {journal} {\bibinfo  {journal} {Phys. Rev. B}\ }\textbf {\bibinfo {volume} {109}},\ \bibinfo {pages} {115137} (\bibinfo {year} {2024})}\BibitemShut {NoStop}%
\bibitem [{\citenamefont {Lignier}\ \emph {et~al.}(2007)\citenamefont {Lignier}, \citenamefont {Sias}, \citenamefont {Ciampini}, \citenamefont {Singh}, \citenamefont {Zenesini}, \citenamefont {Morsch},\ and\ \citenamefont {Arimondo}}]{PhysRevLett.99.220403}%
  \BibitemOpen
  \bibfield  {author} {\bibinfo {author} {\bibfnamefont {H.}~\bibnamefont {Lignier}}, \bibinfo {author} {\bibfnamefont {C.}~\bibnamefont {Sias}}, \bibinfo {author} {\bibfnamefont {D.}~\bibnamefont {Ciampini}}, \bibinfo {author} {\bibfnamefont {Y.}~\bibnamefont {Singh}}, \bibinfo {author} {\bibfnamefont {A.}~\bibnamefont {Zenesini}}, \bibinfo {author} {\bibfnamefont {O.}~\bibnamefont {Morsch}},\ and\ \bibinfo {author} {\bibfnamefont {E.}~\bibnamefont {Arimondo}},\ }\bibfield  {title} {\bibinfo {title} {Dynamical control of matter-wave tunneling in periodic potentials},\ }\href {https://doi.org/10.1103/PhysRevLett.99.220403} {\bibfield  {journal} {\bibinfo  {journal} {Phys. Rev. Lett.}\ }\textbf {\bibinfo {volume} {99}},\ \bibinfo {pages} {220403} (\bibinfo {year} {2007})}\BibitemShut {NoStop}%
\bibitem [{\citenamefont {Hensgens}\ \emph {et~al.}(2017)\citenamefont {Hensgens}, \citenamefont {Fujita}, \citenamefont {Janssen}, \citenamefont {Li}, \citenamefont {Van~Diepen}, \citenamefont {Reichl}, \citenamefont {Wegscheider}, \citenamefont {Das~Sarma},\ and\ \citenamefont {Vandersypen}}]{hensgens2017quantum}%
  \BibitemOpen
  \bibfield  {author} {\bibinfo {author} {\bibfnamefont {T.}~\bibnamefont {Hensgens}}, \bibinfo {author} {\bibfnamefont {T.}~\bibnamefont {Fujita}}, \bibinfo {author} {\bibfnamefont {L.}~\bibnamefont {Janssen}}, \bibinfo {author} {\bibfnamefont {X.}~\bibnamefont {Li}}, \bibinfo {author} {\bibfnamefont {C.}~\bibnamefont {Van~Diepen}}, \bibinfo {author} {\bibfnamefont {C.}~\bibnamefont {Reichl}}, \bibinfo {author} {\bibfnamefont {W.}~\bibnamefont {Wegscheider}}, \bibinfo {author} {\bibfnamefont {S.}~\bibnamefont {Das~Sarma}},\ and\ \bibinfo {author} {\bibfnamefont {L.~M.}\ \bibnamefont {Vandersypen}},\ }\bibfield  {title} {\bibinfo {title} {Quantum simulation of a fermi--hubbard model using a semiconductor quantum dot array},\ }\href {https://www.nature.com/articles/nature23022} {\bibfield  {journal} {\bibinfo  {journal} {Nature}\ }\textbf {\bibinfo {volume} {548}},\ \bibinfo {pages} {70} (\bibinfo {year} {2017})}\BibitemShut {NoStop}%
\bibitem [{\citenamefont {D'Alessio}\ and\ \citenamefont {Rigol}(2014)}]{PhysRevX.4.041048}%
  \BibitemOpen
  \bibfield  {author} {\bibinfo {author} {\bibfnamefont {L.}~\bibnamefont {D'Alessio}}\ and\ \bibinfo {author} {\bibfnamefont {M.}~\bibnamefont {Rigol}},\ }\bibfield  {title} {\bibinfo {title} {Long-time behavior of isolated periodically driven interacting lattice systems},\ }\href {https://doi.org/10.1103/PhysRevX.4.041048} {\bibfield  {journal} {\bibinfo  {journal} {Phys. Rev. X}\ }\textbf {\bibinfo {volume} {4}},\ \bibinfo {pages} {041048} (\bibinfo {year} {2014})}\BibitemShut {NoStop}%
\bibitem [{\citenamefont {Dunlap}\ and\ \citenamefont {Kenkre}(1986)}]{PhysRevB.34.3625}%
  \BibitemOpen
  \bibfield  {author} {\bibinfo {author} {\bibfnamefont {D.~H.}\ \bibnamefont {Dunlap}}\ and\ \bibinfo {author} {\bibfnamefont {V.~M.}\ \bibnamefont {Kenkre}},\ }\bibfield  {title} {\bibinfo {title} {Dynamic localization of a charged particle moving under the influence of an electric field},\ }\href {https://doi.org/10.1103/PhysRevB.34.3625} {\bibfield  {journal} {\bibinfo  {journal} {Phys. Rev. B}\ }\textbf {\bibinfo {volume} {34}},\ \bibinfo {pages} {3625} (\bibinfo {year} {1986})}\BibitemShut {NoStop}%
\bibitem [{\citenamefont {Eckardt}\ \emph {et~al.}(2005)\citenamefont {Eckardt}, \citenamefont {Weiss},\ and\ \citenamefont {Holthaus}}]{PhysRevLett.95.260404}%
  \BibitemOpen
  \bibfield  {author} {\bibinfo {author} {\bibfnamefont {A.}~\bibnamefont {Eckardt}}, \bibinfo {author} {\bibfnamefont {C.}~\bibnamefont {Weiss}},\ and\ \bibinfo {author} {\bibfnamefont {M.}~\bibnamefont {Holthaus}},\ }\bibfield  {title} {\bibinfo {title} {Superfluid-insulator transition in a periodically driven optical lattice},\ }\href {https://doi.org/10.1103/PhysRevLett.95.260404} {\bibfield  {journal} {\bibinfo  {journal} {Phys. Rev. Lett.}\ }\textbf {\bibinfo {volume} {95}},\ \bibinfo {pages} {260404} (\bibinfo {year} {2005})}\BibitemShut {NoStop}%
\bibitem [{\citenamefont {Tsuji}\ \emph {et~al.}(2008)\citenamefont {Tsuji}, \citenamefont {Oka},\ and\ \citenamefont {Aoki}}]{PhysRevB.78.235124}%
  \BibitemOpen
  \bibfield  {author} {\bibinfo {author} {\bibfnamefont {N.}~\bibnamefont {Tsuji}}, \bibinfo {author} {\bibfnamefont {T.}~\bibnamefont {Oka}},\ and\ \bibinfo {author} {\bibfnamefont {H.}~\bibnamefont {Aoki}},\ }\bibfield  {title} {\bibinfo {title} {Correlated electron systems periodically driven out of equilibrium: $\text{Floquet}+\text{DMFT}$ formalism},\ }\href {https://doi.org/10.1103/PhysRevB.78.235124} {\bibfield  {journal} {\bibinfo  {journal} {Phys. Rev. B}\ }\textbf {\bibinfo {volume} {78}},\ \bibinfo {pages} {235124} (\bibinfo {year} {2008})}\BibitemShut {NoStop}%
\bibitem [{\citenamefont {Madison}\ \emph {et~al.}(1998)\citenamefont {Madison}, \citenamefont {Fischer}, \citenamefont {Diener}, \citenamefont {Niu},\ and\ \citenamefont {Raizen}}]{PhysRevLett.81.5093}%
  \BibitemOpen
  \bibfield  {author} {\bibinfo {author} {\bibfnamefont {K.~W.}\ \bibnamefont {Madison}}, \bibinfo {author} {\bibfnamefont {M.~C.}\ \bibnamefont {Fischer}}, \bibinfo {author} {\bibfnamefont {R.~B.}\ \bibnamefont {Diener}}, \bibinfo {author} {\bibfnamefont {Q.}~\bibnamefont {Niu}},\ and\ \bibinfo {author} {\bibfnamefont {M.~G.}\ \bibnamefont {Raizen}},\ }\bibfield  {title} {\bibinfo {title} {Dynamical bloch band suppression in an optical lattice},\ }\href {https://doi.org/10.1103/PhysRevLett.81.5093} {\bibfield  {journal} {\bibinfo  {journal} {Phys. Rev. Lett.}\ }\textbf {\bibinfo {volume} {81}},\ \bibinfo {pages} {5093} (\bibinfo {year} {1998})}\BibitemShut {NoStop}%
\bibitem [{\citenamefont {Dolcini}\ and\ \citenamefont {Montorsi}(2013)}]{PhysRevB.88.115115}%
  \BibitemOpen
  \bibfield  {author} {\bibinfo {author} {\bibfnamefont {F.}~\bibnamefont {Dolcini}}\ and\ \bibinfo {author} {\bibfnamefont {A.}~\bibnamefont {Montorsi}},\ }\bibfield  {title} {\bibinfo {title} {Quantum phases of one-dimensional hubbard models with three- and four-body couplings},\ }\href {https://doi.org/10.1103/PhysRevB.88.115115} {\bibfield  {journal} {\bibinfo  {journal} {Phys. Rev. B}\ }\textbf {\bibinfo {volume} {88}},\ \bibinfo {pages} {115115} (\bibinfo {year} {2013})}\BibitemShut {NoStop}%
\bibitem [{\citenamefont {Duan}(2007)}]{Duan_2008}%
  \BibitemOpen
  \bibfield  {author} {\bibinfo {author} {\bibfnamefont {L.-M.}\ \bibnamefont {Duan}},\ }\bibfield  {title} {\bibinfo {title} {General hubbard model for strongly interacting fermions in an optical lattice and its phase detection},\ }\href {https://dx.doi.org/10.1209/0295-5075/81/20001} {\bibfield  {journal} {\bibinfo  {journal} {Europhys. Lett.}\ }\textbf {\bibinfo {volume} {81}},\ \bibinfo {pages} {20001} (\bibinfo {year} {2007})}\BibitemShut {NoStop}%
\bibitem [{\citenamefont {Kiffner}\ \emph {et~al.}(2019)\citenamefont {Kiffner}, \citenamefont {Coulthard}, \citenamefont {Schlawin}, \citenamefont {Ardavan},\ and\ \citenamefont {Jaksch}}]{PhysRevB.99.085116}%
  \BibitemOpen
  \bibfield  {author} {\bibinfo {author} {\bibfnamefont {M.}~\bibnamefont {Kiffner}}, \bibinfo {author} {\bibfnamefont {J.~R.}\ \bibnamefont {Coulthard}}, \bibinfo {author} {\bibfnamefont {F.}~\bibnamefont {Schlawin}}, \bibinfo {author} {\bibfnamefont {A.}~\bibnamefont {Ardavan}},\ and\ \bibinfo {author} {\bibfnamefont {D.}~\bibnamefont {Jaksch}},\ }\bibfield  {title} {\bibinfo {title} {Manipulating quantum materials with quantum light},\ }\href {https://doi.org/10.1103/PhysRevB.99.085116} {\bibfield  {journal} {\bibinfo  {journal} {Phys. Rev. B}\ }\textbf {\bibinfo {volume} {99}},\ \bibinfo {pages} {085116} (\bibinfo {year} {2019})}\BibitemShut {NoStop}%
\bibitem [{\citenamefont {Mark}\ and\ \citenamefont {Motrunich}(2020)}]{PhysRevB.102.075132}%
  \BibitemOpen
  \bibfield  {author} {\bibinfo {author} {\bibfnamefont {D.~K.}\ \bibnamefont {Mark}}\ and\ \bibinfo {author} {\bibfnamefont {O.~I.}\ \bibnamefont {Motrunich}},\ }\bibfield  {title} {\bibinfo {title} {$\ensuremath{\eta}$-pairing states as true scars in an extended hubbard model},\ }\href {https://doi.org/10.1103/PhysRevB.102.075132} {\bibfield  {journal} {\bibinfo  {journal} {Phys. Rev. B}\ }\textbf {\bibinfo {volume} {102}},\ \bibinfo {pages} {075132} (\bibinfo {year} {2020})}\BibitemShut {NoStop}%
\bibitem [{\citenamefont {Micnas}\ \emph {et~al.}(1989)\citenamefont {Micnas}, \citenamefont {Ranninger},\ and\ \citenamefont {Robaszkiewicz}}]{PhysRevB.39.11653}%
  \BibitemOpen
  \bibfield  {author} {\bibinfo {author} {\bibfnamefont {R.}~\bibnamefont {Micnas}}, \bibinfo {author} {\bibfnamefont {J.}~\bibnamefont {Ranninger}},\ and\ \bibinfo {author} {\bibfnamefont {S.}~\bibnamefont {Robaszkiewicz}},\ }\bibfield  {title} {\bibinfo {title} {Superconductivity in a narrow-band system with intersite electron pairing in two dimensions. ii. effects of nearest-neighbor exchange and correlated hopping},\ }\href {https://doi.org/10.1103/PhysRevB.39.11653} {\bibfield  {journal} {\bibinfo  {journal} {Phys. Rev. B}\ }\textbf {\bibinfo {volume} {39}},\ \bibinfo {pages} {11653} (\bibinfo {year} {1989})}\BibitemShut {NoStop}%
\bibitem [{\citenamefont {Gardas}\ \emph {et~al.}(2017)\citenamefont {Gardas}, \citenamefont {Dziarmaga},\ and\ \citenamefont {Zurek}}]{PhysRevB.95.104306}%
  \BibitemOpen
  \bibfield  {author} {\bibinfo {author} {\bibfnamefont {B.}~\bibnamefont {Gardas}}, \bibinfo {author} {\bibfnamefont {J.}~\bibnamefont {Dziarmaga}},\ and\ \bibinfo {author} {\bibfnamefont {W.~H.}\ \bibnamefont {Zurek}},\ }\bibfield  {title} {\bibinfo {title} {Dynamics of the quantum phase transition in the one-dimensional bose-hubbard model: Excitations and correlations induced by a quench},\ }\href {https://doi.org/10.1103/PhysRevB.95.104306} {\bibfield  {journal} {\bibinfo  {journal} {Phys. Rev. B}\ }\textbf {\bibinfo {volume} {95}},\ \bibinfo {pages} {104306} (\bibinfo {year} {2017})}\BibitemShut {NoStop}%
\bibitem [{\citenamefont {Kennes}\ \emph {et~al.}(2018)\citenamefont {Kennes}, \citenamefont {de~la Torre}, \citenamefont {Ron}, \citenamefont {Hsieh},\ and\ \citenamefont {Millis}}]{PhysRevLett.120.127601}%
  \BibitemOpen
  \bibfield  {author} {\bibinfo {author} {\bibfnamefont {D.~M.}\ \bibnamefont {Kennes}}, \bibinfo {author} {\bibfnamefont {A.}~\bibnamefont {de~la Torre}}, \bibinfo {author} {\bibfnamefont {A.}~\bibnamefont {Ron}}, \bibinfo {author} {\bibfnamefont {D.}~\bibnamefont {Hsieh}},\ and\ \bibinfo {author} {\bibfnamefont {A.~J.}\ \bibnamefont {Millis}},\ }\bibfield  {title} {\bibinfo {title} {Floquet engineering in quantum chains},\ }\href {https://doi.org/10.1103/PhysRevLett.120.127601} {\bibfield  {journal} {\bibinfo  {journal} {Phys. Rev. Lett.}\ }\textbf {\bibinfo {volume} {120}},\ \bibinfo {pages} {127601} (\bibinfo {year} {2018})}\BibitemShut {NoStop}%
\bibitem [{\citenamefont {Hauschild}\ and\ \citenamefont {Pollmann}(2018)}]{tenpy}%
  \BibitemOpen
  \bibfield  {author} {\bibinfo {author} {\bibfnamefont {J.}~\bibnamefont {Hauschild}}\ and\ \bibinfo {author} {\bibfnamefont {F.}~\bibnamefont {Pollmann}},\ }\bibfield  {title} {\bibinfo {title} {{Efficient numerical simulations with Tensor Networks: Tensor Network Python (TeNPy)}},\ }\href {https://doi.org/10.21468/SciPostPhysLectNotes.5} {\bibfield  {journal} {\bibinfo  {journal} {SciPost Phys. Lect. Notes}\ ,\ \bibinfo {pages} {5}} (\bibinfo {year} {2018})},\ \Eprint {https://arxiv.org/abs/1805.00055} {arXiv:1805.00055} \BibitemShut {NoStop}%
\bibitem [{\citenamefont {Essler}\ \emph {et~al.}(2005)\citenamefont {Essler}, \citenamefont {Frahm}, \citenamefont {G{\"o}hmann}, \citenamefont {Kl{\"u}mper},\ and\ \citenamefont {Korepin}}]{essler2005one}%
  \BibitemOpen
  \bibfield  {author} {\bibinfo {author} {\bibfnamefont {F.~H.}\ \bibnamefont {Essler}}, \bibinfo {author} {\bibfnamefont {H.}~\bibnamefont {Frahm}}, \bibinfo {author} {\bibfnamefont {F.}~\bibnamefont {G{\"o}hmann}}, \bibinfo {author} {\bibfnamefont {A.}~\bibnamefont {Kl{\"u}mper}},\ and\ \bibinfo {author} {\bibfnamefont {V.~E.}\ \bibnamefont {Korepin}},\ }\href {https://www.cambridge.org/core/books/onedimensional-hubbard-model/7F70AFCF477126DD8C9C9D1310910E5A} {\emph {\bibinfo {title} {The one-dimensional Hubbard model}}}\ (\bibinfo  {publisher} {Cambridge University Press},\ \bibinfo {year} {2005})\BibitemShut {NoStop}%
\bibitem [{\citenamefont {Mendoza-Arenas}(2022)}]{Mendoza-Arenas_2022}%
  \BibitemOpen
  \bibfield  {author} {\bibinfo {author} {\bibfnamefont {J.~J.}\ \bibnamefont {Mendoza-Arenas}},\ }\bibfield  {title} {\bibinfo {title} {Dynamical quantum phase transitions in the one-dimensional extended fermi–hubbard model},\ }\href {https://doi.org/10.1088/1742-5468/ac6031} {\bibfield  {journal} {\bibinfo  {journal} {J. Stat. Mech: Theory Exp.}\ }\textbf {\bibinfo {volume} {2022}},\ \bibinfo {pages} {043101} (\bibinfo {year} {2022})}\BibitemShut {NoStop}%
\bibitem [{Note3()}]{Note3}%
  \BibitemOpen
  \bibinfo {note} {The Floquet micro-motion will be ignored in the following, which is a good approximation since the return rate $\protect \mathcal {L}_{(t)}$ we simulate represents an expectation value (of the observable $ \vert \psi _{CDW} \rangle \langle \psi _{CDW} \vert $). If we instead simulated the Floquet fidelity, i.e. the wavefunction overlap between the Floquet evolution and the exact dynamics, then we would have to take the micro-motion into account.}\BibitemShut {Stop}%
\bibitem [{\citenamefont {Sharma}\ \emph {et~al.}(2014)\citenamefont {Sharma}, \citenamefont {Russomanno}, \citenamefont {Santoro},\ and\ \citenamefont {Dutta}}]{Sharma_2014}%
  \BibitemOpen
  \bibfield  {author} {\bibinfo {author} {\bibfnamefont {S.}~\bibnamefont {Sharma}}, \bibinfo {author} {\bibfnamefont {A.}~\bibnamefont {Russomanno}}, \bibinfo {author} {\bibfnamefont {G.~E.}\ \bibnamefont {Santoro}},\ and\ \bibinfo {author} {\bibfnamefont {A.}~\bibnamefont {Dutta}},\ }\bibfield  {title} {\bibinfo {title} {Loschmidt echo and dynamical fidelity in periodically driven quantum systems},\ }\href {https://doi.org/10.1209/0295-5075/106/67003} {\bibfield  {journal} {\bibinfo  {journal} {Europhys. Lett.}\ }\textbf {\bibinfo {volume} {106}},\ \bibinfo {pages} {67003} (\bibinfo {year} {2014})}\BibitemShut {NoStop}%
\bibitem [{Note4()}]{Note4}%
  \BibitemOpen
  \bibinfo {note} {We note that the $\protect \mathcal {O}(J^2)$ part of the FSWT Floquet Hamiltonian (\ref {Floquet-H'}), given by $\protect \frac {1}{2}([\protect \hat {y}_2,\protect \hat {H}^{(1)}_{-1}]+ H.c.)$ in Eq.~(\ref {Floquet-H'2-t2}), is vital for accurately describing the dynamics involving the doublon excitations.}\BibitemShut {Stop}%
\bibitem [{\citenamefont {Okamoto}\ and\ \citenamefont {Peronaci}(2021)}]{okamoto2021floquet}%
  \BibitemOpen
  \bibfield  {author} {\bibinfo {author} {\bibfnamefont {J.}~\bibnamefont {Okamoto}}\ and\ \bibinfo {author} {\bibfnamefont {F.}~\bibnamefont {Peronaci}},\ }\bibfield  {title} {\bibinfo {title} {Floquet prethermalization and rabi oscillations in optically excited hubbard clusters},\ }\href {https://www.nature.com/articles/s41598-021-97104-x} {\bibfield  {journal} {\bibinfo  {journal} {Sci. Rep.}\ }\textbf {\bibinfo {volume} {11}},\ \bibinfo {pages} {1} (\bibinfo {year} {2021})}\BibitemShut {NoStop}%
\bibitem [{\citenamefont {Haegeman}\ \emph {et~al.}(2011)\citenamefont {Haegeman}, \citenamefont {Cirac}, \citenamefont {Osborne}, \citenamefont {Pi\ifmmode~\check{z}\else \v{z}\fi{}orn}, \citenamefont {Verschelde},\ and\ \citenamefont {Verstraete}}]{PhysRevLett.107.070601}%
  \BibitemOpen
  \bibfield  {author} {\bibinfo {author} {\bibfnamefont {J.}~\bibnamefont {Haegeman}}, \bibinfo {author} {\bibfnamefont {J.~I.}\ \bibnamefont {Cirac}}, \bibinfo {author} {\bibfnamefont {T.~J.}\ \bibnamefont {Osborne}}, \bibinfo {author} {\bibfnamefont {I.}~\bibnamefont {Pi\ifmmode~\check{z}\else \v{z}\fi{}orn}}, \bibinfo {author} {\bibfnamefont {H.}~\bibnamefont {Verschelde}},\ and\ \bibinfo {author} {\bibfnamefont {F.}~\bibnamefont {Verstraete}},\ }\bibfield  {title} {\bibinfo {title} {Time-dependent variational principle for quantum lattices},\ }\href {https://link.aps.org/doi/10.1103/PhysRevLett.107.070601} {\bibfield  {journal} {\bibinfo  {journal} {Phys. Rev. Lett.}\ }\textbf {\bibinfo {volume} {107}},\ \bibinfo {pages} {070601} (\bibinfo {year} {2011})}\BibitemShut {NoStop}%
\bibitem [{\citenamefont {Haegeman}\ \emph {et~al.}(2016)\citenamefont {Haegeman}, \citenamefont {Lubich}, \citenamefont {Oseledets}, \citenamefont {Vandereycken},\ and\ \citenamefont {Verstraete}}]{PhysRevB.94.165116}%
  \BibitemOpen
  \bibfield  {author} {\bibinfo {author} {\bibfnamefont {J.}~\bibnamefont {Haegeman}}, \bibinfo {author} {\bibfnamefont {C.}~\bibnamefont {Lubich}}, \bibinfo {author} {\bibfnamefont {I.}~\bibnamefont {Oseledets}}, \bibinfo {author} {\bibfnamefont {B.}~\bibnamefont {Vandereycken}},\ and\ \bibinfo {author} {\bibfnamefont {F.}~\bibnamefont {Verstraete}},\ }\bibfield  {title} {\bibinfo {title} {Unifying time evolution and optimization with matrix product states},\ }\href {https://link.aps.org/doi/10.1103/PhysRevB.94.165116} {\bibfield  {journal} {\bibinfo  {journal} {Phys. Rev. B}\ }\textbf {\bibinfo {volume} {94}},\ \bibinfo {pages} {165116} (\bibinfo {year} {2016})}\BibitemShut {NoStop}%
\bibitem [{Note5()}]{Note5}%
  \BibitemOpen
  \bibinfo {note} {We can treat $\protect \hat {f}^{(1)}_1$ as a matrix product operator, and solve the Sylvester equation (\ref {FSWT-f^1_1}) using tensor network methods. The corresponding lowest order Floquet Hamiltonian $\protect \hat {H}^{0} + \protect \hat {H}'^{(2)}$ is directly given in an MPO form, which has large bond dimension so that it cannot be explicitly written down analytically, but this MPO can be directly used in the DMRG to predict the steady state under weak drive. This Sylvester-based MPO approach for Floquet Hamiltonian is free from the many-body quasi-energy issue faced by Sambe space DMRG methods like Ref~\cite {sahoo2019periodically}.}\BibitemShut {Stop}%
\bibitem [{\citenamefont {Bukov}\ \emph {et~al.}(2016{\natexlab{b}})\citenamefont {Bukov}, \citenamefont {Heyl}, \citenamefont {Huse},\ and\ \citenamefont {Polkovnikov}}]{PhysRevB.93.155132}%
  \BibitemOpen
  \bibfield  {author} {\bibinfo {author} {\bibfnamefont {M.}~\bibnamefont {Bukov}}, \bibinfo {author} {\bibfnamefont {M.}~\bibnamefont {Heyl}}, \bibinfo {author} {\bibfnamefont {D.~A.}\ \bibnamefont {Huse}},\ and\ \bibinfo {author} {\bibfnamefont {A.}~\bibnamefont {Polkovnikov}},\ }\bibfield  {title} {\bibinfo {title} {Heating and many-body resonances in a periodically driven two-band system},\ }\href {https://doi.org/10.1103/PhysRevB.93.155132} {\bibfield  {journal} {\bibinfo  {journal} {Phys. Rev. B}\ }\textbf {\bibinfo {volume} {93}},\ \bibinfo {pages} {155132} (\bibinfo {year} {2016}{\natexlab{b}})}\BibitemShut {NoStop}%
\bibitem [{\citenamefont {Herrmann}\ \emph {et~al.}(2018)\citenamefont {Herrmann}, \citenamefont {Murakami}, \citenamefont {Eckstein},\ and\ \citenamefont {Werner}}]{Herrmann_2017}%
  \BibitemOpen
  \bibfield  {author} {\bibinfo {author} {\bibfnamefont {A.}~\bibnamefont {Herrmann}}, \bibinfo {author} {\bibfnamefont {Y.}~\bibnamefont {Murakami}}, \bibinfo {author} {\bibfnamefont {M.}~\bibnamefont {Eckstein}},\ and\ \bibinfo {author} {\bibfnamefont {P.}~\bibnamefont {Werner}},\ }\bibfield  {title} {\bibinfo {title} {Floquet prethermalization in the resonantly driven hubbard model},\ }\href {https://doi.org/10.1209/0295-5075/120/57001} {\bibfield  {journal} {\bibinfo  {journal} {Europhys. Lett.}\ }\textbf {\bibinfo {volume} {120}},\ \bibinfo {pages} {57001} (\bibinfo {year} {2018})}\BibitemShut {NoStop}%
\bibitem [{\citenamefont {Mendoza-Arenas}\ \emph {et~al.}(2017)\citenamefont {Mendoza-Arenas}, \citenamefont {Gómez-Ruiz}, \citenamefont {Eckstein}, \citenamefont {Jaksch},\ and\ \citenamefont {Clark}}]{https://doi.org/10.1002/andp.201700024}%
  \BibitemOpen
  \bibfield  {author} {\bibinfo {author} {\bibfnamefont {J.~J.}\ \bibnamefont {Mendoza-Arenas}}, \bibinfo {author} {\bibfnamefont {F.~J.}\ \bibnamefont {Gómez-Ruiz}}, \bibinfo {author} {\bibfnamefont {M.}~\bibnamefont {Eckstein}}, \bibinfo {author} {\bibfnamefont {D.}~\bibnamefont {Jaksch}},\ and\ \bibinfo {author} {\bibfnamefont {S.~R.}\ \bibnamefont {Clark}},\ }\bibfield  {title} {\bibinfo {title} {Ultra-fast control of magnetic relaxation in a periodically driven hubbard model},\ }\href {https://onlinelibrary.wiley.com/doi/abs/10.1002/andp.201700024} {\bibfield  {journal} {\bibinfo  {journal} {Annalen der Physik}\ }\textbf {\bibinfo {volume} {529}},\ \bibinfo {pages} {1700024} (\bibinfo {year} {2017})}\BibitemShut {NoStop}%
\bibitem [{\citenamefont {Spalek}(2007)}]{spalek2007tj}%
  \BibitemOpen
  \bibfield  {author} {\bibinfo {author} {\bibfnamefont {J.}~\bibnamefont {Spalek}},\ }\href@noop {} {\bibinfo {title} {tj model then and now: a personal perspective from the pioneering times}} (\bibinfo {year} {2007}),\ \Eprint {https://arxiv.org/abs/0706.4236} {arXiv:0706.4236 [cond-mat.str-el]} \BibitemShut {NoStop}%
\bibitem [{\citenamefont {Mentink}\ \emph {et~al.}(2015)\citenamefont {Mentink}, \citenamefont {Balzer},\ and\ \citenamefont {Eckstein}}]{mentink2015ultrafast}%
  \BibitemOpen
  \bibfield  {author} {\bibinfo {author} {\bibfnamefont {J.}~\bibnamefont {Mentink}}, \bibinfo {author} {\bibfnamefont {K.}~\bibnamefont {Balzer}},\ and\ \bibinfo {author} {\bibfnamefont {M.}~\bibnamefont {Eckstein}},\ }\bibfield  {title} {\bibinfo {title} {Ultrafast and reversible control of the exchange interaction in mott insulators},\ }\href {https://www.nature.com/articles/ncomms7708} {\bibfield  {journal} {\bibinfo  {journal} {Nat. Commun.}\ }\textbf {\bibinfo {volume} {6}},\ \bibinfo {pages} {1} (\bibinfo {year} {2015})}\BibitemShut {NoStop}%
\bibitem [{\citenamefont {Coulthard}\ \emph {et~al.}(2017)\citenamefont {Coulthard}, \citenamefont {Clark}, \citenamefont {Al-Assam}, \citenamefont {Cavalleri},\ and\ \citenamefont {Jaksch}}]{PhysRevB.96.085104}%
  \BibitemOpen
  \bibfield  {author} {\bibinfo {author} {\bibfnamefont {J.~R.}\ \bibnamefont {Coulthard}}, \bibinfo {author} {\bibfnamefont {S.~R.}\ \bibnamefont {Clark}}, \bibinfo {author} {\bibfnamefont {S.}~\bibnamefont {Al-Assam}}, \bibinfo {author} {\bibfnamefont {A.}~\bibnamefont {Cavalleri}},\ and\ \bibinfo {author} {\bibfnamefont {D.}~\bibnamefont {Jaksch}},\ }\bibfield  {title} {\bibinfo {title} {Enhancement of superexchange pairing in the periodically driven hubbard model},\ }\href {https://doi.org/10.1103/PhysRevB.96.085104} {\bibfield  {journal} {\bibinfo  {journal} {Phys. Rev. B}\ }\textbf {\bibinfo {volume} {96}},\ \bibinfo {pages} {085104} (\bibinfo {year} {2017})}\BibitemShut {NoStop}%
\bibitem [{\citenamefont {Sie}\ \emph {et~al.}(2017)\citenamefont {Sie}, \citenamefont {Lui}, \citenamefont {Lee}, \citenamefont {Fu}, \citenamefont {Kong},\ and\ \citenamefont {Gedik}}]{Sie2018}%
  \BibitemOpen
  \bibfield  {author} {\bibinfo {author} {\bibfnamefont {E.~J.}\ \bibnamefont {Sie}}, \bibinfo {author} {\bibfnamefont {C.~H.}\ \bibnamefont {Lui}}, \bibinfo {author} {\bibfnamefont {Y.-H.}\ \bibnamefont {Lee}}, \bibinfo {author} {\bibfnamefont {L.}~\bibnamefont {Fu}}, \bibinfo {author} {\bibfnamefont {J.}~\bibnamefont {Kong}},\ and\ \bibinfo {author} {\bibfnamefont {N.}~\bibnamefont {Gedik}},\ }\bibfield  {title} {\bibinfo {title} {Large, valley-exclusive bloch-siegert shift in monolayer ws2},\ }\href {https://doi.org/10.1126/science.aal2241} {\bibfield  {journal} {\bibinfo  {journal} {Science}\ }\textbf {\bibinfo {volume} {355}},\ \bibinfo {pages} {1066} (\bibinfo {year} {2017})}\BibitemShut {NoStop}%
\bibitem [{\citenamefont {Oliver}\ \emph {et~al.}(2005)\citenamefont {Oliver}, \citenamefont {Yu}, \citenamefont {Lee}, \citenamefont {Berggren}, \citenamefont {Levitov},\ and\ \citenamefont {Orlando}}]{doi:10.1126/science.1119678}%
  \BibitemOpen
  \bibfield  {author} {\bibinfo {author} {\bibfnamefont {W.~D.}\ \bibnamefont {Oliver}}, \bibinfo {author} {\bibfnamefont {Y.}~\bibnamefont {Yu}}, \bibinfo {author} {\bibfnamefont {J.~C.}\ \bibnamefont {Lee}}, \bibinfo {author} {\bibfnamefont {K.~K.}\ \bibnamefont {Berggren}}, \bibinfo {author} {\bibfnamefont {L.~S.}\ \bibnamefont {Levitov}},\ and\ \bibinfo {author} {\bibfnamefont {T.~P.}\ \bibnamefont {Orlando}},\ }\bibfield  {title} {\bibinfo {title} {Mach-zehnder interferometry in a strongly driven superconducting qubit},\ }\href {https://doi.org/10.1126/science.1119678} {\bibfield  {journal} {\bibinfo  {journal} {Science}\ }\textbf {\bibinfo {volume} {310}},\ \bibinfo {pages} {1653} (\bibinfo {year} {2005})}\BibitemShut {NoStop}%
\bibitem [{\citenamefont {Nocera}\ \emph {et~al.}(2017)\citenamefont {Nocera}, \citenamefont {Polkovnikov},\ and\ \citenamefont {Feiguin}}]{PhysRevA.95.023601}%
  \BibitemOpen
  \bibfield  {author} {\bibinfo {author} {\bibfnamefont {A.}~\bibnamefont {Nocera}}, \bibinfo {author} {\bibfnamefont {A.}~\bibnamefont {Polkovnikov}},\ and\ \bibinfo {author} {\bibfnamefont {A.~E.}\ \bibnamefont {Feiguin}},\ }\bibfield  {title} {\bibinfo {title} {Unconventional fermionic pairing states in a monochromatically tilted optical lattice},\ }\href {https://link.aps.org/doi/10.1103/PhysRevA.95.023601} {\bibfield  {journal} {\bibinfo  {journal} {Phys. Rev. A}\ }\textbf {\bibinfo {volume} {95}},\ \bibinfo {pages} {023601} (\bibinfo {year} {2017})}\BibitemShut {NoStop}%
\bibitem [{\citenamefont {Gao}\ \emph {et~al.}(2020)\citenamefont {Gao}, \citenamefont {Coulthard}, \citenamefont {Jaksch},\ and\ \citenamefont {Mur-Petit}}]{PhysRevLett.125.195301}%
  \BibitemOpen
  \bibfield  {author} {\bibinfo {author} {\bibfnamefont {H.}~\bibnamefont {Gao}}, \bibinfo {author} {\bibfnamefont {J.~R.}\ \bibnamefont {Coulthard}}, \bibinfo {author} {\bibfnamefont {D.}~\bibnamefont {Jaksch}},\ and\ \bibinfo {author} {\bibfnamefont {J.}~\bibnamefont {Mur-Petit}},\ }\bibfield  {title} {\bibinfo {title} {Anomalous spin-charge separation in a driven hubbard system},\ }\href {https://doi.org/10.1103/PhysRevLett.125.195301} {\bibfield  {journal} {\bibinfo  {journal} {Phys. Rev. Lett.}\ }\textbf {\bibinfo {volume} {125}},\ \bibinfo {pages} {195301} (\bibinfo {year} {2020})}\BibitemShut {NoStop}%
\bibitem [{\citenamefont {Baum}\ \emph {et~al.}(2018)\citenamefont {Baum}, \citenamefont {van Nieuwenburg},\ and\ \citenamefont {Refael}}]{10.21468/SciPostPhys.5.2.017}%
  \BibitemOpen
  \bibfield  {author} {\bibinfo {author} {\bibfnamefont {Y.}~\bibnamefont {Baum}}, \bibinfo {author} {\bibfnamefont {E.~P.~L.}\ \bibnamefont {van Nieuwenburg}},\ and\ \bibinfo {author} {\bibfnamefont {G.}~\bibnamefont {Refael}},\ }\bibfield  {title} {\bibinfo {title} {From dynamical localization to bunching in interacting floquet systems},\ }\href {https://doi.org/10.21468/SciPostPhys.5.2.017} {\bibfield  {journal} {\bibinfo  {journal} {SciPost Phys.}\ }\textbf {\bibinfo {volume} {5}},\ \bibinfo {pages} {017} (\bibinfo {year} {2018})}\BibitemShut {NoStop}%
\bibitem [{\citenamefont {McCaul}\ \emph {et~al.}(2020)\citenamefont {McCaul}, \citenamefont {Orthodoxou}, \citenamefont {Jacobs}, \citenamefont {Booth},\ and\ \citenamefont {Bondar}}]{PhysRevA.101.053408}%
  \BibitemOpen
  \bibfield  {author} {\bibinfo {author} {\bibfnamefont {G.}~\bibnamefont {McCaul}}, \bibinfo {author} {\bibfnamefont {C.}~\bibnamefont {Orthodoxou}}, \bibinfo {author} {\bibfnamefont {K.}~\bibnamefont {Jacobs}}, \bibinfo {author} {\bibfnamefont {G.~H.}\ \bibnamefont {Booth}},\ and\ \bibinfo {author} {\bibfnamefont {D.~I.}\ \bibnamefont {Bondar}},\ }\bibfield  {title} {\bibinfo {title} {Controlling arbitrary observables in correlated many-body systems},\ }\href {https://link.aps.org/doi/10.1103/PhysRevA.101.053408} {\bibfield  {journal} {\bibinfo  {journal} {Phys. Rev. A}\ }\textbf {\bibinfo {volume} {101}},\ \bibinfo {pages} {053408} (\bibinfo {year} {2020})}\BibitemShut {NoStop}%
\bibitem [{\citenamefont {Chaudhary}\ \emph {et~al.}(2019)\citenamefont {Chaudhary}, \citenamefont {Hsieh},\ and\ \citenamefont {Refael}}]{PhysRevB.100.220403}%
  \BibitemOpen
  \bibfield  {author} {\bibinfo {author} {\bibfnamefont {S.}~\bibnamefont {Chaudhary}}, \bibinfo {author} {\bibfnamefont {D.}~\bibnamefont {Hsieh}},\ and\ \bibinfo {author} {\bibfnamefont {G.}~\bibnamefont {Refael}},\ }\bibfield  {title} {\bibinfo {title} {Orbital floquet engineering of exchange interactions in magnetic materials},\ }\href {https://doi.org/10.1103/PhysRevB.100.220403} {\bibfield  {journal} {\bibinfo  {journal} {Phys. Rev. B}\ }\textbf {\bibinfo {volume} {100}},\ \bibinfo {pages} {220403(R)} (\bibinfo {year} {2019})}\BibitemShut {NoStop}%
\bibitem [{\citenamefont {Rapp}\ \emph {et~al.}(2012)\citenamefont {Rapp}, \citenamefont {Deng},\ and\ \citenamefont {Santos}}]{PhysRevLett.109.203005}%
  \BibitemOpen
  \bibfield  {author} {\bibinfo {author} {\bibfnamefont {A.}~\bibnamefont {Rapp}}, \bibinfo {author} {\bibfnamefont {X.}~\bibnamefont {Deng}},\ and\ \bibinfo {author} {\bibfnamefont {L.}~\bibnamefont {Santos}},\ }\bibfield  {title} {\bibinfo {title} {Ultracold lattice gases with periodically modulated interactions},\ }\href {https://doi.org/10.1103/PhysRevLett.109.203005} {\bibfield  {journal} {\bibinfo  {journal} {Phys. Rev. Lett.}\ }\textbf {\bibinfo {volume} {109}},\ \bibinfo {pages} {203005} (\bibinfo {year} {2012})}\BibitemShut {NoStop}%
\bibitem [{\citenamefont {Zhao}\ \emph {et~al.}(2023)\citenamefont {Zhao}, \citenamefont {Lee}, \citenamefont {Aliyu},\ and\ \citenamefont {Loh}}]{zhao2023floquet}%
  \BibitemOpen
  \bibfield  {author} {\bibinfo {author} {\bibfnamefont {L.}~\bibnamefont {Zhao}}, \bibinfo {author} {\bibfnamefont {M.~D.~K.}\ \bibnamefont {Lee}}, \bibinfo {author} {\bibfnamefont {M.~M.}\ \bibnamefont {Aliyu}},\ and\ \bibinfo {author} {\bibfnamefont {H.}~\bibnamefont {Loh}},\ }\bibfield  {title} {\bibinfo {title} {Floquet-tailored rydberg interactions},\ }\href {https://www.nature.com/articles/s41467-023-42899-8} {\bibfield  {journal} {\bibinfo  {journal} {Nat. Commun.}\ }\textbf {\bibinfo {volume} {14}},\ \bibinfo {pages} {7128} (\bibinfo {year} {2023})}\BibitemShut {NoStop}%
\bibitem [{\citenamefont {Bluvstein}\ \emph {et~al.}(2021)\citenamefont {Bluvstein}, \citenamefont {Omran}, \citenamefont {Levine}, \citenamefont {Keesling}, \citenamefont {Semeghini}, \citenamefont {Ebadi}, \citenamefont {Wang}, \citenamefont {Michailidis}, \citenamefont {Maskara}, \citenamefont {Ho}, \citenamefont {Choi}, \citenamefont {Serbyn}, \citenamefont {Greiner}, \citenamefont {Vuletić},\ and\ \citenamefont {Lukin}}]{doi:10.1126/science.abg2530}%
  \BibitemOpen
  \bibfield  {author} {\bibinfo {author} {\bibfnamefont {D.}~\bibnamefont {Bluvstein}}, \bibinfo {author} {\bibfnamefont {A.}~\bibnamefont {Omran}}, \bibinfo {author} {\bibfnamefont {H.}~\bibnamefont {Levine}}, \bibinfo {author} {\bibfnamefont {A.}~\bibnamefont {Keesling}}, \bibinfo {author} {\bibfnamefont {G.}~\bibnamefont {Semeghini}}, \bibinfo {author} {\bibfnamefont {S.}~\bibnamefont {Ebadi}}, \bibinfo {author} {\bibfnamefont {T.~T.}\ \bibnamefont {Wang}}, \bibinfo {author} {\bibfnamefont {A.~A.}\ \bibnamefont {Michailidis}}, \bibinfo {author} {\bibfnamefont {N.}~\bibnamefont {Maskara}}, \bibinfo {author} {\bibfnamefont {W.~W.}\ \bibnamefont {Ho}}, \bibinfo {author} {\bibfnamefont {S.}~\bibnamefont {Choi}}, \bibinfo {author} {\bibfnamefont {M.}~\bibnamefont {Serbyn}}, \bibinfo {author} {\bibfnamefont {M.}~\bibnamefont {Greiner}}, \bibinfo {author} {\bibfnamefont {V.}~\bibnamefont {Vuletić}},\ and\ \bibinfo {author} {\bibfnamefont {M.~D.}\ \bibnamefont {Lukin}},\ }\bibfield  {title} {\bibinfo {title}
  {Controlling quantum many-body dynamics in driven rydberg atom arrays},\ }\href {https://doi.org/10.1126/science.abg2530} {\bibfield  {journal} {\bibinfo  {journal} {Science}\ }\textbf {\bibinfo {volume} {371}},\ \bibinfo {pages} {1355} (\bibinfo {year} {2021})}\BibitemShut {NoStop}%
\bibitem [{\citenamefont {Su}\ \emph {et~al.}(2018)\citenamefont {Su}, \citenamefont {Shen}, \citenamefont {Liang},\ and\ \citenamefont {Zhang}}]{PhysRevA.98.032306}%
  \BibitemOpen
  \bibfield  {author} {\bibinfo {author} {\bibfnamefont {S.~L.}\ \bibnamefont {Su}}, \bibinfo {author} {\bibfnamefont {H.~Z.}\ \bibnamefont {Shen}}, \bibinfo {author} {\bibfnamefont {E.}~\bibnamefont {Liang}},\ and\ \bibinfo {author} {\bibfnamefont {S.}~\bibnamefont {Zhang}},\ }\bibfield  {title} {\bibinfo {title} {One-step construction of the multiple-qubit rydberg controlled-phase gate},\ }\href {https://doi.org/10.1103/PhysRevA.98.032306} {\bibfield  {journal} {\bibinfo  {journal} {Phys. Rev. A}\ }\textbf {\bibinfo {volume} {98}},\ \bibinfo {pages} {032306} (\bibinfo {year} {2018})}\BibitemShut {NoStop}%
\bibitem [{\citenamefont {Levine}\ \emph {et~al.}(2019)\citenamefont {Levine}, \citenamefont {Keesling}, \citenamefont {Semeghini}, \citenamefont {Omran}, \citenamefont {Wang}, \citenamefont {Ebadi}, \citenamefont {Bernien}, \citenamefont {Greiner}, \citenamefont {Vuleti\ifmmode~\acute{c}\else \'{c}\fi{}}, \citenamefont {Pichler},\ and\ \citenamefont {Lukin}}]{PhysRevLett.123.170503}%
  \BibitemOpen
  \bibfield  {author} {\bibinfo {author} {\bibfnamefont {H.}~\bibnamefont {Levine}}, \bibinfo {author} {\bibfnamefont {A.}~\bibnamefont {Keesling}}, \bibinfo {author} {\bibfnamefont {G.}~\bibnamefont {Semeghini}}, \bibinfo {author} {\bibfnamefont {A.}~\bibnamefont {Omran}}, \bibinfo {author} {\bibfnamefont {T.~T.}\ \bibnamefont {Wang}}, \bibinfo {author} {\bibfnamefont {S.}~\bibnamefont {Ebadi}}, \bibinfo {author} {\bibfnamefont {H.}~\bibnamefont {Bernien}}, \bibinfo {author} {\bibfnamefont {M.}~\bibnamefont {Greiner}}, \bibinfo {author} {\bibfnamefont {V.}~\bibnamefont {Vuleti\ifmmode~\acute{c}\else \'{c}\fi{}}}, \bibinfo {author} {\bibfnamefont {H.}~\bibnamefont {Pichler}},\ and\ \bibinfo {author} {\bibfnamefont {M.~D.}\ \bibnamefont {Lukin}},\ }\bibfield  {title} {\bibinfo {title} {Parallel implementation of high-fidelity multiqubit gates with neutral atoms},\ }\href {https://link.aps.org/doi/10.1103/PhysRevLett.123.170503} {\bibfield  {journal} {\bibinfo  {journal} {Phys. Rev. Lett.}\ }\textbf {\bibinfo
  {volume} {123}},\ \bibinfo {pages} {170503} (\bibinfo {year} {2019})}\BibitemShut {NoStop}%
\bibitem [{\citenamefont {Sahoo}\ \emph {et~al.}(2019)\citenamefont {Sahoo}, \citenamefont {Schneider},\ and\ \citenamefont {Eggert}}]{sahoo2019periodically}%
  \BibitemOpen
  \bibfield  {author} {\bibinfo {author} {\bibfnamefont {S.}~\bibnamefont {Sahoo}}, \bibinfo {author} {\bibfnamefont {I.}~\bibnamefont {Schneider}},\ and\ \bibinfo {author} {\bibfnamefont {S.}~\bibnamefont {Eggert}},\ }\href@noop {} {\bibinfo {title} {Periodically driven many-body systems: A floquet density matrix renormalization group study}} (\bibinfo {year} {2019}),\ \Eprint {https://arxiv.org/abs/1906.00004} {arXiv:1906.00004 [cond-mat.str-el]} \BibitemShut {NoStop}%
\bibitem [{Note6()}]{Note6}%
  \BibitemOpen
  \bibinfo {note} {For example, in a monochromatically driven system described by $\protect \hat {H}_t = \protect \hat {H}^{(0)} + \protect \hat {H}^{(1)}_{-1}e^{-i\omega t} + \protect \hat {H}^{(1)}_1e^{i\omega t}$, by expanding the expectation value $\langle \protect \hat {\protect \mathcal {U}}_{t,t_0}^{\protect \dag } \protect \hat {A} \protect \hat {\protect \mathcal {U}}_{t,t_0} \rangle $ in orders of $g$ and picking out terms oscillating at $e^{ij{\omega }t}$ where $j\in \protect \mathds {Z}$, we find that the $\protect \mathcal {O}(g)$ linear response of an observable $\protect \hat {A}$ is given by $\delta ^{(1)} \langle \protect \hat {A} \rangle = \langle [\protect \hat {F}^{(1)}_t,\protect \hat {A}] \rangle $, and the $\protect \mathcal {O}(g^2)$ response is given by $\delta ^{(2)} \langle \protect \hat {A} \rangle = \langle [\protect \hat {F}^{(2)}_t,\protect \hat {A}] \rangle + \protect \frac {1}{2} \langle [\protect \hat {F}^{(1)}_t,[\protect \hat {F}^{(1)}_t,\protect \hat {A}]] \rangle $. This
  means we can directly compute these responses when $\protect \hat {f}^{(n)}_j$ is found.}\BibitemShut {Stop}%
\bibitem [{\citenamefont {Jaksch}\ \emph {et~al.}(1998)\citenamefont {Jaksch}, \citenamefont {Bruder}, \citenamefont {Cirac}, \citenamefont {Gardiner},\ and\ \citenamefont {Zoller}}]{jaksch1998cold}%
  \BibitemOpen
  \bibfield  {author} {\bibinfo {author} {\bibfnamefont {D.}~\bibnamefont {Jaksch}}, \bibinfo {author} {\bibfnamefont {C.}~\bibnamefont {Bruder}}, \bibinfo {author} {\bibfnamefont {J.~I.}\ \bibnamefont {Cirac}}, \bibinfo {author} {\bibfnamefont {C.~W.}\ \bibnamefont {Gardiner}},\ and\ \bibinfo {author} {\bibfnamefont {P.}~\bibnamefont {Zoller}},\ }\bibfield  {title} {\bibinfo {title} {Cold bosonic atoms in optical lattices},\ }\href {https://link.aps.org/doi/10.1103/PhysRevLett.81.3108} {\bibfield  {journal} {\bibinfo  {journal} {Phys. Rev. Lett}\ }\textbf {\bibinfo {volume} {81}},\ \bibinfo {pages} {3108} (\bibinfo {year} {1998})}\BibitemShut {NoStop}%
\bibitem [{\citenamefont {Imada}\ \emph {et~al.}(1998)\citenamefont {Imada}, \citenamefont {Fujimori},\ and\ \citenamefont {Tokura}}]{RevModPhys.70.1039}%
  \BibitemOpen
  \bibfield  {author} {\bibinfo {author} {\bibfnamefont {M.}~\bibnamefont {Imada}}, \bibinfo {author} {\bibfnamefont {A.}~\bibnamefont {Fujimori}},\ and\ \bibinfo {author} {\bibfnamefont {Y.}~\bibnamefont {Tokura}},\ }\bibfield  {title} {\bibinfo {title} {Metal-insulator transitions},\ }\href {https://doi.org/10.1103/RevModPhys.70.1039} {\bibfield  {journal} {\bibinfo  {journal} {Rev. Mod. Phys.}\ }\textbf {\bibinfo {volume} {70}},\ \bibinfo {pages} {1039} (\bibinfo {year} {1998})}\BibitemShut {NoStop}%
\end{thebibliography}%

\appendix

\section{The formal solution to the Sylvester equation} \label{linear-response}

Here, we formally solve the Sylvester equations in our FSWT, based on the following mathematical relation (see Theorem 9.2 in Ref.~\cite{bhatia1997and}): For 3 arbitrary operators $\hat{a}$, $\hat{b}$ and $\hat{c}$, where each eigenvalue of $\hat{a}$ has positive real part, and each eigenvalue of $\hat{b}$ is purely imaginary, the solution to the Sylvester equation $\hat{a} \hat{f} - \hat{f} \hat{b} = \hat{c} $ is given by
\begin{equation}\label{math-solution}
\begin{split}
\hat{f} &= - e^{-t \hat{a}} \hat{f} e^{t \hat{b}} \big\vert_{t=0}^{\infty} = - \int_{0}^{\infty} dt ~ \partial_t \big( e^{-t \hat{a}} \hat{f} e^{t \hat{b}} \big) \\
&= - \int_{0}^{\infty} dt ~ \partial_t \big( e^{-t \hat{a}} \big) \hat{f} e^{t \hat{b}} + e^{-t \hat{a}} \hat{f} \partial_t \big( e^{t \hat{b}} \big) \\
&=  \int_{0}^{\infty} dt ~   e^{-t \hat{a}}~ \hat{a} \hat{f} ~ e^{t \hat{b}} - e^{-t \hat{a}} ~ \hat{f} \hat{b} ~ e^{t \hat{b}}  \\
&= \int_{0}^{\infty} dt ~   e^{-t \hat{a}}~ \hat{c} ~ e^{t \hat{b}} 
\end{split}
\end{equation}
If we take $\hat{a} = -i ( \hat{H}^{(0)} + j \omL ) + 0^+ $, $\hat{b} = -i  \hat{H}^{(0)} $ and $\hat{c} = -i  \hat{H}^{(1)}_j $, we can identify the lowest-order Sylvester equation (\ref{formula-for-f_1^1}) with this form, and thus we can directly write down its solution $\hat{f}^{(1)}_j$ using Eq.~(\ref{math-solution}): $\forall j \neq 0$,  
\begin{equation}\label{formal-solu-f^1_j}
    \hat{f}_j^{(1)} = -i \int_0^{\infty} dt ~ e^{i j \omL t} ~ e^{- 0^+ t} ~ e^{i \hat{H}^{(0)} t} \hat{H}_j^{(1)}  e^{-i \hat{H}^{(0)} t}
\end{equation}

Eq.~(\ref{math-solution}) further provides the formal solution to the Sylvester equation at higher orders. For example, to solve the second-order equation (\ref{formula-for-f_1^2}), we take $\hat{a} = -i ( \hat{H}^{(0)} + j \omL ) + 0^+ $, $\hat{b} = -i  \hat{H}^{(0)} $ and $\hat{c} = -i  \hat{S}^{(2)}_j $ where 
\begin{equation*}
\hat{S}^{(2)}_j = \hat{H}^{(2)}_j + \frac{1}{2} \sum\limits_{j' \neq 0} [\hat{f}_{j'}^{(1)},\hat{H}_{j-j'}^{(1)}] + \frac{1}{2} [\hat{f}_j^{(1)},\hat{H}_0^{(1)}]
\end{equation*}
denotes the source term in the Sylvester equation (\ref{formula-for-f_1^2}). Eq.~(\ref{math-solution}) then gives, $\forall j \neq 0$,
\begin{equation}
\hat{f}^{(2)}_j = -i \int_0^{\infty} dt ~ e^{i j \omL t} ~ e^{- 0^+ t} ~ e^{i \hat{H}^{(0)} t} \hat{S}_j^{(2)}  e^{-i \hat{H}^{(0)} t}
\end{equation}
In general, once the $n$th order Sylvester equation is constructed from which source term $\hat{S}_j^{(n)}$ is identified, the formal solution is directly given by,  $\forall j \neq 0$,
\begin{equation}
\hat{f}^{(n)}_j = -i \int_0^{\infty} dt ~ e^{i j \omL t} ~ e^{- 0^+ t} ~ e^{i \hat{H}^{(0)} t} \hat{S}_j^{(n)}  e^{-i \hat{H}^{(0)} t}
\end{equation}
which always takes the form of a frequency-domain (i.e. Laplace-transformed) Heisenberg operator under the undriven Hamiltonian $\hat{H}^{(0)}$.

\subsection{Link to the retarded Green function}
The above formal solution links $\hat{f}^{(n)}_j$ to the retarded Green function of the undriven system.
To show this, we consider the correlator $\langle [\hat{f}^{(1)}_1,\hat{A}] \rangle$ where $\hat{A}$ is an observable and the expectation value is taken over the thermal/ground state of the undriven system $\hat{H}^{(0)}$. Then according to Eq.~(\ref{formal-solu-f^1_j}) we have
\begin{equation}\label{appendixA.lin.res}
\langle [\hat{f}^{(1)}_1,\hat{A}] \rangle = -i \int_{-\infty}^{\infty} dt~ e^{i (\omL + i 0^+) t} ~\theta_t  \langle [\big( \hat{H}_1^{(1)} \big)^{H}_t ,\hat{A}] \rangle
\end{equation}
where $\theta_t$ is the step function at $t=0$, $ \big( \hat{H}_1^{(1)} \big)^{H}_t = e^{i \hat{H}^{(0)} t} \hat{H}_1^{(1)}  e^{-i \hat{H}^{(0)} t}$ represents the Heisenberg operator of $\hat{H}_1^{(1)}$. 
Hence, $\langle [\hat{f}^{(1)}_1,\hat{A}] \rangle$ is a frequency-domain retarded Green function of the undriven system $\hat{H}^{(0)}$. 
This direct link between $\hat{f}^{(n)}_j$ and the retarded Green function suggests that we can obtain the linear and non-linear responses of the system from the solutions of the Sylvester equations~
\footnote{For example, in a monochromatically driven system described by $\hat{H}_t = \hat{H}^{(0)} + \hat{H}^{(1)}_{-1}e^{-i\omega t} + \hat{H}^{(1)}_1e^{i\omega t}$, by expanding the expectation value $\langle  \hat{\mathcal{U}}_{t,t_0}^{\dag} \hat{A} \hat{\mathcal{U}}_{t,t_0}  \rangle$ in orders of $g$ and picking out terms oscillating at $e^{ij\omL t}$ where $j\in \mathds{Z}$, we find that the $\mathcal{O}(g)$ linear response of an observable $\hat{A}$ is given by $\delta^{(1)} \langle \hat{A} \rangle = \langle  [\hat{F}^{(1)}_t,\hat{A}] \rangle$, and the $\mathcal{O}(g^2)$ response is given by $\delta^{(2)} \langle \hat{A} \rangle = \langle  [\hat{F}^{(2)}_t,\hat{A}]  \rangle  + \frac{1}{2} \langle [\hat{F}^{(1)}_t,[\hat{F}^{(1)}_t,\hat{A}]]  \rangle$. This means we can directly compute these responses when $\hat{f}^{(n)}_j$ is found.}.
Practically, to evaluate Eq.~(\ref{appendixA.lin.res}) for strongly correlated systems, we need to obtain the thermal density operator numerically, which we leave for future work.

\section{Comparison to Floquet Magnus High Frequency Expansion}\label{appendix:compare_HFE}
Here, we assume a simple single-frequency driving Hamiltonian,
\begin{equation}
\hat{H}_t = \hat{H}^{(0)} + \hat{H}_0^{(1)} + \hat{H}^{(1)}_{-1}e^{-i\omega t} + \hat{H}^{(1)}_1e^{i\omega t}
\end{equation}
and compare the Floquet high-frequency expansion (HFE) with the solutions of the FSWT. In HFE, Eq.~(\ref{H'}) which describes the relation between $\hat{H}'$, $\hat{H}_t$ and $\hat{F}_t$, is equivalently rewritten as the Magnus equation \cite{casas2001floquet,PhysRevB.93.144307}
\begin{equation}\label{van-Vlek-eqn}
\begin{split}
i\partial_t \hat{F}_t &= \sum\limits_{k=0}^{\infty} \frac{B_k}{k!} (ad_{\hat{F}_t})^k \big[(-1)^{(k+1)} \hat{H}_t + \hat{H}' \big], 
\end{split}
\end{equation}
where $ad_{\hat{F}_t}$ is an adjoint operator defined by $ad_{\hat{F}_t} \hat{X} = [\hat{F}_t,\hat{X}]$, and $B_k$ is the $k$-th Bernoulli number.

In the high-frequency limit, we expand $\hat{F}_t$ and $\hat{H}'$ in Eq.~(\ref{van-Vlek-eqn}) in orders of $\omega^{-1}$, 
which yields the following Floquet Hamiltonian 
\begin{equation}\label{van-Vlek-result}
\begin{split}
&\hat{H}'_{HFE} 
= \hat{H}^{(0)} + \hat{H}_{0}^{(1)} 
+ \frac{1}{\omega} [ \hat{H}^{(1)}_1 , \hat{H}^{(1)}_{-1} ] \\
&+ [\frac{[ \hat{H}^{(1)}_1 , \hat{H}^{(0)} + \hat{H}_{0}^{(1)} ]}{2\omega^2} , \hat{H}^{(1)}_{-1} ] 
+ [\frac{[ \hat{H}^{(1)}_{-1} , \hat{H}^{(0)} + \hat{H}_{0}^{(1)} ]}{2\omega^2} , \hat{H}^{(1)}_1 ]  \\
&+ \mathcal{O} (\omL^{-2}).
\end{split}
\end{equation}
In FSWT, if we assume the driving frequency $\omL$ to be the largest energy scale, and expand $\hat{f}_{j}^{(n)}$ in the Sylvester Eqs.~(\ref{formula-for-f_1^1}) and ({\ref{formula-for-f_1^2}}) in orders of $\omega^{-1}$, we find the solution
\begin{equation}
\begin{split}
\hat{f}_{1}^{(1)} &= \frac{1}{\omega} \hat{H}^{(1)}_1 + \frac{1}{\omega^2} [\hat{H}^{(1)}_1,\hat{H}^{(0)}] + O[\frac{1}{\omega^3}] \\
\hat{f}_{1}^{(2)} &=  \frac{1}{\omega^2} [\hat{H}^{(1)}_1,\hat{H}_{0}^{(1)}] + O[\frac{1}{\omega^3}] 
\end{split}
\end{equation}
Inserting this high-frequency solution back into Eqs.~(\ref{H'2}) and (\ref{H'3}), we find our Floquet Hamiltonian $\hat{H}'=\hat{H}'^{(0)} +\hat{H}'^{(1)} +\hat{H}'^{(2)} +\hat{H}'^{(3)} $ indeed reduces to $\hat{H}'_{HFE}$ in Eq.~(\ref{van-Vlek-result}). 
This confirms that the FSWT reduces to the HFE if the Sylvester equations are solved in orders of inverse driving frequency $1/\omL$.

In the driven Hubbard system studied in Section \ref{example}, we additionally have $\hat{H}_0^{(1)} =0$, $\hat{H}^{(1)}_{-1} = \hat{H}^{(1)}_{1}$ and $[ \hat{H}^{(1)}_{-1}, \hat{U} ] = 0$ where $\hat{U}$ is the interaction term in $\hat{H}^{(0)}$. For this specific system, we find that in the Floquet Hamiltonian given by the Magnus expansion [see e.g. Eq.~(33) in Ref.~\cite{PhysRevB.93.144307}], the leading order $\sim \omL^{-1}$ vanishes, the order $\mathcal{O}(\omL^{-2})$ only contains the driving-induced bandwidth renormalisation effect. 
The order $\mathcal{O}(\omL^{-3})$ vanishes, such that correlation effects only appear starting from order $\omL^{-4}$. This exactly agrees with our FSWT result Eq.~(\ref{Floquet-H'}) once we expand it in orders of $1/\omL$.   

\section{The equivalence between FSWT and the Sambe space block-diagonalisation}\label{appendix:equivalence}
\begin{figure}
\includegraphics[width=0.45\textwidth]{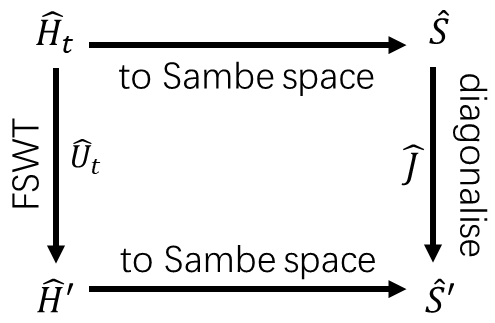}
\caption{
The relation between our FSWT and the Sambe-space van Vleck block-diagonalisation in Ref.~\cite{Eckardt_2015}. The van Vleck method requires knowledge of the eigenbasis of the undriven system. This can be circumvented with the  
Sylvester equations in the FWST method.
}
\label{fig.relation}
\end{figure}

Suppose the time-dependent unitary operator $\hat{U}_t$ transforms the Hamiltonian~(\ref{H_t})
into a time-independent effective Hamiltonian $\hat{H}'$ according to Eq.~(\ref{eq.propagator}), then we define its Fourier coefficients as
\begin{equation}\label{Ut-Fourier}
    \hat{U}_t = \sum\limits_{j=-\infty}^{\infty} \hat{U}_j e^{i j \omL t}.
\end{equation}
We can directly check that the Sambe space Hamiltonian of $\hat{H}_t$, given by
\begin{equation}\label{Sambe}
    \hat{\mathcal{S}} = \sum\limits_{j,j'} (\delta_{j,j'} j \omL + \hat{H}_{j-j'} ) \vert j \rangle \langle j' \vert,
\end{equation}
where 
$\vert j \rangle$ represents the $j$th Floquet sector,
becomes block-diagonalized in every Floquet sector after a transform $ \hat{\mathcal{S}}' \to \hat{J} \hat{\mathcal{S}} \hat{J}^{\dag} $, where the matrix $\hat{J}$ is defined as
\begin{equation}\label{Sambe-diag-J}
    \hat{J} = \sum\limits_{j,j'} \hat{U}_{j-j'} \vert j \rangle \langle j' \vert.  
\end{equation}
To see this, we consider the matrix element
\begin{equation}
\begin{split}
&\langle j \vert \hat{\mathcal{S}}' \vert j' \rangle = \langle j \vert \hat{J} \hat{\mathcal{S}} \hat{J}^{\dag} \vert j' \rangle \\
&= \sum_{j_1,j_2} \hat{U}_{j-j_1} (\delta_{j_1,j_2} j_1 \omL + \hat{H}_{j_1 - j_2} ) (\hat{U}_{j'-j_2})^{\dag} \\
&= \int_{0}^{T} \frac{dt}{T} e^{i(j'-j)\omL t} \left( \hat{U}_t \hat{H}_t \hat{U}_t^{\dag} + i (\partial_t \hat{U}_t ) \hat{U}_t^{\dag} + j \omL \right) \\
&= \hat{H}'_{j-j'} + j \omL ~ \delta_{j,j'}
\end{split}
\end{equation}
where $T=2\pi/\omL$ is the driving period, and in the third line we have inserted the Fourier series expansion $ \hat{H}_j = \int_0^T \frac{dt}{T} e^{-ij\omL t} \hat{H}_t $ and $ \hat{U}_j = \int_0^T \frac{dt}{T} e^{-ij\omL t} \hat{U}_t $. The last line identifies the Fourier coefficients of the transformed Hamiltonian $\hat{H}'_t \equiv \hat{U}_t \hat{H}_t \hat{U}_t^{\dag} + i (\partial_t \hat{U}_t ) \hat{U}_t^{\dag}$. Since the $\hat{U}_t$ in our FSWT makes $\hat{H}'_t$ time-independent, $\hat{H}'_{j-j'} \sim \delta_{j,j'}$, we see that $\hat{\mathcal{S}}'$ is indeed block-diagonalised. This shows that our Floquet Schrieffer Wolff transform is perturbatively block-diagonalising the Sambe space Hamiltonian, as represented in Fig.\ref{fig.relation}.

\section{The Sylvester Equation of the driven Hubbard chain}

\subsection{The $\mathcal{O}(g)$ lowest order} \label{appendix:first-order-example}
Here we solve the lowest-order Sylvester equation (\ref{FSWT-f^1_1}) for the driven $L$-site Hubbard chain described by Eqs.~(\ref{H^0-example}) and (\ref{H_1-example}), based on the approach used for the driven Hubbard dimer in Section~\ref{subsec:driven-dimer}. We expand the solution $\hat{f}^{(1)}_{1}$ in orders of hopping $J$, i.e. $\hat{f}^{(1)}_1 = \sum_{n=0}^{\infty} \hat{y}_n$ where $\hat{y}_n \sim J^n$, and then Eq.~(\ref{FSWT-f^1_1}) is decoupled into the set of Eqs.~(\ref{y}). 

Due to the commutation relation $[\hat{H}^{(1)}_{1},\hat{U}]=0$, the solution to Eq.~(\ref{y0}) is exactly given by
\begin{equation}\label{result-y0}
\hat{y}_0 = \frac{\hat{H}^{(1)}_{1}}{\omL} = \frac{g}{\omL} \sum\limits_{s} \sum\limits_{j=1}^{L} j ~ \hat{n}_{j,s}
\end{equation}
Then, to solve Eq.~(\ref{y1}) for $\hat{y}_1$, we need to calculate the commutator
\begin{equation}\label{[y0,h]}
[\hat{y}_0,\hat{h}] = \frac{J g}{\omL} \sum\limits_{s} \sum\limits_{i,j =1}^{L} \left( \delta_{i-j,1} - \delta_{j-i,1} \right) \hat{c}_{j,s}^{\dag} \hat{c}_{i,s}
\end{equation}
which acts as the \textit{source} of the $J^1$-order in Eq.~(\ref{y1}). Since the Sylvester equation is linear, 
we only need to solve the following part (which sums to the final result for $\hat{y}_1$),
\begin{equation}\label{y1-part}
    \hat{c}_{j,s}^{\dag} \hat{c}_{i,s} + [ \hat{x}_1 , \hat{U} ] - \omL \hat{x}_1 = 0.
\end{equation}
Since $\hat{U}$ only contains local operators, 
the solution $\hat{x}_1$ can only contain the degree of freedom of sites $i$ and $j$, and thus we need to solve (for $i\neq j$)
\begin{equation}\label{x1-equation}
    \hat{c}_{j,s}^{\dag} \hat{c}_{i,s} + [ \hat{x}_1 , U \hat{n}_{i,\uparrow} \hat{n}_{i,\downarrow} + U \hat{n}_{j,\uparrow} \hat{n}_{j,\downarrow}  ] - \omL \hat{x}_1 = 0
\end{equation}
which is no longer a many-body problem. 
According to the symmetry argument in Section~\ref{subsec:driven-dimer}, the solution $\hat{x}_1$ can only take the following form
\begin{equation}\label{x-form}
\hat{x}_1 = \hat{c}_{j,s}^{\dag} \hat{c}_{i,s} \frac{1}{\omL} ( 1 + \beta' \hat{n}_{j,\bar{s}} + \gamma' \hat{n}_{i,\bar{s}} + \delta' \hat{n}_{j,\bar{s}} \hat{n}_{i,\bar{s}} )
\end{equation}
whose parameters can be determined by inserting Eq.~(\ref{x-form}) into Eq.~(\ref{x1-equation}) and then requiring the prefactors of each operator in Eq.~(\ref{x1-equation}) to vanish. The result is given in Eq.~(\ref{parameters'}).
From this result Eq.~(\ref{x-form}), we know that the Sylvester equation Eq.~(\ref{y1}) with source term Eq.~(\ref{[y0,h]}) has the solution
\begin{equation}\label{y1-solu}
\begin{split}
\hat{y}_1 &= \frac{J g}{\omL^2} \sum\limits_{s} \sum\limits_{i,j =1}^{L} \left( \delta_{i-j,1} - \delta_{j-i,1} \right) \hat{c}_{j,s}^{\dag} \hat{c}_{i,s} \\
&~~~~~~~~~~~ \times ( 1 + \beta' \hat{n}_{j,\bar{s}} + \gamma' \hat{n}_{i,\bar{s}} + \delta' \hat{n}_{j,\bar{s}} \hat{n}_{i,\bar{s}} )
\end{split}
\end{equation}
Collecting $\hat{y}_0$ and $\hat{y}_1$ gives the solution Eq.~(\ref{FSWT-H'-chain-g^2}) shown in the main part.
Next we further outline how to solve Eq.~(\ref{y2}) for $\hat{y}_2 \sim J^2$. 
Its source term reads
\begin{widetext}
\begin{equation}\label{[y1,h]}
\begin{split}
[\hat{y}_1,\hat{h}] 
&= \frac{2 J^2 g}{\omL^2}  \sum\limits_{j =1}^{L-1}
\bigg\{ ~ (\beta'-\gamma') (\hat{c}_{j \uparrow}^{\dag} \hat{c}_{j \downarrow}^{\dag}\hat{c}_{j+1 \uparrow} \hat{c}_{j+1 \downarrow} - \hat{c}_{j+1 \uparrow}^{\dag} \hat{c}_{j+1 \downarrow}^{\dag}\hat{c}_{j \uparrow} \hat{c}_{j \downarrow} ) ~~ - \hat{n}_j + \hat{n}_{j+1}  \\
&~~~~~~~~~~~~~~~~~~~~~~~~ 
+ (\beta'+\gamma') \big( ~ \hat{n}_{j+1\uparrow} \hat{n}_{j+1\downarrow} (1-\hat{n}_j) ~ - ~ \hat{n}_{j\uparrow} \hat{n}_{j\downarrow} (1-\hat{n}_{j+1})  ~ \big) \bigg\}
\\
&+ \frac{J^2 g}{\omL^2}  \sum\limits_{j =2}^{L-1} \sum\limits_{s} \bigg\{ ~
\hat{c}_{j-1 s}^{\dag} \hat{c}_{j+1 s} \big( (\beta' - \gamma') \hat{n}_{j,\bar{s}} - \beta' \hat{n}_{j-1,\bar{s}} + \gamma' \hat{n}_{j+1,\bar{s}} + (\beta' + \gamma') \hat{n}_{j,\bar{s}} ( \hat{n}_{j-1,\bar{s}} - \hat{n}_{j+1,\bar{s}} ) \big) \\
&~~~~~~~~~~~~~~~~~~~~ 
+ \hat{c}_{j+1 s}^{\dag} \hat{c}_{j-1 s} \big( (\gamma' - \beta') \hat{n}_{j,\bar{s}} - \gamma' \hat{n}_{j-1,\bar{s}} + \beta' \hat{n}_{j+1,\bar{s}} + (\beta' + \gamma') \hat{n}_{j,\bar{s}} ( \hat{n}_{j-1,\bar{s}} - \hat{n}_{j+1,\bar{s}} ) \big) \\
&~~~~~~~~~~~~~~~~~~~~
+ \hat{c}_{j s}^{\dag}  \hat{c}_{j-1 \bar{s}}^{\dag} \hat{c}_{j \bar{s}} \hat{c}_{j+1 s} ~ \big( (\beta' - \gamma') + (\beta' + \gamma') (\hat{n}_{j-1 s} - \hat{n}_{j+1 \bar{s}} )  \big) \\
&~~~~~~~~~~~~~~~~~~~~
+ \hat{c}_{j+1 s}^{\dag}  \hat{c}_{j \bar{s}}^{\dag} \hat{c}_{j-1 \bar{s}} \hat{c}_{j s} ~ \big( ( \gamma' - \beta' ) + (\beta' + \gamma') (\hat{n}_{j-1 s} - \hat{n}_{j+1 \bar{s}} )  \big) \\
&~~~~~~~~~~~~~~~~~~~~ 
+ ( \hat{c}_{j s}^{\dag}  \hat{c}_{j \bar{s}}^{\dag} \hat{c}_{j-1 \bar{s}} \hat{c}_{j+1 s} 
+ \hat{c}_{j+1 s}^{\dag}  \hat{c}_{j-1 \bar{s}}^{\dag} \hat{c}_{j \bar{s}} \hat{c}_{j s} ) 
~ (\beta' + \gamma') (\hat{n}_{j+1 \bar{s}} - \hat{n}_{j-1 s} )   ~ \bigg\} 
\end{split}
\end{equation}
\end{widetext}
where we use $\delta'=-\beta'-\gamma'$ and $\hat{n}_j = \sum_s \hat{n}_{j,s}$ to simplify the result.
In the non-interacting limit $U\to0$ (where $\beta'=\gamma'=\delta'=0$), we find the commutator $[\hat{y}_1,\hat{h}]$ reduces completely to boundary effects, which only contain the operators at the boundary sites $j=1,L$. 
However, for $U\neq0$, $[\hat{y}_1,\hat{h}]$ will become non-vanishing in the bulk. Since the Sylvester equation (\ref{y2}) is linear, we can find the solution for each individual term in this $[\hat{y}_1,\hat{h}]$, and then sum them up to get $\hat{y}_2$, just as what we did above to find $\hat{y}_1$. This means that to find $\hat{y}_2$, we need to solve 3-site Sylvester equations, such as 
\begin{equation}\label{y2-part}
\begin{split}
\hat{c}_{j,s}^{\dag} \hat{c}_{k,\bar{s}}^{\dag} \hat{c}_{j,\bar{s}} \hat{c}_{i,s}  + [ \hat{x}_2 , \hat{U} ] - \omL \hat{x}_2 &= 0, \\
\hat{c}_{j,s}^{\dag} \hat{c}_{j,\bar{s}}^{\dag} \hat{c}_{k,\bar{s}} \hat{c}_{i,s}  + [ \hat{x}_3 , \hat{U} ] - \omL \hat{x}_3 &= 0,
\end{split}
\end{equation}
which is similar to Eq.~(\ref{y1-part}). According to the same symmetry argument, the solution to Eq.~(\ref{y2-part}) takes the form 
\begin{equation}
\begin{split}
\hat{x}_2 &= \hat{c}_{j,s}^{\dag} \hat{c}_{k,\bar{s}}^{\dag} \hat{c}_{j,\bar{s}} \hat{c}_{i,s} \frac{1}{\omL}  (1 + \beta' \hat{n}_{k,s} + \gamma' \hat{n}_{i,\bar{s}} + \delta' \hat{n}_{k,s} \hat{n}_{i,\bar{s}}  ) \\
\hat{x}_3 &= \hat{c}_{j,s}^{\dag} \hat{c}_{j,\bar{s}}^{\dag} \hat{c}_{k,\bar{s}} \hat{c}_{i,s} \frac{1}{\omL}
\big( \frac{\omL}{\omL+U} -\beta' (\hat{n}_{k,s} + \hat{n}_{i,\bar{s}}) -\delta' \hat{n}_{k,s} \hat{n}_{i,\bar{s}}  \big) 
\end{split}
\end{equation}
where the coefficients are decided by Eq.~(\ref{y2-part}). The solution of $\hat{y}_2$ is then constructed from $\hat{x}_2$, according to the source term $[\hat{y}_1,\hat{h}]$. In the final solution of $\hat{y}_2$, each term will contain a common factor $J^2 g / \omL^3$, together with a term-specific factor that makes $\hat{y}_2$ accurate at arbitrary $U/J$ ratio. In our driven Hubbard chain, $\hat{y}_2$ is found to be
\begin{widetext}
\begin{equation}\label{y2-full-result}
\begin{split}
\hat{y}_2 &= \frac{2 J^2 g}{\omL^3} \sum\limits_{j =1}^{L-1}
\bigg\{ ~ (\beta'-\gamma') (\hat{c}_{j \uparrow}^{\dag} \hat{c}_{j \downarrow}^{\dag}\hat{c}_{j+1 \uparrow} \hat{c}_{j+1 \downarrow} - \hat{c}_{j+1 \uparrow}^{\dag} \hat{c}_{j+1 \downarrow}^{\dag}\hat{c}_{j \uparrow} \hat{c}_{j \downarrow} ) ~~ - \hat{n}_j + \hat{n}_{j+1}  \\
&~~~~~~~~~~~~~~~~~~~~~~~~ 
+ (\beta'+\gamma') \big( ~ \hat{n}_{j+1\uparrow} \hat{n}_{j+1\downarrow} (1-\hat{n}_j) ~ - ~ \hat{n}_{j\uparrow} \hat{n}_{j\downarrow} (1-\hat{n}_{j+1})  ~ \big) \bigg\} \\
&+ \frac{J^2 g}{\omL^3}  \sum\limits_{j =2}^{L-1} \sum\limits_{s} \bigg\{ ~
\hat{c}_{j-1 s}^{\dag} \hat{c}_{j+1 s} \big( (\beta' - \gamma') \hat{n}_{j,\bar{s}} - \beta' \hat{n}_{j-1,\bar{s}} + \gamma' \hat{n}_{j+1,\bar{s}} - \delta' \hat{n}_{j,\bar{s}} ( \hat{n}_{j-1,\bar{s}} - \hat{n}_{j+1,\bar{s}} ) \big)
( 1 + \beta' \hat{n}_{ j-1 \bar{s}} + \gamma' \hat{n}_{j+1 \bar{s}} + \delta' \hat{n}_{ j-1 \bar{s}} \hat{n}_{j+1 \bar{s}} )
\\
&~~~~~~~~~~~~~~~~~~~~ 
+ \hat{c}_{j+1 s}^{\dag} \hat{c}_{j-1 s} \big( (\gamma' - \beta') \hat{n}_{j,\bar{s}} - \gamma' \hat{n}_{j-1,\bar{s}} + \beta' \hat{n}_{j+1,\bar{s}} - \delta' \hat{n}_{j,\bar{s}} ( \hat{n}_{j-1,\bar{s}} - \hat{n}_{j+1,\bar{s}} ) \big)
( 1 + \beta' \hat{n}_{ j+1 \bar{s}} + \gamma' \hat{n}_{j-1 \bar{s}} + \delta' \hat{n}_{ j+1 \bar{s}} \hat{n}_{j-1 \bar{s}} )
\\
&~~~~~~~~~~~~~~~~~~~~ 
+ \hat{c}_{j s}^{\dag}  \hat{c}_{j-1 \bar{s}}^{\dag} \hat{c}_{j \bar{s}} \hat{c}_{j+1 s} ~ \big( (\beta' - \gamma') -\delta' (\hat{n}_{j-1 s} - \hat{n}_{j+1 \bar{s}} )  \big)
( 1 + \beta' \hat{n}_{ j-1 s} + \gamma' \hat{n}_{j+1 \bar{s}} + \delta' \hat{n}_{ j-1 s} \hat{n}_{j+1 \bar{s}} )
\\
&~~~~~~~~~~~~~~~~~~~~ 
+ \hat{c}_{j+1 s}^{\dag}  \hat{c}_{j \bar{s}}^{\dag} \hat{c}_{j-1 \bar{s}} \hat{c}_{j s} ~ \big( ( \gamma' - \beta' ) -\delta' (\hat{n}_{j-1 s} - \hat{n}_{j+1 \bar{s}} )  \big) 
( 1 + \beta' \hat{n}_{j+1 \bar{s}}  + \gamma' \hat{n}_{ j-1 s}  + \delta' \hat{n}_{j+1 \bar{s}} \hat{n}_{ j-1 s}  )
\\
&~~~~~~~~~~~~~~~~~~~~ 
+  \hat{c}_{j s}^{\dag}  \hat{c}_{j \bar{s}}^{\dag} \hat{c}_{j-1 \bar{s}} \hat{c}_{j+1 s} 
~ \delta' ( \hat{n}_{j-1 s} - \hat{n}_{j+1 \bar{s}}  )
\big( \frac{\omL}{\omL+U} - \beta' (\hat{n}_{j-1 s} + \hat{n}_{j+1 \bar{s}} ) - \delta' \hat{n}_{j-1 s} \hat{n}_{j+1 \bar{s}}  \big)
\\
&~~~~~~~~~~~~~~~~~~~~ 
+ \hat{c}_{j+1 s}^{\dag}  \hat{c}_{j-1 \bar{s}}^{\dag} \hat{c}_{j \bar{s}} \hat{c}_{j s} 
~ \delta' ( \hat{n}_{j-1 s} - \hat{n}_{j+1 \bar{s}}  )
\big( \frac{\omL}{\omL-U} - \gamma' (\hat{n}_{j-1 s} + \hat{n}_{j+1 \bar{s}} ) - \delta' \hat{n}_{j-1 s} \hat{n}_{j+1 \bar{s}}  \big)
~ \bigg\} 
\end{split}
\end{equation}   
In this $\mathcal{O}(J^2)$ order micro-motion $\hat{y}_2$, we find driving-induced two-site correlated processes, including doublon-holon exchange and doublon-holon density-density interactions, as well as three-site correlated processes, including next-nearest neighbour hopping, two-electron hopping, doublon formation and dissociation.
This $\hat{y}_2$ contributes to the following $\mathcal{O}(J^2)$ term in the lowest order Floquet Hamiltonian $\hat{H}'^{(2)}$ in Eq.~(\ref{Floquet-H'}), which reads
\begin{equation}\label{Floquet-H'2-t2}
\begin{split}
& \frac{1}{2}([\hat{y}_2,\hat{H}^{(1)}_{-1}]+ H.c.) \\
&= \frac{4 J^2 g^2}{\omL^3} (\beta'-\gamma') \sum\limits_{j =1}^{L-1} \big(
\hat{c}_{j \uparrow}^{\dag} \hat{c}_{j \downarrow}^{\dag}\hat{c}_{j+1 \uparrow} \hat{c}_{j+1 \downarrow} + \hat{c}_{j+1 \uparrow}^{\dag} \hat{c}_{j+1 \downarrow}^{\dag}\hat{c}_{j \uparrow} \hat{c}_{j \downarrow} \big) \\
&+ \frac{J^2 g^2}{\omL^3} (\beta' - \gamma')  \sum\limits_{j =2}^{L-1} \sum\limits_{s} \bigg\{ ~
(\hat{c}_{j-1 s}^{\dag} \hat{c}_{j+1 s} + \hat{c}_{j+1 s}^{\dag} \hat{c}_{j-1 s} )  \big( 2 \hat{n}_{j,\bar{s}} - \hat{n}_{j-1,\bar{s}} - \hat{n}_{j+1,\bar{s}} + \delta' (1- 2\hat{n}_{j,\bar{s}}) ( \hat{n}_{j-1,\bar{s}} + \hat{n}_{j+1,\bar{s}} - 2 \hat{n}_{ j-1 \bar{s}} \hat{n}_{j+1 \bar{s}} ) \big)
\\
&~~~~~~~~~~~~~~~~~~~~~~~~~~~~~~~~~~~~ 
+ 2  (\hat{c}_{j s}^{\dag}  \hat{c}_{j-1 \bar{s}}^{\dag} \hat{c}_{j \bar{s}} \hat{c}_{j+1 s} + \hat{c}_{j+1 s}^{\dag}  \hat{c}_{j \bar{s}}^{\dag} \hat{c}_{j-1 \bar{s}} \hat{c}_{j s})
( 1 - \delta' \hat{n}_{ j-1 s} - \delta' \hat{n}_{j+1 \bar{s}} + 2 \delta' \hat{n}_{ j-1 s} \hat{n}_{j+1 \bar{s}} )
~ \bigg\} 
\end{split}
\end{equation}
\end{widetext}
where the coefficients are given in Eq.~(\ref{parameters'}). We find this $\mathcal{O}(J^2)$ Floquet Hamiltonian contains fewer terms than the micro-motion $\hat{y}_2$: According to our driving term $\hat{H}^{(1)}_{-1}$, if a term in $\hat{y}_2$ conserves the center-of-mass position, it will commute with $\hat{H}^{(1)}_{-1}$, and thus it will have no impact on the Floquet Hamiltonian (\ref{Floquet-H'2-t2}).

\subsection{The $\mathcal{O}(g^2)$ second-lowest order}\label{appendix:higher-order-example}

For the second-lowest order ($n=2$) driving effects, we need to solve two Sylvester equations according to Eq.~(\ref{formula-for-f_1^2}) for the system described by Eq.~(\ref{Ht-example}),
\begin{subequations}\label{FSWT-f^2}
\begin{align}
&  [\hat{f}^{(2)}_1,\hat{H}^{(0)}] - \omL \hat{f}^{(2)}_1 = 0  \label{FSWT-f^2_1} \\
&  \frac{1}{2} [\hat{f}^{(1)}_1,\hat{H}^{(1)}_1] + [\hat{f}^{(2)}_2,\hat{H}^{(0)}] - 2 \omL \hat{f}^{(2)}_2 = 0.  \label{FSWT-f^2_2}
\end{align}
\end{subequations}
Form these, we find $\hat{f}^{(2)}_{\pm1} =0$ in this model, and thus according to Eq.~(\ref{H'3}), there is no $\mathcal{O}(g^3)$-correction to the Floquet Hamiltonian, $\hat{H}'^{(3)}=0$. More generally, only the even-order $\hat{H}'^{(2n)}$ Floquet Hamiltonians do not vanish in the driven system described by the Hamiltonian Eq.~(\ref{Ht-example}). The vanishing of odd orders
can be verified using the Floquet method \cite{PhysRevB.101.024303} based on Gaussian elimination. 

Here we solve Eq.~(\ref{FSWT-f^2_2}) in the $J\ll \omL$ limit, where we expand the solution $\hat{f}^{(2)}_2$ in orders of hopping $J$, i.e. $\hat{f}^{(2)}_2 = \sum_{n=0}^{\infty} \hat{z}_n$ where $\hat{z}_n \sim J^n$. This decomposes the Sylvester equation  (\ref{FSWT-f^2_2}) in orders of $J$, e.g. 
\begin{subequations}\label{z}
    \begin{align}
& \frac{1}{2}[\hat{y}_0,\hat{H}^{(1)}_{1}] + [\hat{z}_0,\hat{U}] - 2\omL \hat{z}_0 = 0  \label{z0} \\
& \frac{1}{2}[\hat{y}_1,\hat{H}^{(1)}_{1}] +  [\hat{z}_0,\hat{h}] + [\hat{z}_1,\hat{U}] - 2\omL \hat{z}_1 = 0 \label{z1} \\
& \frac{1}{2}[\hat{y}_2,\hat{H}^{(1)}_{1}] +  [\hat{z}_1,\hat{h}] + [\hat{z}_2,\hat{U}] - 2\omL \hat{z}_2 = 0 
    \end{align}
\end{subequations}
and so on. Here we also used the expansion $\hat{f}^{(1)}_1 = \sum_{n=0}^{\infty} \hat{y}_n$ where $\hat{y}_0$ and $\hat{y}_1$ is solved in Appendix \ref{appendix:first-order-example}.
According to the result Eq.~(\ref{result-y0}), we find $[\hat{y}_0,\hat{H}^{(1)}_{1}]=0$, and thus Eq.~(\ref{z0}) gives $\hat{z}_0=0$. The source term in Eq.~(\ref{z1}) is then given by
\begin{widetext}
\begin{equation}
\begin{split}
& \frac{1}{2}[\hat{y}_1,\hat{H}^{(1)}_{1}]  = \frac{J g^2}{2 \omL^2} \sum\limits_{s} \sum\limits_{i,j =1}^{L} \left( \delta_{i-j,1} + \delta_{j-i,1} \right) 
\hat{c}_{j,s}^{\dag} \hat{c}_{i,s}  ( 1 + \beta' \hat{n}_{j,\bar{s}} + \gamma' \hat{n}_{i,\bar{s}} + \delta' \hat{n}_{j,\bar{s}} \hat{n}_{i,\bar{s}} ).
\end{split}
\end{equation}
Similar to what we did in Appendix \ref{appendix:first-order-example}, we solve the Sylvester equation for each term in $\frac{1}{2}[\hat{y}_1,\hat{H}^{(1)}_{1}]$ and then sum them up, which gives the solution to Eq.~(\ref{z1}). It reads explicitly
\begin{equation}\label{z1-solu}
\begin{split}
\hat{z}_1 &= \frac{J g^2}{4 \omL^3} \sum\limits_{s} \sum\limits_{i,j =1}^{L} \left( \delta_{i-j,1} + \delta_{j-i,1} \right) 
\hat{c}_{j,s}^{\dag} \hat{c}_{i,s}  ( 1 + \beta' \hat{n}_{j,\bar{s}} + \gamma' \hat{n}_{i,\bar{s}} + \delta' \hat{n}_{j,\bar{s}} \hat{n}_{i,\bar{s}} ) \times
( 1 + \beta'' \hat{n}_{j,\bar{s}} + \gamma'' \hat{n}_{i,\bar{s}} + \delta'' \hat{n}_{j,\bar{s}} \hat{n}_{i,\bar{s}} ) \\
&\equiv \frac{J g^2}{4 \omL^3} \sum\limits_{s} \sum\limits_{i,j =1}^{L} \left( \delta_{i-j,1} + \delta_{j-i,1} \right) \hat{c}_{j,s}^{\dag} \hat{c}_{i,s} 
( 1 + \beta_2 \hat{n}_{j,\bar{s}} + \gamma_2 \hat{n}_{i,\bar{s}} + \delta_2 \hat{n}_{j,\bar{s}} \hat{n}_{i,\bar{s}} ) 
\end{split}
\end{equation}    
\end{widetext}
where in the first line $\beta'',\gamma'', \delta''$ is are simply the coefficients $\beta',\gamma', \delta'$ with $\omL$ replaced by $2\omL$. Since $\hat{f}^{(2)}_2 = \hat{z}_1 + \mathcal{O}(J^2)$, we thus see from Eq.~(\ref{z1-solu}) that $\hat{f}^{(2)}_2$ diverges not only when $\omL=U$ (when $\gamma'$ diverges), but also when $\omL=U/2$ (when $\gamma''$ diverges). The divergence at $\omL=U/2$ is understood as the two-photon resonance to excite a doublon. In the second line of Eq.~(\ref{z1-solu}), we have evaluated the multiplication in the first line, from which the parameters $\beta_2,\gamma_2,\delta_2$ are defined for future use, i.e. $\beta_2\equiv \beta'+\beta''+\beta'\beta''$, $\gamma_2\equiv \gamma'+\gamma''+\gamma'\gamma''$ and $\delta_2 \equiv \delta' + \delta'' + \gamma'\beta'' + \gamma''\beta' + \gamma'\delta'' + \gamma''\delta' + \beta'\delta'' + \beta''\delta' + \delta'\delta'' $.

\subsection{The $\mathcal{O}(g^3)$ third-lowest order}\label{appendix:3rd-order-example}
To find the third lowest order ($n=3$) driving effect in the driving systems described by Eq.~(\ref{Ht-example}), we need to solve 3 Sylvester equations
\begin{widetext}

\begin{subequations}\label{FSWT-f^3}
\begin{align}
 \frac{1}{2} [\hat{f}^{(2)}_2,\hat{H}^{(1)}_{-1}] 
+ \frac{1}{12} [\hat{f}^{(1)}_{-1},[\hat{f}^{(1)}_{1},\hat{H}^{(1)}_{1}]] 
+ \frac{2}{3} [\hat{f}^{(1)}_1,\hat{H}'^{(2)}] +  [\hat{f}^{(3)}_1,\hat{H}^{(0)}] - \omL \hat{f}^{(3)}_1 
&= 0  \label{FSWT-f^3_1} \\
[\hat{f}^{(3)}_2,\hat{H}^{(0)}] - 2 \omL \hat{f}^{(3)}_2 &= 0  \label{FSWT-f^3_2} \\
\frac{1}{2} [\hat{f}^{(2)}_2,\hat{H}^{(1)}_1] + \frac{1}{12} [\hat{f}^{(1)}_{1},[\hat{f}^{(1)}_{1},\hat{H}^{(1)}_{1}]] 
+ [\hat{f}^{(3)}_3,\hat{H}^{(0)}] - 3 \omL \hat{f}^{(3)}_3 &= 0  \label{FSWT-f^3_3} 
\end{align}
\end{subequations}
whose solutions determine the $\mathcal{O}(g^4)$ order Floquet Hamiltonian correction, 
\begin{equation}\label{H'4}
\hat{H}'^{(4)} = 
\left( \frac{1}{2} [\hat{f}^{(3)}_1, \hat{H}^{(1)}_{-1}] 
+ \frac{1}{12} [\hat{f}^{(2)}_2,[\hat{f}^{(1)}_{-1}, \hat{H}^{(1)}_{-1}]] 
+ \frac{1}{12} [\hat{f}^{(1)}_{-1},[\hat{f}^{(2)}_{2}, \hat{H}^{(1)}_{-1}]] 
- \frac{1}{12} [\hat{f}^{(1)}_{1},[\hat{f}^{(1)}_{-1}, \hat{H}'^{(2)}]]  \right)
+ H.c.
\end{equation}    
These higher-order solutions can be derived in the same way. We find $\hat{f}^{(3)}_1$ diverges at $\omL=U, U/2$, $\hat{f}^{(3)}_2 = 0$, and $\hat{f}^{(3)}_3$ diverges at $\omL=U, U/2, U/3$. This means $\hat{H}'^{(4)}$ diverges at $\omL=U, U/2$.
Explicitly, by solving Eq.~(\ref{FSWT-f^3_1}), we find
\begin{equation}\label{f^3_1-solu}
\begin{split}
\hat{f}^{(3)}_1 &= \frac{J g^3}{\omL^4} \sum\limits_{s} \sum\limits_{i,j =1}^{L} \left( \delta_{i-j,1} - \delta_{j-i,1} \right) \hat{c}_{j,s}^{\dag} \hat{c}_{i,s} 
( -\frac{11}{24} + \beta_3 \hat{n}_{j,\bar{s}} + \gamma_3 \hat{n}_{i,\bar{s}} + \delta_3 \hat{n}_{j,\bar{s}} \hat{n}_{i,\bar{s}} ) ~~ + \mathcal{O}(J^2)
\end{split}
\end{equation}
where the parameters $\beta_3,\gamma_3,\delta_3$ are determined from
\begin{equation}
\begin{split}
( -\frac{11}{24} + \beta_3 \hat{n}_{j,\bar{s}} + \gamma_3 \hat{n}_{i,\bar{s}} + \delta_3 \hat{n}_{j,\bar{s}} \hat{n}_{i,\bar{s}} ) 
&= \frac{1}{8} ( 1 + \beta' \hat{n}_{j,\bar{s}} + \gamma' \hat{n}_{i,\bar{s}} + \delta' \hat{n}_{j,\bar{s}} \hat{n}_{i,\bar{s}} )^2  ~ 
( -1 + \beta'' \hat{n}_{j,\bar{s}} + \gamma'' \hat{n}_{i,\bar{s}} + \delta'' \hat{n}_{j,\bar{s}} \hat{n}_{i,\bar{s}} ) \\
& - \frac{1}{3} ( 1 + \beta' \hat{n}_{j,\bar{s}} + \gamma' \hat{n}_{i,\bar{s}} + \delta' \hat{n}_{j,\bar{s}} \hat{n}_{i,\bar{s}} ) ~ 
( 1 + \beta' \hat{n}_{j,\bar{s}} + \gamma' \hat{n}_{i,\bar{s}} + \delta' \hat{n}_{j,\bar{s}} \hat{n}_{i,\bar{s}} ).
\end{split}
\end{equation}    
Inserting the solution of $\hat{f}^{(2)}_2$ and $\hat{f}^{(3)}_1$, i.e. Eqs.~(\ref{z1-solu}) and (\ref{f^3_1-solu}), into Eq.~(\ref{H'4}), we find
\begin{equation}\label{H'4-result}
\begin{split}
\hat{H}'^{(4)} &= \frac{g^4 J}{2\omL^4} \sum\limits_{s} \sum\limits_{i,j =1}^{L} 
\left( \delta_{i-j,1} + \delta_{j-i,1} \right) 
(\hat{c}_{j,s}^{\dag} \hat{c}_{i,s} + \hat{c}_{i,s}^{\dag} \hat{c}_{j,s} )
\big(
-\frac{1}{4} + \beta_4 \hat{n}_{j,\bar{s}} + \gamma_4 \hat{n}_{i,\bar{s}} + \delta_4 \hat{n}_{j,\bar{s}} \hat{n}_{i,\bar{s}}
\big)
\\
&= \frac{g^4 J}{\omL^4} \sum\limits_{s} \sum\limits_{j=1}^{L-1}  ( \hat{c}_{j,s}^{\dag} \hat{c}_{j+1,s} + \hat{c}_{j+1,s}^{\dag} \hat{c}_{j,s}   )  
\left( -\frac{1}{4} + \frac{\beta_4 + \gamma_4}{2} (\hat{n}_{j,\bar{s}} +  \hat{n}_{j+1,\bar{s}} ) +  \delta_4 \hat{n}_{j,\bar{s}} \hat{n}_{j+1,\bar{s}} \right) ~~ + \mathcal{O}(J^2)
\end{split}
\end{equation}
where we have defined $\beta_4 \equiv \beta_3 + \beta_2 / 24 + \beta' / 6 $, $\gamma_4 \equiv \gamma_3 + \gamma_2 / 24 + \gamma' / 6 $ and $\delta_4 \equiv \delta_3 + \delta_2 / 24 + \delta' / 6 $. In the non-interacting case $U=0$, these 3 parameters vanish, and then $\hat{H}'^{(4)}$ reduces exactly to the non-interacting dynamical localisation effect predicted before in Ref.~\cite{PhysRevB.34.3625}: The bandwidth renormalisation factor $\frac{1}{4} g^4/\omL^4$ in Eq.~(\ref{H'4-result}) matches the Taylor expansion of the Bessel function $\mathcal{J}_0 (2g/\omL) = 1 - g^2/\omL^2 + \frac{1}{4} g^4/\omL^4 + \mathcal{O}(g^6)$.

\end{widetext}
\subsection{FSWT in the strong driving frame}\label{FSWT-Peierls}

To apply FSWT to the strongly driven Hubbard model (i.e., $g\gg \omL$ ) it is favourable to change into a strong driving frame as, otherwise, we would require to solve the Sylvester equations to higher orders of $g$. The change into a strong driving frame is a direct route to observe the Bessel function renormalization in the FSWT Hamiltonian, as we explain below. To change frames, we apply the Peierls transform \cite{doi:10.1126/science.1119678,PhysRevLett.125.195301,10.21468/SciPostPhys.5.2.017}
\begin{equation}
\hat{U}^{\text{r}}_t = \exp \left( i  \frac{2g}{\omL} \sin(\omL t) \sum_{j=1}^{L} j \sum_s \hat{n}_{j,s}  \right)
\end{equation}
to the lab-frame Hamiltonian in Eq. (\ref{Ht-example}).
The transformed Hamiltonian acquires a Peierls phase coupling, replacing the driving strength $g$ with $J$. It reads
\begin{equation}\label{H_t-strong-drive}
\begin{split}
\hat{H}_{P,t} = \hat{H}_P^{(0)} + \sum_{j=-\infty}^{\infty} \hat{H}^{(1)}_{P,j} e^{i j \omega t},
\end{split}
\end{equation}
where the superscript now represents the orders of hopping $J$. In Eq.~(\ref{H_t-strong-drive}), the undriven part no longer contains hopping, as it is given by
\begin{equation}
\hat{H}_P^{(0)} =  U \sum_{{\bf j}} \hat{n}_{{\bf j},\uparrow} \hat{n}_{{\bf j},\downarrow},
\end{equation}
and the driving term oscillating at $e^{i j \omega t}$ reads
\begin{equation}
\hat{H}^{(1)}_{P,j} = -J ~ \mathcal{J}_j ( 2g/ \omega ) 
\sum\limits_{s} \sum\limits_{j=1}^{L-1} (\hat{c}_{j+1,s}^{\dag} \hat{c}_{j,s} + \hat{c}_{j,s}^{\dag} \hat{c}_{j+1,s} ),
\end{equation}
where $\mathcal{J}_j$ is the $j$th kind Bessel function.
In this rotating frame, the driving term has a static component $\hat{H}^{(1)}_{P,0}$ which accounts for the well known $\mathcal{J}_0 (2g/\omL)$ bandwidth renormalisation. The $\mathcal{O}(J)$ lowest order Sylvester equation for the Hamiltonian in Eq.~(\ref{H_t-strong-drive}) is, for all $j\neq0$,
\begin{equation}\label{FSWT-f^1_j}
\hat{H}^{(1)}_{P,j} + [\hat{f}^{(1)}_j,\hat{H}_P^{(0)}] - j \omega \hat{f}^{(1)}_j = 0.
\end{equation}
We thus see that, solving Eq.~(\ref{FSWT-f^1_j}) is equivalent to solving Eq.~(\ref{y1}), which results in the parameters akin to $\beta'$, $\gamma'$ and  $\delta'$. The corresponding FSWT Hamiltonian in this strong driving frame will be explored in future work.

\section{Simulations on the driven Hubbard model}
\subsection{DMRG simulation of the metal-insulator phase transition}
\label{appendix:MIT}
Here, we present details of the iDMRG simulation in Fig.~\ref{fig:MIT}. 
Unlike the Bose-Hubbard model \cite{jaksch1998cold}, the one-dimensional Fermi-Hubbard model shows a metal-insulator transition at $U/J=0$ at half-filling \cite{RevModPhys.70.1039,essler2005one}.
This makes it difficult to witness the phase transition by modifying the ratio $U/J$ and calculating the single-particle gap. Therefore, we instead include a non-zero chemical potential $\mu$ in Eq.~(\ref{H^0-example}),  
and witness the (filling-control) phase transition by decreasing the chemical potential $\mu$, during which the occupancy changes from $\langle \hat{n}_j \rangle = 1$ to incommensurate values \cite{essler2005one}. 
We fix the parameters to $ U=4J, \omL=12J, g=3J$, which is within the applicability range of our FSWT result Eq.~(\ref{Floquet-H'}).
We use infinite-DMRG to simulate the Floquet Hamiltonian's ground state, truncating the bond dimension to $\chi=600$. The exploration around the transition shows a shift in non-commensurate occupation values and critical behaviour due to the additional correlated hopping terms compared to the HFE. To obtain the change in the critical point of the transition, we interpolated each point (in the FSWT curve) near the transition with a function $1-A\sqrt{\mu_c-\mu}$, typical of the commensurate-incommensurate phase transition of the Hubbard model \cite{essler2005one}. The interpolation of HFE and the undriven curve is based on the exact Lieb-Wu equation \cite{essler2005one}.

\begin{figure}[t]
    \centering
    \includegraphics[width=0.45\textwidth]{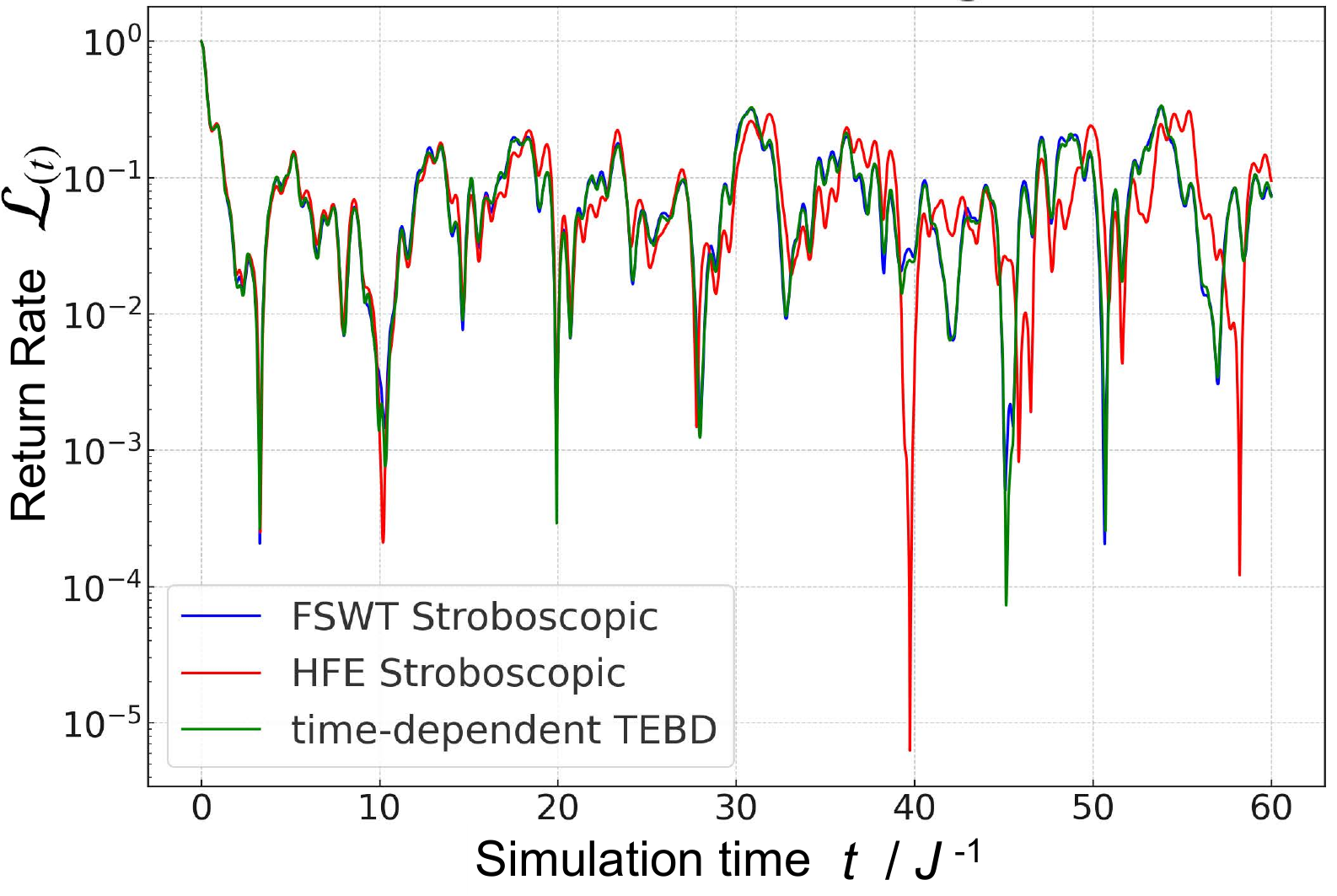}
    \caption{The return rate, Eq.~(\ref{eq.Loschmidt}) of a $L=6$ lattice for the parameters in Fig.~\ref{fig:MIT}, i.e., $\omL=12 J$, $U=4 J$ and $g=3 J$.}
    \label{fig:assure}
\end{figure}

To confirm the applicability of our FSWT Hamiltonian (\ref{Floquet-H'}) for Fig.~\ref{fig:MIT}, we show in Fig.~\ref{fig:assure} the return rate defined in Eq.~(\ref{eq.Loschmidt}) on a finite $L=6$ lattice, using the same parameters chosen in Fig.~\ref{fig:MIT}. We see that the FSWT result Eq.~(\ref{Floquet-H'}), which includes the $\mathcal{O}(J)$ and $\mathcal{O}(J^2)$ correlated hopping terms, matches very well with the exact dynamics throughout the simulation time, while the HFE result obviously deviates from the exact dynamics at early simulation time. 
We find that the $\mathcal{O}(J^2)$ term, Eq.~(\ref{Floquet-H'2-t2}), is vital for the FSWT Hamiltonian (\ref{Floquet-H'}) to describe the ground state and evolution accurately.  

\subsection{The accuracy of FSWT Hamiltonian at low frequencies}\label{appendix:breakdown}
\begin{figure}[h!]
    \centering
    \includegraphics[width=0.45\textwidth]{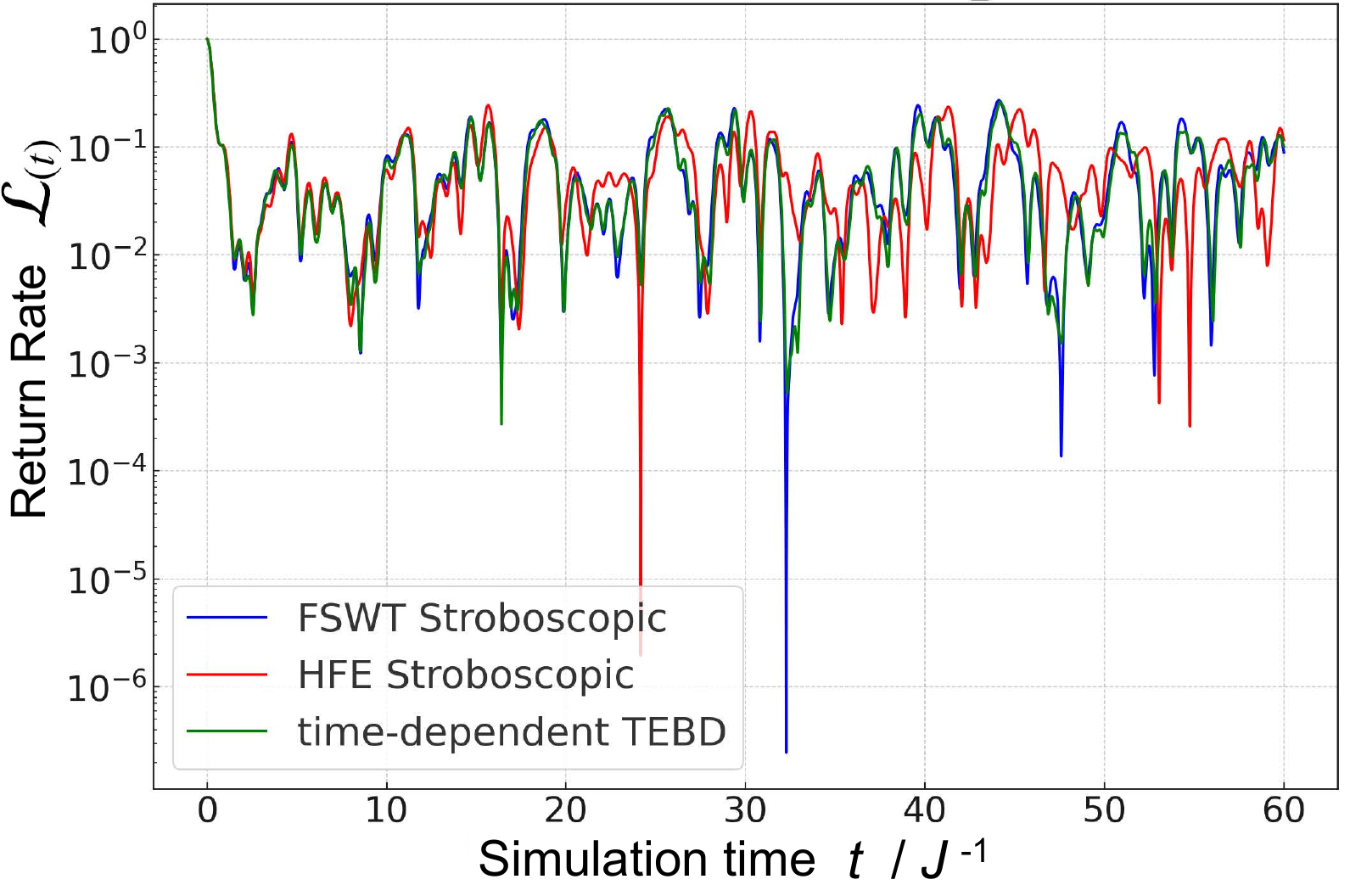}
    \caption{The return rate Eq.~(\ref{eq.Loschmidt}) for $\omega=8.5 J$, $U=3J$ and $g=\omL/4$. Noticeable deviations of FSWT dynamics appear at, e.g., time = 51 $J^{-1}$. }
    \label{fig:breakdown}
\end{figure}

Here in Fig.~\ref{fig:breakdown} we show the return rate dynamics in Fig.~\ref{fig:Evo}(c) for $\omega=8.5 J$, where the FSWT stroboscopic dynamics (which ignores the $\hat{y}_3$ correction) starts to noticeably deviate from the exact dynamics, showing a relative error $\mathcal{E}\sim0.2$ according to Eq.~(\ref{eq.NRMSE}). 

In addition, in Fig.~\ref{fig:Evo}(c), the relative error $\mathcal{E}$ of FSWT stroboscopic dynamics shows several local maxima around $\omL\approx 7J, 7.5J, 8J$ and $9J$. Since the Mott gap is $U=3 J$ for the Hubbard chain simulated in Fig.~\ref{fig:Evo}(c), these local maxima correspond to driving resonance above the Mott gap. Similar above-Mott-gap driving resonance has been reported previously in the linear absorption spectrum of the driven Hubbard clusters \cite{okamoto2021floquet}. For the parameters considered in Fig.~\ref{fig:Evo}(c), these resonances do not significantly affect the accuracy of our FSWT Hamiltonian (\ref{Floquet-H'}).

\begin{figure}[h]
    \centering
    \includegraphics[width=0.45\textwidth]{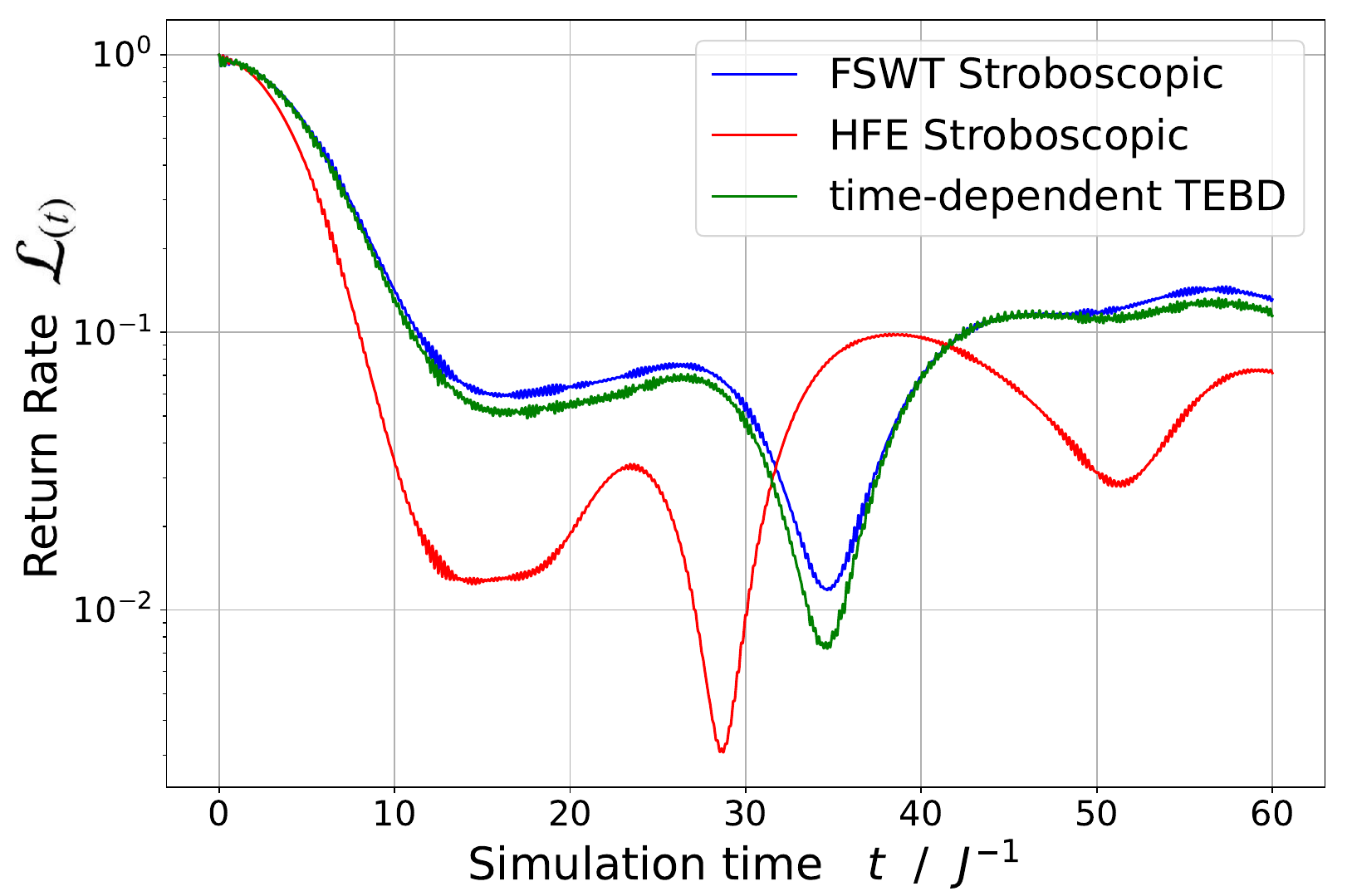}
    \caption{The return rate Eq.~(\ref{eq.Loschmidt}) in the in-gap driving regime, with $L=10$, $\omega=16 J$, $U=28J$ and $g=\omL/4$. }
    \label{fig:in-gap}
\end{figure}

Besides, to benchmark the accuracy of FSWT in the in-gap driving regime where $\omL<U$, in Fig.~\ref{fig:in-gap} we simulate the return rate for $U=28 J$ with other parameters matching Fig.~\ref{fig:Evo}(b). In this in-gap regime, our FSWT result (\ref{Floquet-H'}) again outperforms the lowest-order HFE result, while higher-order lab-frame HFE corrections lead to divergence, rendering them ineffective.

\begin{figure}[h]
    \centering
    \includegraphics[width=0.45\textwidth]{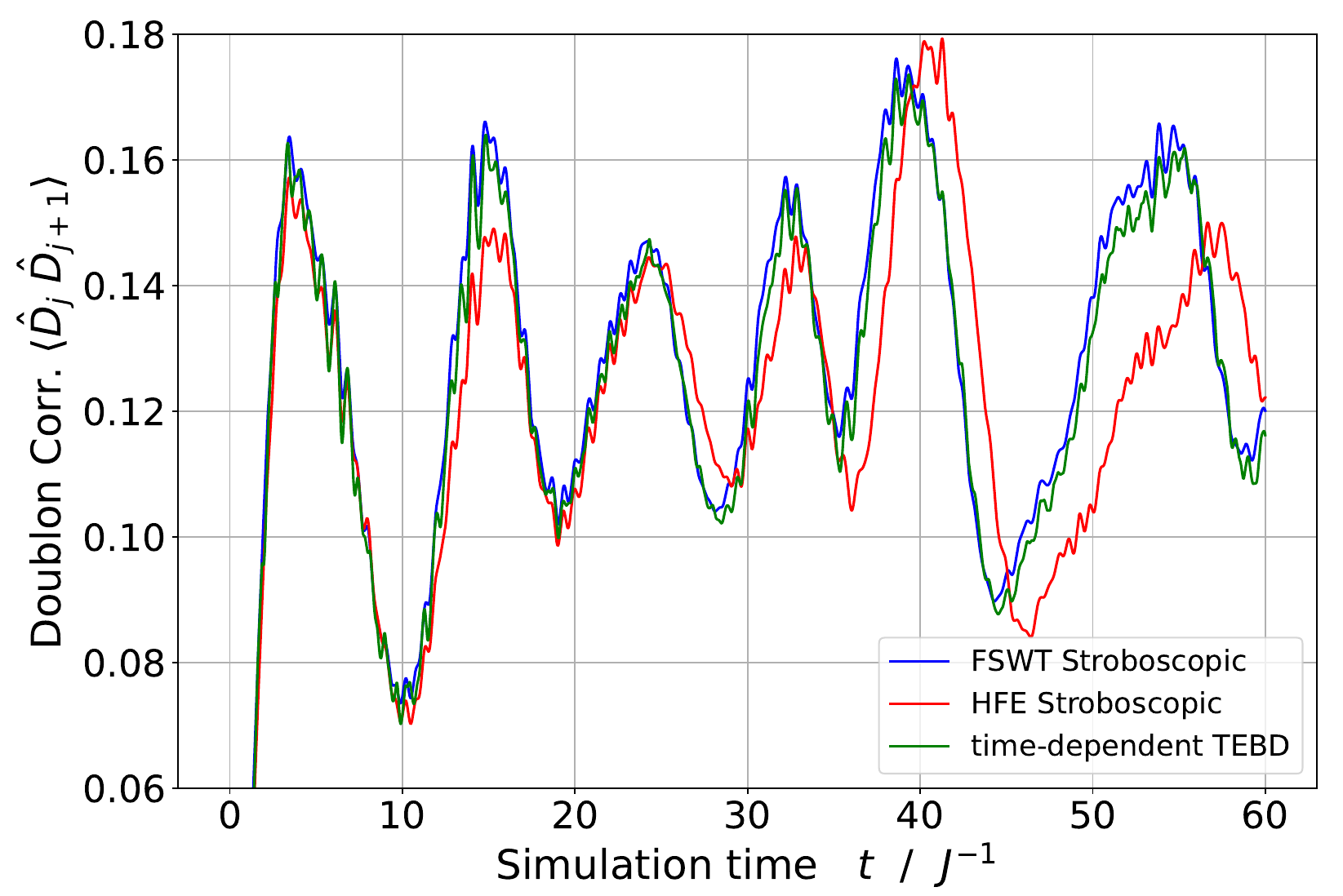}
    \caption{The doublon-doublon correlation $\langle \hat{D}_j \hat{D}_{j+1}\rangle$ for the parameters considered in Fig.~\ref{fig:Evo}(b), with $L=10$, $\omega=16 J$, $U=8J$ and $g=\omL/4$. Here $\hat{D}_{j}=\hat{n}_{j\uparrow}\hat{n}_{j\downarrow}$ is the doublon density operator, and we take $j=L/2$ to observe the doublon-doublon correlation at the centre of the chain. }
    \label{fig:d-d-correlation}
\end{figure}

Finally, we benchmark the accuracy of FSWT in predicting the response of physically observable quantities. In Fig.~\ref{fig:d-d-correlation}, we plot the driven dynamics of the nearest-neighbour doublon-doublon correlation at the centre of the chain, using the same parameters as Fig.~\ref{fig:Evo}(b). We again start the simulation from the product CDW state $\vert \psi_{CDW} \rangle$, thus the nearest-neighbour doublon correlation is initially zero. Compared to the exact TEBD simulation, the FSWT stroboscopic Hamiltonian (\ref{Floquet-H'}) predicts the correlation dynamics with significantly improved accuracy than HFE. This improvement arises solely from the driving-induced $\mathcal{O}(J)$ and $\mathcal{O}(J^2)$ correlated hoppings in Eq.~(\ref{Floquet-H'}).

\subsection{The accuracy of FSWT for other initial states}\label{appendix:other-ini}
\begin{figure}[h]
    \centering
    \includegraphics[width=0.45\textwidth]{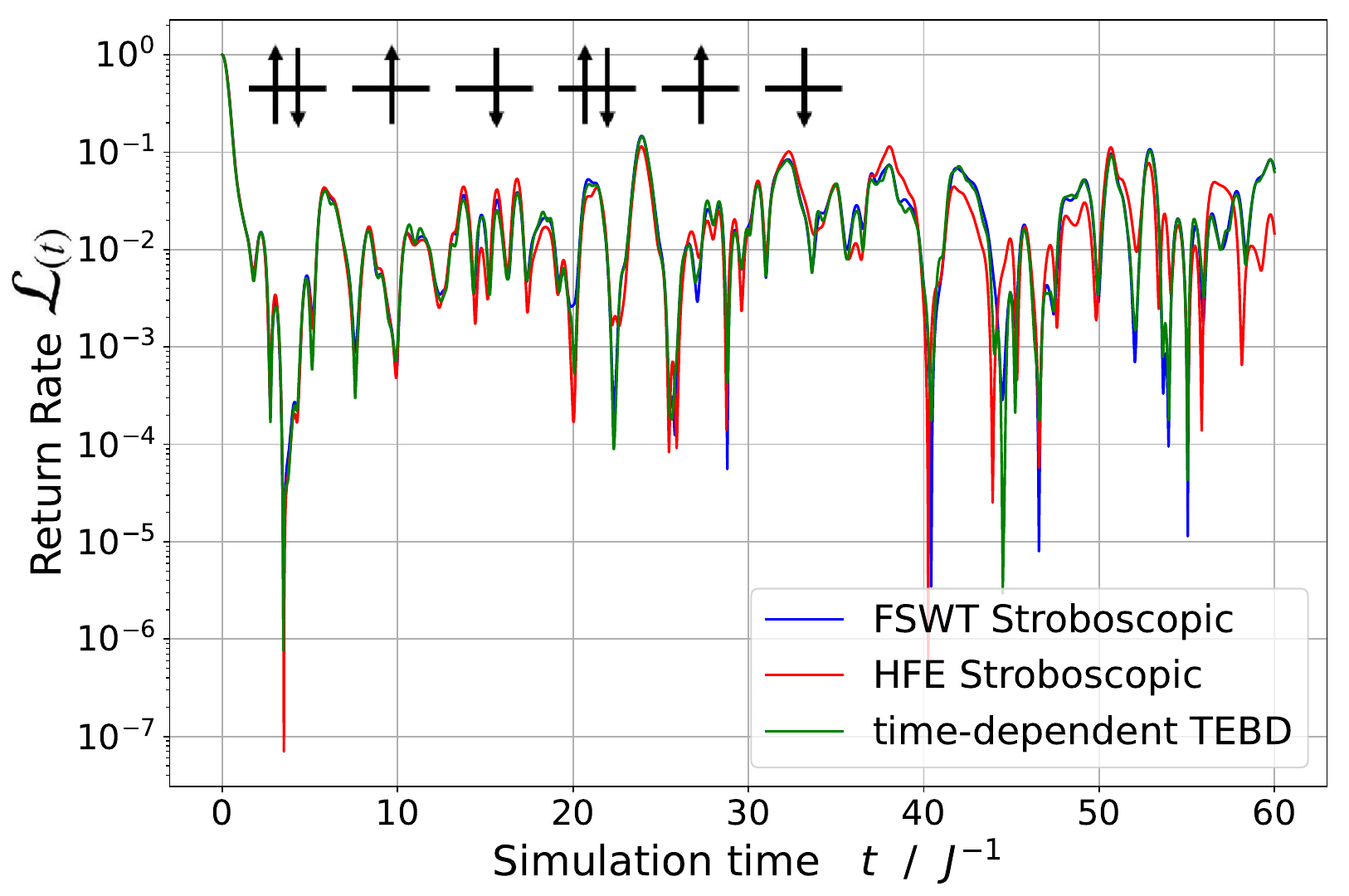}
    \caption{The return rate Eq.~(\ref{eq.Loschmidt}) starting from a non-half-filling state. The parameters match those used in Fig.~\ref{fig:Evo}(c), with $L=6$, $\omega=10 J$, $U=3J$ and $g=\omL/4$. }
    \label{fig:other-initial-states}
\end{figure}

The FSWT result remains valid in predicting the dynamics starting from other initial states. For example, the return rate starting from a far-from-equilibrium product state with 2/3-filling is shown in Fig.~\ref{fig:other-initial-states}. Again, the FSWT stroboscopic dynamics outperforms the HFE result throughout the dynamics.

\newpage
\begin{widetext}
\section{Evaluation of the projection Eq.~(\ref{project_large_U_limit})}\label{appendix:project_large_U_limit}
In the derivation of Eq.~(\ref{project_large_U_limit}), we need to evaluate the projected Hamiltonian $\mathcal{P} \hat{h} \hat{H}'^{(2)} \mathcal{P}$, where $\mathcal{P} \equiv \Pi_{j=1}^{N} (1-\hat{n}_{j,\uparrow}\hat{n}_{j,\downarrow}) $ is the projector on the zero-doublon manifold. According to the definition of projector $\mathcal{P}$, we have 
\begin{equation}
\begin{split}
&\hat{c}_{j,s}^{\dag} \hat{c}_{j+1,s} \hat{n}_{j+1,\bar{s}} \mathcal{P} 
= \hat{c}_{j,s}^{\dag} \hat{c}_{j+1,s} \hat{n}_{j+1,\bar{s}} (1- \hat{n}_{j+1,s} \hat{n}_{j+1,\bar{s}}) \mathcal{P} =0, 
\end{split}
\end{equation}
and similarly $\hat{c}_{j,s}^{\dag} \hat{c}_{j+1,s} \hat{n}_{j,\bar{s}} \hat{n}_{j+1,\bar{s}} \mathcal{P} =0 $. 
Furthermore, since we consider the half-filling case, we have $\hat{n}_{j,\bar{s}} \to 1 - \hat{n}_{j,s} $, which gives
\begin{equation}
\begin{split}
\hat{c}_{j,s}^{\dag} \hat{c}_{j+1,s} \hat{n}_{j,\bar{s}} \mathcal{P} &\to \hat{c}_{j,s}^{\dag} \hat{c}_{j+1,s} (1 - \hat{n}_{j,s}) \mathcal{P} 
= \hat{c}_{j,s}^{\dag} \hat{c}_{j+1,s} \mathcal{P}.
\end{split}
\end{equation}
Thus, when the projector $\mathcal{P} $ is applied to the 
Floquet Hamiltonian~(\ref{Floquet-H'}), correct to $\mathcal{O}(J)$, we have
\begin{equation}
\begin{split}
\hat{H}'^{(2)} \mathcal{P} &= \frac{g^2 J}{\omL^2}
\sum\limits_{s} \sum\limits_{j=1}^{L-1} ( \hat{c}_{j,s}^{\dag} \hat{c}_{j+1,s} + \hat{c}_{j+1,s}^{\dag} \hat{c}_{j,s}   ) 
\left( 1 + \frac{\beta' + \gamma'}{2} (\hat{n}_{j,\bar{s}} +  \hat{n}_{j+1,\bar{s}} ) +  \delta' \hat{n}_{j,\bar{s}} \hat{n}_{j+1,\bar{s}} \right) \mathcal{P} \\
&= \frac{g^2 J}{\omL^2} \sum\limits_{s} \sum\limits_{j=1}^{L-1}
\hat{c}_{j,s}^{\dag} \hat{c}_{j+1,s}  \left( 1 + \frac{\beta' + \gamma'}{2} (\hat{n}_{j,\bar{s}} +  \hat{n}_{j+1,\bar{s}} ) +  \delta' \hat{n}_{j,\bar{s}} \hat{n}_{j+1,\bar{s}} \right) \mathcal{P} \\
&~~~~~~~~~~~~~~~~~   + \hat{c}_{j+1,s}^{\dag} \hat{c}_{j,s}   \left( 1 + \frac{\beta' + \gamma'}{2} (\hat{n}_{j,\bar{s}} +  \hat{n}_{j+1,\bar{s}} ) +  \delta' \hat{n}_{j,\bar{s}} \hat{n}_{j+1,\bar{s}} \right) \mathcal{P} \\
&= \frac{g^2 J}{\omL^2} \sum\limits_{s} \sum\limits_{j=1}^{L-1} (\hat{c}_{j,s}^{\dag} \hat{c}_{j+1,s} + \hat{c}_{j+1,s}^{\dag} \hat{c}_{j,s})
\left( 1 + \frac{\beta' + \gamma'}{2}  \right) \mathcal{P} \\
&= -\frac{g^2 }{\omL^2} \left( 1 + \frac{\beta' + \gamma'}{2}  \right) ~~ \times ~~ \hat{h}
 \mathcal{P}, \\
\end{split}
\end{equation}
and therefore
\begin{equation}\label{appendix-project_large_U_limit}
\begin{split}
\mathcal{P} (\hat{h}+\hat{H}'^{(2)}) \frac{1}{-U} (\hat{h}+\hat{H}'^{(2)}) \mathcal{P} 
&\approx \frac{1}{-U} \mathcal{P} \hat{h}^2 \mathcal{P} + \frac{1}{-U} \mathcal{P} \hat{h} \hat{H}'^{(2)} \mathcal{P} + \frac{1}{-U} \mathcal{P}  \hat{H}'^{(2)} \hat{h} \mathcal{P}  \\
&\approx \frac{1}{-U} \mathcal{P} \hat{h}^2 \mathcal{P} 
\times \left( 1 -2 \frac{g^2 }{\omL^2} \left( 1 + \frac{\beta' + \gamma'}{2}  \right)  \right)  \\
&= \bigg(  \frac{4 J^2}{U} (1-2\frac{g^2}{\omL^2}) + 4  \frac{g^2 J^2 }{\omL^2} \left( \frac{1}{U-\omL} + \frac{1}{\omL+U} \right)  \bigg) 
\times \sum\limits_{j=1}^{L-1} {\bf S}_j \cdot {\bf S}_{j+1}. 
\end{split} 
\end{equation}
\end{widetext}

\end{document}